\documentclass[nonacm]{acmart}

\usepackage{xcolor}  
\usepackage{epigraph}
\usepackage{amsthm,amsmath,amssymb}
\usepackage{hyperref}
\usepackage{afterpage,array}
\usepackage[title]{appendix}
\usepackage[shortlabels]{enumitem}
\usepackage{multirow,rotating,rotfloat}
\usepackage{subcaption}
\usepackage[flushleft]{threeparttable}
\usepackage{tcolorbox}
\usepackage{booktabs}
\usepackage{lscape}
\usepackage{soul}
\usepackage{natbib}
\usepackage{titlecaps}
\usepackage{float}
\usepackage{tikz}
\usetikzlibrary{shapes.geometric}
\usepackage{xspace}
\usepackage{tabularx}
\usepackage{array}

\newcommand{\ai}{AI\xspace}
\newcommand{\xai}{XAI\xspace}
\newcommand{\ml}{ML\xspace}
\newcommand{\llm}{LLM\xspace}
\newcommand{\llms}{LLMs\xspace}
\newcommand{\hci}{HCI\xspace}

\newcommand{\lime}{LIME\xspace}
\newcommand{\shap}{SHAP\xspace}

\newcommand{\mlLong}{machine learning\xspace}

\newcommand{\llmLong}{large language model\xspace}

\newcommand{\pdpsLong}{partial dependence plots\xspace}

\newcommand{\XaiFull}{Explainable artificial intelligence (XAI)\xspace}

\newcommand{\pdpsFull}{partial dependence plots (PDPs)\xspace}
\newcommand{\limeFull}{local interpretable model-agnostic explanations (LIME)\xspace}

\mathchardef\mhyphen="2D

\newenvironment{hypothesis}{\trivlist\item\relax}{\endtrivlist}

\definecolor{NoExplColor}{HTML}{B0B0B0}      
\definecolor{WithExplColor}{HTML}{5B9BD5}    

\DeclareRobustCommand{\NoExpl}{\tikz[baseline=-0.5ex]\node[draw=NoExplColor!80!black, fill=NoExplColor!20, rounded corners=1pt, inner sep=2pt, line width=0.6pt, font=\footnotesize\bfseries] {\textcolor{black}{N}};\xspace}
\DeclareRobustCommand{\WithExpl}{\tikz[baseline=-0.5ex]\node[draw=WithExplColor!80!black, fill=WithExplColor!20, rounded corners=1pt, inner sep=2pt, line width=0.6pt, font=\footnotesize\bfseries] {\textcolor{black}{E}};\xspace}

\definecolor{ScenOneColor}{HTML}{56B4E9}     
\definecolor{ScenTwoColor}{HTML}{E69F00}     
\definecolor{ScenThreeColor}{HTML}{CC79A7}   
\definecolor{ScenFourColor}{HTML}{009E73}    
\definecolor{ScenFiveColor}{HTML}{D55E00}    
\definecolor{ScenSixColor}{HTML}{0072B2}     
\definecolor{ScenSevenColor}{HTML}{F0E442}   

\DeclareRobustCommand{\ScenOne}{\tikz[baseline=-0.5ex]\node[draw=ScenOneColor!80!black, fill=ScenOneColor!20, rounded corners=1pt, inner sep=2pt, line width=0.6pt, font=\footnotesize\bfseries] {\textcolor{black}{1}};\xspace}
\DeclareRobustCommand{\ScenTwo}{\tikz[baseline=-0.5ex]\node[draw=ScenTwoColor!80!black, fill=ScenTwoColor!20, rounded corners=1pt, inner sep=2pt, line width=0.6pt, font=\footnotesize\bfseries] {\textcolor{black}{2}};\xspace}
\DeclareRobustCommand{\ScenThree}{\tikz[baseline=-0.5ex]\node[draw=ScenThreeColor!80!black, fill=ScenThreeColor!20, rounded corners=1pt, inner sep=2pt, line width=0.6pt, font=\footnotesize\bfseries] {\textcolor{black}{3}};\xspace}
\DeclareRobustCommand{\ScenFour}{\tikz[baseline=-0.5ex]\node[draw=ScenFourColor!80!black, fill=ScenFourColor!20, rounded corners=1pt, inner sep=2pt, line width=0.6pt, font=\footnotesize\bfseries] {\textcolor{black}{4}};\xspace}
\DeclareRobustCommand{\ScenFive}{\tikz[baseline=-0.5ex]\node[draw=ScenFiveColor!80!black, fill=ScenFiveColor!20, rounded corners=1pt, inner sep=2pt, line width=0.6pt, font=\footnotesize\bfseries] {\textcolor{black}{5}};\xspace}
\DeclareRobustCommand{\ScenSix}{\tikz[baseline=-0.5ex]\node[draw=ScenSixColor!80!black, fill=ScenSixColor!20, rounded corners=1pt, inner sep=2pt, line width=0.6pt, font=\footnotesize\bfseries] {\textcolor{black}{6}};\xspace}
\DeclareRobustCommand{\ScenSeven}{\tikz[baseline=-0.5ex]\node[draw=ScenSevenColor!80!black, fill=ScenSevenColor!20, rounded corners=1pt, inner sep=2pt, line width=0.6pt, font=\footnotesize\bfseries] {\textcolor{black}{7}};\xspace}

\newcommand{\supported}{\textcolor{green!60!black}{\checkmark}}

\acmDOI{}
\acmISBN{}
\acmConference[]{}{}{}
\acmYear{}
\copyrightyear{}

\title{Can AI Help Explain AI? Effects of Multimodal LLM-Generated Explanations on Perceived Understanding and Perceived Trust in XAI Visualizations}

\author{Mario Truss}
\affiliation{%
  \institution{Adobe}
  \city{Hamburg}
  \country{Germany}
}
\email{mtruss@adobe.com}

\author{Marc Schmitt}
\affiliation{%
  \institution{University of Oxford}
  \city{Oxford}
  \country{UK}
}
\email{marc.schmitt@cs.ox.ac.uk}

\author{Stefan Feuerriegel}
\affiliation{%
  \institution{LMU Munich \& Munich Center for Machine Learning}
  \city{Munich}
  \country{Germany}
}
\email{feuerriegel@lmu.de}

\begin{document}

\begin{teaserfigure}
    \includegraphics[width=\textwidth]{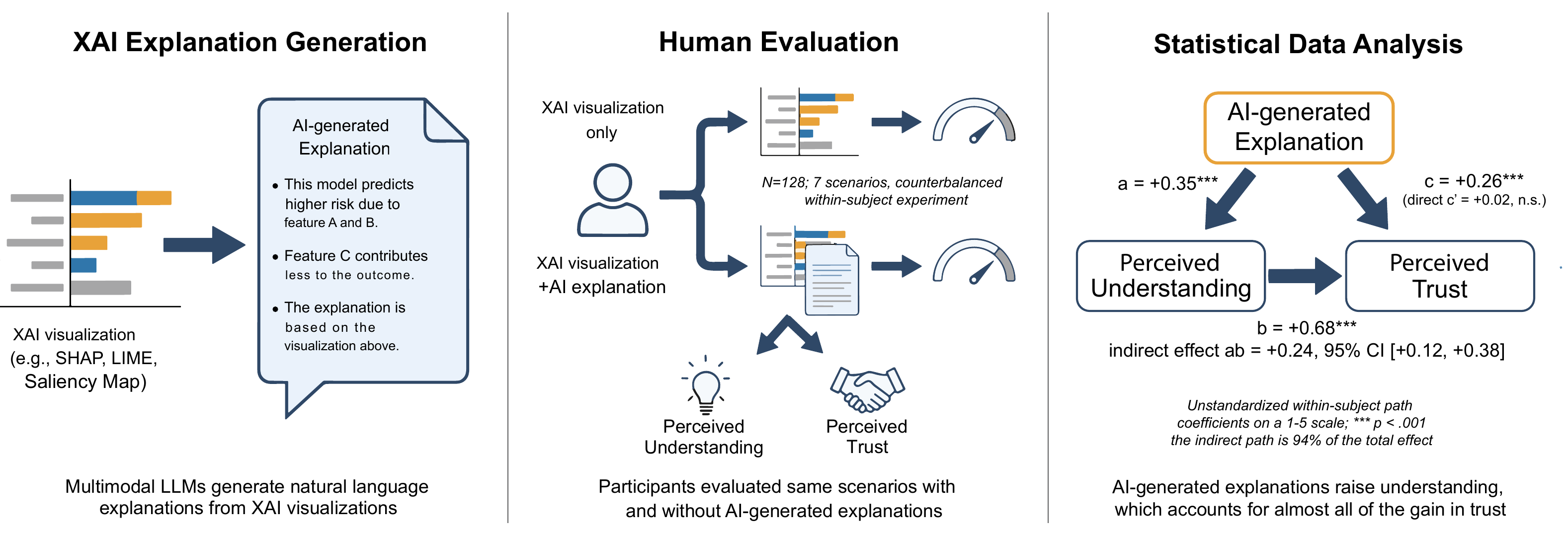}
    \caption{Overview of the study approach. Multimodal \llms generate natural-language explanations from \xai visualizations together with source-derived context (panel~1); a pre-registered within-subject experiment then compares user responses to the same visualizations with and without those explanations (panel~2); panel~3 decomposes the resulting \textit{Perceived Trust} gain into the part that runs through \textit{Perceived Understanding}, in points of the five-point response scale (via a mediation analysis).}
    \Description{A three-panel schematic summarizing the study. The left panel shows an \xai visualization, such as a \shap, \lime, or saliency map output, feeding into a speech bubble that contains a short bulleted natural-language explanation; a multimodal \llmLong produces that explanation from the visualization together with the source-derived task and model context. The middle panel shows a single participant icon branching into two paths, one presenting the \xai visualization alone and one presenting the same visualization together with an AI-generated explanation, each path ending in a rating dial, with icons below for the two primary outcome measures: Perceived Understanding and Perceived Trust. A note states that the experiment was a counterbalanced within-subject design with 128 participants and seven scenarios. The right panel shows a mediation diagram in which a box labelled AI-generated explanation points to both Perceived Understanding and Perceived Trust, and Perceived Understanding in turn points to Perceived Trust. The paths are labelled: a equals plus 0.35 from the explanation to Perceived Understanding, b equals plus 0.68 from Perceived Understanding to Perceived Trust, and a total effect c of plus 0.26 from the explanation to Perceived Trust, all three significant at p below .001, alongside a remaining direct effect c prime of plus 0.02 that is not significant. A line below reports the indirect effect of the explanation on Perceived Trust through Perceived Understanding as plus 0.24 with a 95 percent bootstrap confidence interval from 0.12 to 0.38, and a footnote states that these are unstandardized within-subject path coefficients on the one-to-five response scale and that this indirect path represents 94 percent of the total effect descriptively. Read together, explanations raise Perceived Understanding, and the association between explanations and Perceived Trust runs mainly through that gain in Perceived Understanding.}
    \label{fig:teaser}
\end{teaserfigure}

\begin{abstract}
\XaiFull aims to make \mlLong models understandable, yet many \xai visualizations remain difficult for end-users to interpret. Here, we evaluate whether augmenting XAI outputs with textual explanations from multimodal \llms improves perceived understanding and perceived trust. We conducted a pre-registered within-subject experiment ($N$ = 128) across seven decision scenarios and five \xai methods, with and without LLM-generated explanations. The explanations improved perceived understanding and perceived trust by 0.35 and 0.26 points on five-point scales (both $p$ < .001). The gain appeared in all seven scenarios, did not differ between local and global explanations, and was largest for participants with the least prior \xai exposure. A mediation analysis indicates an indirect effect on perceived trust through perceived understanding. These results suggest that multimodal \llms can provide a useful interpretive layer for \xai beyond expert audiences.
\end{abstract}

\maketitle



\section{Introduction}

\XaiFull is now widely used in practice to provide insights into how models reach conclusions \cite{ehsan2021who, Wangetal2021, zhang2021relatable, Bilaletal2025}. While originally developed to support model debugging and validation, \xai outputs are increasingly used by managers, auditors, and decision-makers to support oversight, justify automated recommendations, build trust, and enable accountable decision-making. Yet, the outputs of many \xai approaches remain difficult to interpret in practice, especially for end-users \cite{bobekUsercentricEvaluationExplainability2025}.

Over the past decade, many \xai methods have been developed, including \shap \cite{lundbergUnifiedApproachInterpreting2017, lundbergLocalExplanationsGlobal2020,senonerUsingExplainableArtificial2022}, \lime \cite{ribeiroWhyShouldTrust2016, salihPerspectiveExplainableArtificial2025}, \pdpsLong \cite{friedmanGreedyFunctionApproximation2001, kruschelChallengingPerformanceInterpretabilityTradeOff2025}, and saliency-based approaches
\cite{selvarajuGradCAMVisualExplanations2020, senonerExplainableAIImproves2024a}. These techniques can broadly be categorized into (i) local explanations, which focus on individual predictions, and (ii) global explanations, which describe overall model behavior. However, many of these approaches have been developed primarily with an emphasis on model fidelity and technical faithfulness rather than human interpretability. \emph{As a result, explanations that are mathematically rigorous are not necessarily understandable to end-users.}

In practice, \xai visualizations can be challenging to understand and interpret, especially for end-users such as managers and other decision-makers. Many \xai approaches combine feature importance rankings, instance-level attributions, and baseline comparisons within a single visualization. Further, such visualizations span different aggregation levels, such as instance-level predictions and global patterns, which require users to reconcile multiple layers of abstraction. For example, a \shap summary plot (``beeswarm'') displays the distribution of feature attributions across an entire dataset, while encoding feature importance, directionality, magnitude, and data density simultaneously through position and color \cite{lundbergLocalExplanationsGlobal2020}. Integrating these signals into a coherent mental model can be challenging for end-users. Empirical research shows that users frequently misinterpret attribution magnitudes, confuse explanation levels, or overgeneralize from complex visual encodings \cite{Wangetal2021, Wangetal2022, ehsan2021who, poursabzi2021manipulating, humerReassuringMisleadingDebunking2024}.

In this paper, we use multimodal \llms to generate explanations of \xai visualizations in natural language. Our work is motivated by prior work applying generative \ai to \mlLong models, yet with \emph{different} objectives; existing works are primarily \textit{methodological}. One stream employs \llms to explain the model behavior directly. For instance, systems such as \textit{TalkToModel} \cite{slackExplainingMachineLearning2023} allow users to query \mlLong models through a conversational interface. However, such systems operate on the underlying model and data in response to user-initiated queries, so the interpretive work starts from the model rather than from an \xai visualization that a user has already been given. Another stream integrates \llms with explanation techniques to translate \xai outputs into text. For example, \textit{LangXAI} demonstrates that vision language models can generate textual descriptions of \xai visualizations \cite{nguyenLangXAI2024}, and related works propose the broader idea of using \llms to “explain explanations” with a methodological focus on preserving the explanatory faithfulness \cite{Zyteketal2024, zytekExplingoExplainingAI2024,belloThreelevelFrameworkLLMenhanced2025}. 

In contrast, our focus is \textit{not} methodological but \emph{behavioral}: we aim to test how users respond to AI-generated explanations of \xai visualizations. To the best of our knowledge, ours is the first controlled study to test the effects of such explanations on Perceived Understanding and Perceived Trust when they are generated by a multimodal \llm from a diverse set of rendered \xai visualizations. 

From a human–computer interaction (\hci) perspective, explanations support sense-making by shaping users’ mental models and calibrating perceived trust. Research in human–\ai interaction suggests that perceived understanding is a key antecedent of appropriately calibrated perceived trust and effective oversight \cite{ibrahimExplanationsImproveQuality2023, chenWhenExplainExploring2024}. By translating complex visual encodings into structured narratives, multimodal \llms may reduce cognitive load and help users integrate multiple explanation signals into a coherent interpretation. Whether this translation meaningfully improves perceived understanding and perceived trust, particularly for end-users, remains an open empirical question. We thus address the following \textbf{research questions (RQs):}
\begin{itemize}
    \item[\textbf{RQ1:}] \emph{How do LLM-generated textual explanations of \xai visualizations influence end-users’ perceived understanding of how an \ai system makes decisions?}
    \item[\textbf{RQ2:}] \emph{How do LLM-generated explanations of \xai visualizations affect end-users’ perceived trust in AI-assisted decisions?}
    \item[\textbf{RQ3:}] \emph{To what extent does the perceived understanding mediate the relationship between LLM-generated explanations and end-users’ perceived trust?}
\end{itemize}

To answer these RQs, we conducted a pre-registered within-subject experiment ($N$ = 128). Our research design is shown in Figure~\ref{fig:teaser}. Participants evaluated \xai visualizations under two conditions: (a)~\xai visualizations alone and (b)~\xai visualizations augmented with LLM-generated explanations generated in natural language. We measured \emph{Perceived Understanding} and \emph{Perceived Trust} in the \ai system, page interaction time, and perceived helpfulness of the explanations. Explanations were generated by a state-of-the-art multimodal \llm that reads the \xai visualization images directly. We compare user responses across different settings with seven use cases inspired by real-world applications, spanning different \xai methods and data modalities. Our main analysis is based on GPT-5.6 Sol; to demonstrate generalizability, we also demonstrate robustness in a validation sample ($N$ = 63; using GPT-4o).

Overall, our \textbf{main findings} are\footnote{Code and data are provided for review via https://anonymous.4open.science/r/canaiexplainai-SEPT2026.}:
\begin{enumerate}
    \item We demonstrate that LLM-generated explanations for \xai visualizations significantly improve \textit{Perceived Understanding} ($\beta$ = 0.35 points on a 5-point-scale, $p$ < .001) and \textit{Perceived Trust} ($\beta$ = 0.26 points, $p$ < .001). As a result, we offer systematic evidence across seven \xai visualization types that automated textual explanations can improve \xai usability.
    \item We observed positive effects in all seven scenarios, which spanned several \xai methods (\shap, \lime, saliency maps, etc.) and data modalities (tabular, text, image); however, we also find heterogeneity across participants (e.g., those with less prior exposure to \xai gained more).
    \item We provide evidence that the pathway linking explanations to \textit{Perceived Trust} is  through \textit{Perceived Understanding}.
\end{enumerate}

\section{Related Work}

Prior work on explanations in \ai systems can be organized into three main research streams: (1)~human evaluation of traditional \xai methods; (2) LLM-based systems that explain model behavior directly; and (3)~LLM-based systems that translate \xai visualizations into natural language. Yet, our literature review below shows that substantial gaps remain in bridging technical \xai capabilities with end-user comprehension.

\subsection{The Tension Between XAI and Perceived Understanding}

Our work builds upon a growing body of research exploring how users perceive and understand \xai.\footnote{For a detailed overview of \xai methods, see recent surveys \cite{Bilaletal2025, wilkinsonExplainingExplainabilityComprehensive2026}.} In principle, the goal of \xai is to increase transparency; yet, substantial evidence demonstrates the challenges users face in interpreting these explanations. A mixed-methods study of users with and without an \ai background \cite{ehsan2021who} found that both groups placed unwarranted faith in numerical explanations and derived value from explanations in ways their designers had not intended. This reveals a gap between the development of \xai methods and perceived understanding of their real-world utility. Recent surveys on human-centered \xai \cite{haqueExplainableArtificialIntelligence2023,severesHumanSideXAI2023} provide further supporting evidence of this pattern and highlight that the majority of \xai research prioritizes technical performance metrics over user comprehension. But the complexity of many \xai visualizations can lead to misunderstandings, particularly among end-users, because the effectiveness of an explanation is highly dependent on the user's background and expertise \cite{zhang2021relatable}.

Research on explanations in AI-assisted decision-making has produced mixed findings regarding the effectiveness of such explanations. A meta-analysis synthesizing results across multiple \xai studies found a statistically significant, positive impact of explanations on user performance \cite{schemmerMetaAnalysisUtilityExplainable2022}, though the effect seems to depend on explanation type, task characteristics, and user expertise \cite{Wangetal2021, Wangetal2022, schoeffer2023explanations}. Appropriate reliance (i.e., correctly accepting accurate \ai advice while rejecting incorrect advice) is often seen as one potential reason for this \cite{schemmerAppropriateRelianceAI2023}. All in all, this literature stream shows that simply providing explanations does \emph{not} guarantee improved outcomes; rather, the design of explanations matters. 

The utility of explanations further varies with user expertise. Among the few studies that empirically tested \xai with domain experts in real-world tasks is Senoner et al. \cite{senonerExplainableAIImproves2024a}, where the experiments showed that saliency maps improved task performance in manufacturing defect detection and medical imaging. Similarly, recent work in radiology found that text-based explanations from \llms improved diagnostic accuracy, compared to no \llm support \cite{spitzerEffectMedicalExplanations2026}. This is mirrored by current guidelines for responsible \ml deployment in healthcare that underline the need for interpretability to enable clinician oversight and error correction \cite{wiensNoHarmRoadmap2019, beamChallengesReproducibilityMachine2020}. However, these studies focused on domain experts who interpret \xai; yet, what remains unclear is whether end-users without domain expertise can benefit from LLM-generated textual explanations of \xai visualizations.

Research on trust in \ai systems reveals a complex relationship between explanations and user confidence. While transparency through explanations is theorized to support calibrated trust \cite{Wangetal2021}, empirical findings show that trust depends on multiple factors, including explanation accuracy, user expertise, and task context \cite{humerReassuringMisleadingDebunking2024}. Recent work suggests that poorly designed or inaccurate explanations can actually decrease perceived trust and lead to over-reliance or under-reliance on \ai systems \cite{schoeffer2023explanations, morrisonImpactImperfectXAI2024}. Studies examining ``imperfect'' \xai demonstrate that incorrect explanations, even when accompanying correct \ai advice, can create a misinformation effect, thereby causing users to infer flawed reasoning strategies that persist post-collaboration \cite{morrisonImpactImperfectXAI2024}.

Taken together, a central limitation across the above research stream is that most studies focus on technical users (data scientists, engineers) and on the broader range of stakeholders such as managers and domain experts, who act as end-users of \xai and who increasingly interact with \ai systems in practice (which often have different explanation needs; see \cite{liao2024designerly, siposIdentifyingExplanationNeeds2023, sureshExpertiseRolesFramework2021, brennenWhatPeopleReally2020}). Furthermore, studies with \ml practitioners reveal a substantial gap between the outputs of \xai research (i.e., \xai visualizations) and the real-world needs of practitioners (i.e., perceived trust and perceived understanding of \ai models), especially because many existing \xai tools are perceived as too technical for end-users \cite{siposIdentifyingExplanationNeeds2023, ehsan2021who}. While these studies establish that \xai visualizations can be difficult to interpret, they do \emph{not} offer strategies on how to make \xai more accessible to end-users.

\subsection{Methods for Translating Model Predictions into Natural Language}

A growing line of research uses LLMs to explain model behavior directly, without relying on traditional \xai methods. The goal is to translate complex model outputs into summaries in natural language that are easier to understand \cite{yang2022psychological, Bilaletal2025}. Prior work suggests that these approaches can improve how users interact with \ai systems. For example, chain-of-thought explanations have been shown to improve diagnostic performance in medical settings compared to standard model outputs \cite{spitzerEffectMedicalExplanations2026}. Other studies from different contexts show that LLM-generated descriptions of \mlLong models can increase user trust and engagement \cite{Gyawalietal2025, Bokstalleretal2025, fengContextualizingRecommendationExplanations2025} and indicate that narrative explanations help users make sense of the underlying behavior \cite{ehsan2019automated}. In a similar context, \llms are also researched as evaluators of \ai explanation quality  \cite{chan2024large}.

However, this stream faces an important limitation: when \llms explain \mlLong models directly, the explanations are \textbf{not} grounded in established \xai concepts such as feature attribution or feature importance. As a result, the generated explanations lack faithfulness to the underlying model logic. In addition, explanations that are not connected to established \xai frameworks can make communication with \mlLong engineers and other technical stakeholders more difficult, as engineers are typically familiar with widely-used \xai methods, but not the interpretations of \ai model artifacts in a business context. These limitations motivate our approach to develop explanations for established \xai visualization techniques (rather than directly for the underlying ML model). 

\subsection{LLMs as an Explanatory Layer for XAI}

A different stream applies \llms to traditional \xai techniques to translate complex \xai outputs into accessible text. This approach aims to preserve the rigor of established explanation methods while improving comprehensibility for end-users. Part of this work is methodological and demonstrates technical feasibility across different explanation types \cite{Mavrepisetal2024, slackExplainingMachineLearning2023}. For example, \emph{LangXAI} \cite{nguyenLangXAI2024} presents a multimodal approach based on vision-language models that generates textual explanations directly from \xai visualizations.  
\emph{Mind the XAI Gap} \cite{Paraschouetal2025} compares its \llm framework against \lime in a study with 56 participants and reports higher ease of use, efficiency, clarity, and human-friendliness for the generated explanations, but not perceived understanding. \emph{XAIstories} \cite{martensTellMeStory2025} uses LLMs to generate narratives from \shap values and counterfactual explanations and evaluates these with users. However, their work focuses outcomes such as convincingness, likeliness to use, speed, or confidence, but neither perceived understanding nor perceived trust. As such, this work suggests that narrative design elements can support sense-making around XAI use and thus presents the starting point motivating our analysis of whether LLM-generated explanations can improve understanding and trust.

\begin{table}[t]
\centering
\caption{Summary of key literature streams relevant to our study and the resulting research gap in human-centered \xai.}
\label{tab:literature_gap}

\scriptsize
\renewcommand{\arraystretch}{1.08}
\setlength{\tabcolsep}{2.5pt}

\begin{tabularx}{\columnwidth}{
    @{}
    >{\raggedright\arraybackslash}p{1.6cm}
    >{\raggedright\arraybackslash}X
    >{\raggedright\arraybackslash}X
    @{}
}
\toprule
\textbf{Research stream} &
\textbf{What prior works show} &
\textbf{Research gap} \\
\midrule

\textbf{Human evaluation of \xai}
&
Prior work shows that \xai visualizations can be difficult for end-users to interpret
\cite{ehsan2021who, Wangetal2021}. Classical \xai methods have limitations when users lack
domain expertise, as explanations may not improve understanding, uncertainty awareness,
or Trust calibration on unfamiliar tasks \cite{Wangetal2022}.
&
It remains unclear whether LLM-generated explanations in natural language can provide
interpretive scaffolding for end-users and support sense-making of \xai visualizations.
\\
\addlinespace[2pt]

\textbf{LLM-based translation of \xai into text}
&
Recent work demonstrates that \llms (e.g., vision-language models) can translate technical
\xai outputs or visualizations into accessible text
\cite{nguyenLangXAI2024, Mavrepisetal2024, Paraschouetal2025}, establishing the feasibility
of using \llms as an explanatory layer for \xai.
&
Human evaluations in this stream target explanation quality, ease of use, or summary
accuracy. Perceived understanding and perceived trust have not been measured jointly
for the same visualization with and without generated text.
\\
\addlinespace[2pt]

\textbf{Preference, task performance, and perceived outcomes}
&
Studies of LLM-generated explanation systems report that users often prefer natural
language explanations over raw plots or technical outputs \cite{Mavrepisetal2024}, and
conversational systems have been shown to improve objective task performance
\cite{slackExplainingMachineLearning2023}.
&
Preference does not necessarily imply improved perceived understanding or perceived trust calibration.
Controlled studies are needed to compare \xai visualizations with and without additional
explanations in terms of perceived understanding and perceived trust.
\\
\addlinespace[2pt]

\textbf{Passive vs. interactive delivery}
&
Interactive systems such as dialogue-based explanation tools allow users to query models
and obtain explanations in natural language
\cite{slackExplainingMachineLearning2023}. However, such systems assume that users know
which questions to ask.
&
It remains unclear whether passively presented AI-generated explanations are effective
for end-users encountering pre-generated \xai visualizations in decision-support settings.
\\
\addlinespace[2pt]

\textbf{Generalizability across \xai types}
&
Prior work often focuses on specific explanation formats, individual systems, or selected
data types \cite{nguyenLangXAI2024, Paraschouetal2025}.
&
Evidence is needed on whether the benefits of LLM-generated explanations generalize
across established \xai visualization types, domains, and data modalities.
\\

\bottomrule
\end{tabularx}
\end{table}
\Description{A five-row literature table with three columns: Research stream, What prior works show, and Research gap. Rows cover: (1) Human evaluation of XAI — prior work shows XAI visualizations are difficult for end-users to interpret and have limitations for non-experts; gap is whether LLM-generated explanations can provide interpretive scaffolding. (2) LLM-based translation of XAI into text — LLMs can translate XAI outputs into accessible text; gap is that perceived understanding and trust have not been measured jointly for the same visualization with and without generated text. (3) Preference, task performance, and perceived outcomes — users prefer natural-language explanations and conversational systems improve task performance; gap is that preference does not necessarily imply improved understanding or trust calibration. (4) Passive vs. interactive delivery — interactive systems allow users to query models; gap is unclear whether passively presented AI-generated explanations are effective for pre-rendered visualizations in decision-support settings. (5) Generalizability across XAI types — prior work focuses on specific formats and data types; gap is evidence needed on whether benefits generalize across established XAI visualization types, domains, and data modalities.}

\subsection{The Need for Human-Centered Explanations of XAI}

Taken together, prior work establishes that \xai outputs can be difficult for end-users to interpret and that \llms can translate technical model information into more accessible natural language. Existing evaluations, however, primarily examine explanation quality, preference, task performance, or interactive systems in which users actively query the model. It remains unclear whether adding an LLM-generated textual explanation to an already-rendered \xai visualization improves users' perceived understanding and perceived trust. We therefore evaluate these outcomes directly across seven \xai scenarios and additionally examine the role of perceived trust as a mediator. 

\section{Hypotheses}
\label{sec:hypotheses}
Based on the research gaps identified above, we developed three hypotheses that directly address our research questions. Our central assumption is that \llm-generated explanations can act as a boundary artifact \cite{starInstitutionalEcologyTranslations1989, leeBoundaryNegotiatingArtifacts2007, boundaryobjectCHI} between technical \xai visualizations and \xai visualization end-users who might have varying levels of \ai and \xai understanding. The reasoning behind this assumption is that prior research suggests that, in general, textual explanation can help users understand \ai models better \cite{miller2019explanation, liao2020questioning, zhang2021relatable}. However, there seems to be varying effects depending on prior \ai experience and domain knowledge \cite{Wangetal2021}, and there is a risk that more transparency could also lead to cognitive overload \cite{poursabzi2021manipulating, sweller1988cognitive}. Against this background, we propose the following hypotheses:

\begin{hypothesis}
\textbf{H1:} \textit{Perceived Understanding of \ai model is increasing when participants view \xai visualizations with LLM-generated textual explanations compared to visualizations alone.}
\end{hypothesis}

Beyond perceived understanding, perceived trust in \ai systems is deemed to be a prerequisite for successful human-AI collaboration. The relationship between \xai and trust is characterized by mixed findings. On one hand, trust is closely linked to transparency and interpretability. Previous research suggests an increase understanding \cite{chenWhenExplainExploring2024, debritoduarteAITrustCan2023, papagniArtificialAgentsExplainability2023, shinUserPerceptionsAlgorithmic2020}, reliance \cite{Wangetal2021, zhang2020effect, buccinca2021trust}, and over-reliance \cite{buccinca2021trust} through the explainability of \ai via approaches such as \xai visualizations. Further, explanations that clarify the internal decision process of an \ai can calibrate perceived trust by revealing whether predictions are based on meaningful patterns, thereby increasing confidence in valid model reasoning \cite{lai2023towards}. On the other hand, explanations can also \textit{reduce} trust when the explanations expose limitations of the \mlLong model, such as reliance on spurious correlations or high uncertainty in predictions \cite{bansal2021does, green2019principles}. These contradictory effects of explanations (i.e., capable of both increasing and decreasing trust depending on what they reveal) point to a theoretical tension in \xai research. In our case, we hypothesize that LLM-generated explanations will increase trust by because users gain a better understanding through increased explainability, transparency, and interpretability:
\begin{hypothesis}
\textbf{H2:} \textit{Perceived Trust in \ai predictions is increasing when participants view \xai visualizations with LLM-generated textual explanations compared to visualizations alone.}
\end{hypothesis}

The relationship between explainable \ai, perceived trust and perceived understanding is well-motivated, but reflects a complex cognitive process. Previous research shows that a better understanding of \ai can be linked to more trust in the AI model \cite{zhangIntegratedEvaluationFramework2022, zhang2020effect}. Nevertheless trust is a complex process, that lead to \textit{appropriate reliance} \cite{miller2019explanation, lai2023towards, schoefferExplanationsFairnessAppropriate2024} but also over-reliance \cite{buccinca2021trust}, meaning that it could lead to overconfidence. Motivated by this, we later perform a mediation analysis, which allows us to examine the relationship between perceived understanding and perceived trust. If perceived understanding accounts for most of the association with perceived trust, this would be consistent with explanations acting through a comprehension-related pathway, although neither manipulating the mediator nor measuring mental models objectively is possible within this design. Alternatively, a partial mediation would indicate that explanations also affect perceived trust through other behavioral routes. In our analysis, we hypothesize that perceived understanding is the primary mediator of perceived trust:

\begin{hypothesis}
\textbf{H3:} \textit{Perceived Understanding by users mediates the relationship between LLM-generated explanations and Perceived Trust in \ai models.}
\end{hypothesis}

\section{Methodology}
\subsection{Overview}

Our experiment was pre-registered on OSF.\footnote{See  {\url{https://osf.io/9um46/overview?view_only=05fc47b0de284e2da8f8d8d2dec966cb}}} Our methodology is organized along two steps (see Figure~\ref{fig:researchdesign} in Appendix~\ref{app:supp_figures} for a schematic overview) to our answer our hypothesis. First, we generated LLM-based textual explanations from \xai visualizations using a multimodal \llm (Section~\ref{sec:method_explanations}). Second, we conducted a counterbalanced within-subject experiment to assess how LLM-generated explanations affect \textit{Perceived Understanding} and \textit{Perceived Trust} across seven diverse \xai scenarios). The two-step design follows the logic of recent CHI works that evaluate generative \ai interventions in controlled behavioral experiments  \cite{maierPartneringGenerativeAI2026,geisslerDesigningEffectiveDigital2026}. Our reporting of \llm use follows the GUIDE-LLM checklist \cite{feuerriegelReportingChecklistLarge2026}; the completed checklist is provided in Appendix~\ref{app:guide_llm}.

\subsection{Stimuli: LLM-Based Generation of Explanations}
\label{sec:method_explanations}

\textbf{LLM.} \llm-generated explanations were generated with a multimodal vision language model capable of processing \xai visualizations as images and of producing descriptions in natural language. We generated explanations via GPT-5.6 Sol (accessed via the OpenRouter API, \href{OpenRouter API}{https://openrouter.ai/}). GPT-5.6 Sol is a state-of-the-art LLM that achieve state-of-the-art performance in relevant benchmarks \cite{artificialanalysis2026, openrouter2026, arena2026}. We leave parameters at the provider default. Later, we also demonstrate the generalizability of our findings using an alternative LLM (see Section~\ref{sec:validation}). 

\textbf{Prompt.} For explanation generation, we used a structured prompting approach that combined a fixed {six-part-answer structure} with scenario-specific model and task context (see Figure~\ref{fig:prompt}). The prompt design follows best practice for prompt engineering \cite{girayPromptEngineeringChatGPT2023,linHowWriteEffective2024} and prior work on LLM-based explanation generation \cite{Zyteketal2024}. Specifically, the prompt assigns the \llm the role of an expert in \xai communication and asks it to explain each visualization through six components: the \ai use case, the type of \xai visualization, guidance on how to read it, an interpretation of the specific instance, an example or analogy, and the practical implication. The prompt further specifies the intended audience (non-technical users, engineers, and decision-makers), the desired explanatory tone, and a focus on the model's key prediction drivers. 

In addition to the fixed answer structure, we provided scenario-specific context describing the underlying \ml method, prediction task, application domain, data labels, \xai method, and relevant performance metrics. This follows the logic of context engineering, in which task-relevant information is supplied alongside the prompt to improve grounding and reduce unsupported interpretations \cite{meiSurveyContextEngineering2025,huaContextEngineering202025,zhangAGENTICCONTEXTENGINEERING2026}. During prompt development, we also evaluated simpler variants, including a minimal zero-shot prompt and the structured prompt without additional model context. We carefully piloted the prompt and evaluated the results qualitatively; in particular, the context-enriched configuration produced explanations that most closely matched expert-written explanations.

\begin{figure}[htbp]
\begin{tcolorbox}[colback=gray!5!white, colframe=gray!75!black, title=System Prompt: \xai Image Explanation Generation]

You are an expert in Explainable \ai (\xai) communication, tasked with helping non-technical users, engineers, and decision-makers understand how \xai visuals relate to model predictions.

Given an \xai visualization, explain the visualization using the following structure:

\begin{itemize}
    \item \textbf{\ai Model Use Case:} Briefly describe what the \ai is doing (e.g., defect detection in manufacturing).
    
    \item \textbf{Type of \xai Visualization:} Clearly identify the method (e.g., Grad-CAM, heatmap, anomaly map) and define it in simple terms.
    
    \item \textbf{How to Read It:} Describe what the image shows, how colors represent model reasoning (e.g., yellow = important, blue = not relevant), and what this means for the task.
    
    \item \textbf{In This Instance:} Apply the logic to the specific example — what does the highlighted region indicate in this case?
    
    \item \textbf{Example or Analogy:} Provide a practical interpretation or example to help users reason about it (e.g., missing component → flagged as anomaly).
    
    \item \textbf{Implication:} Conclude with how this helps real users (e.g., improved defect detection, human-AI collaboration, fewer false positives).
\end{itemize}

Keep the tone explanatory but precise and concise so it fits into a UI. Use visual literacy cues. Frame the \xai as a decision support tool, not a black-box justification. Support the idea of \ai + human collaboration, not full automation. Include any relevant accuracy improvements (if available from source studies).
\end{tcolorbox}
\caption{The final prompt based on best practices a six-part-answer-structure, which was used for all \xai visualizations; model-specific information was provided in-context per case.}
\Description{A boxed visualization of the system prompt text. It opens by casting the model as an expert in explainable-AI communication addressing non-technical users, engineers, and decision-makers, then asks it to explain a given \xai visualization using six labeled parts: \ai model use case, type of \xai visualization, how to read it, in this instance, example or analogy, and implication. It closes with instructions on tone and stance: keep the tone explanatory but precise, use visual literacy cues, frame the \xai output as a decision-support tool rather than a black-box justification, support human-AI collaboration rather than full automation, and include relevant accuracy improvements where the source study reports them.}
\label{fig:prompt}
\end{figure}

\subsection{Measures}
\label{sec:measures}
We measured two primary outcomes after each scenario using 5-point Likert scales ($1 = $ strongly disagree, $5 = $ strongly agree). (1)~{\textit{Perceived Understanding}} (i.e., \emph{``I understand how the model works''}), and (2)~{\textit{Perceived Trust}} (i.e., \emph{``I have Perceived Trust in the model's prediction''}). Importantly,the \textit{Perceived Trust} item asks about the prediction rather than about the model as a whole or about a decision taken with it. 

We also collected several secondary measures: the time spent for each assessment (i.e., {\textit{PageInteractionTime}}, in seconds) was automatically recorded for each scenario. Participants also rated the perceived helpfulness, education level, prior \ai usage frequency, and \xai tool use frequency. The survey also included an open-ended response for participants to provide qualitative feedback on their perceptions of the explanations and how these influenced their judgment.

\subsection{Participants}
\label{sec:participants}
We recruited $N = 128$ participants through {Prolific} (\url{https://www.prolific.com/}) during August 2026. We applied no country-of-residence restrictions and required fluency in English, a minimum 95\% approval rating, and an education level below the doctoral level, which we applied to filter out highly specialized participants.

We further included two attention checks and an honesty question. As pre-registered, the main analysis focuses on all 128 analyzed participants; results for the subset of $N = 75$ participants who passed all three attention checks is reported as a sensitivity analysis (see Appendix~\ref{app:robustness}), which yields conclusive findings.

The analyzed sample had a mean age of 31.7 years ($SD = 10.5$, range 19--72), with 68.8\% female and 31.3\% male, and was educated predominantly to Bachelor's (57.8\%) or Master's (31.3\%) level. 95.3\% reported using \ai weekly or more, and 59.4\% reported using \xai tools weekly or more (more details can be found in Appendix~\ref{app:participants_demographics}.

\subsection{Scenarios}
\label{sec:scenarios}
We constructed seven experimental scenarios \ScenOne--\ScenSeven by combining a realistic use case with a suitable \xai visualization method. Figure~\ref{fig:survey_screenshots} provides representative examples of all \xai visualizations used in the experiment. As a result, we compare the explanation effects across (i)~various \textbf{real-world application contexts} (job prediction, credit risk, heart attack risk, bike rental demand, mortality risk, manufacturing quality, product defect detection), (ii)~various \textbf{\xai methods} (\shap, \lime, saliency maps, \pdpsLong, variable importance), and (iii)~different \textbf{data modalities} (tabular, text, and image). As such, the scenarios also capture both local explanations (explaining individual predictions) and global explanations (explaining overall model behavior). We ensured that the decision tasks were generally understandable for end-users without explicit domain expertise to ensure that observed effects can be linked to to \xai visualizations and not missing domain knowledge. This holds for the task, but not necessarily for the feature labels, as we have copied them 1:1 from the sources to create realistic conditions.  We derived all scenarios from published studies and established \xai benchmarks (e.g., \cite{schoefferExplanationsFairnessAppropriate2024,senonerExplainableAIImproves2024a}). Table~\ref{tab:scenarios} summarizes all scenarios, including the corresponding \xai method, explanation scope (local vs. global), data modality and sources.

\begin{table}
\centering
\caption{Overview of Experimental Scenarios and \xai Methods}
\label{tab:scenarios}
\small
\begin{tabular}{@{}c
    >{\raggedright\arraybackslash}p{2.3cm}
    >{\raggedright\arraybackslash}p{3.4cm}
    >{\raggedright\arraybackslash}p{1.2cm}
    >{\raggedright\arraybackslash}p{1.7cm}
    >{\raggedright\arraybackslash}p{1.6cm}@{}}
\toprule
\textbf{\#} & \textbf{Scenario} & \textbf{\xai Method} & \textbf{Scope} & \textbf{Data Type} & \textbf{Stimulus source} \\
\midrule
\ScenOne & Job Prediction & Feature Attribution Heatmap & Local & Text & \cite{schoefferExplanationsFairnessAppropriate2024} \\
\ScenTwo & Credit Risk & Variable Importance Plot & Global & Tabular & \cite{schmittAccuracyArtificialIntelligence2023} \\
\ScenThree & Heart Attack & \lime & Local & Tabular & \cite{salihPerspectiveExplainableArtificial2025} \\
\ScenFour & Bike Rental & Partial Dependence Plots & Global & Tabular & \cite{kruschelChallengingPerformanceInterpretabilityTradeOff2025} \\
\ScenFive & Mortality Risk & \shap Attribution (single instance) & Local & Tabular & \cite{lundbergLocalExplanationsGlobal2020} \\
\ScenSix & Quality Driver & \shap Beeswarm Plot & Global & Tabular & \cite{senonerUsingExplainableArtificial2022} \\
\ScenSeven & Product Defect & Anomaly Heatmap (saliency-based) & Local & Image & \cite{senonerExplainableAIImproves2024a} \\
\bottomrule
\end{tabular}
\end{table}
\Description{A seven-row table describing the experimental scenarios with six columns: number, scenario, XAI method, scope, data type, and stimulus source. Scenario 1 (Job Prediction) uses Feature Attribution Heatmap, local scope, text data. Scenario 2 (Credit Risk) uses Variable Importance Plot, global scope, tabular data. Scenario 3 (Heart Attack) uses LIME, local scope, tabular data. Scenario 4 (Bike Rental) uses Partial Dependence Plots, global scope, tabular data. Scenario 5 (Mortality Risk) uses SHAP Attribution for single instance, local scope, tabular data. Scenario 6 (Quality Driver) uses SHAP Beeswarm Plot, global scope, tabular data. Scenario 7 (Product Defect) uses Anomaly Heatmap with saliency-based method, local scope, image data. All seven scenarios are sourced from published studies. The table shows that the design includes four local explanations and three global explanations, and three data modalities (one text, five tabular, one image).}

Table~\ref{tab:scenarios} lists the \xai method, explanation scope and data modality of each. Importantly, all participants were exposed to both local explanations (Scenarios \ScenOne, \ScenThree, \ScenFive, \ScenSeven; $n=4$) and global explanations (Scenarios \ScenTwo, \ScenFour, \ScenSix; $n=3$), enabling a within-subject comparison of explanation scope effects.

\subsection{Experimental Conditions}
\begin{figure}[t]
    \centering
    \begin{subfigure}[b]{0.32\linewidth}
        \centering
        \includegraphics[width=\linewidth]{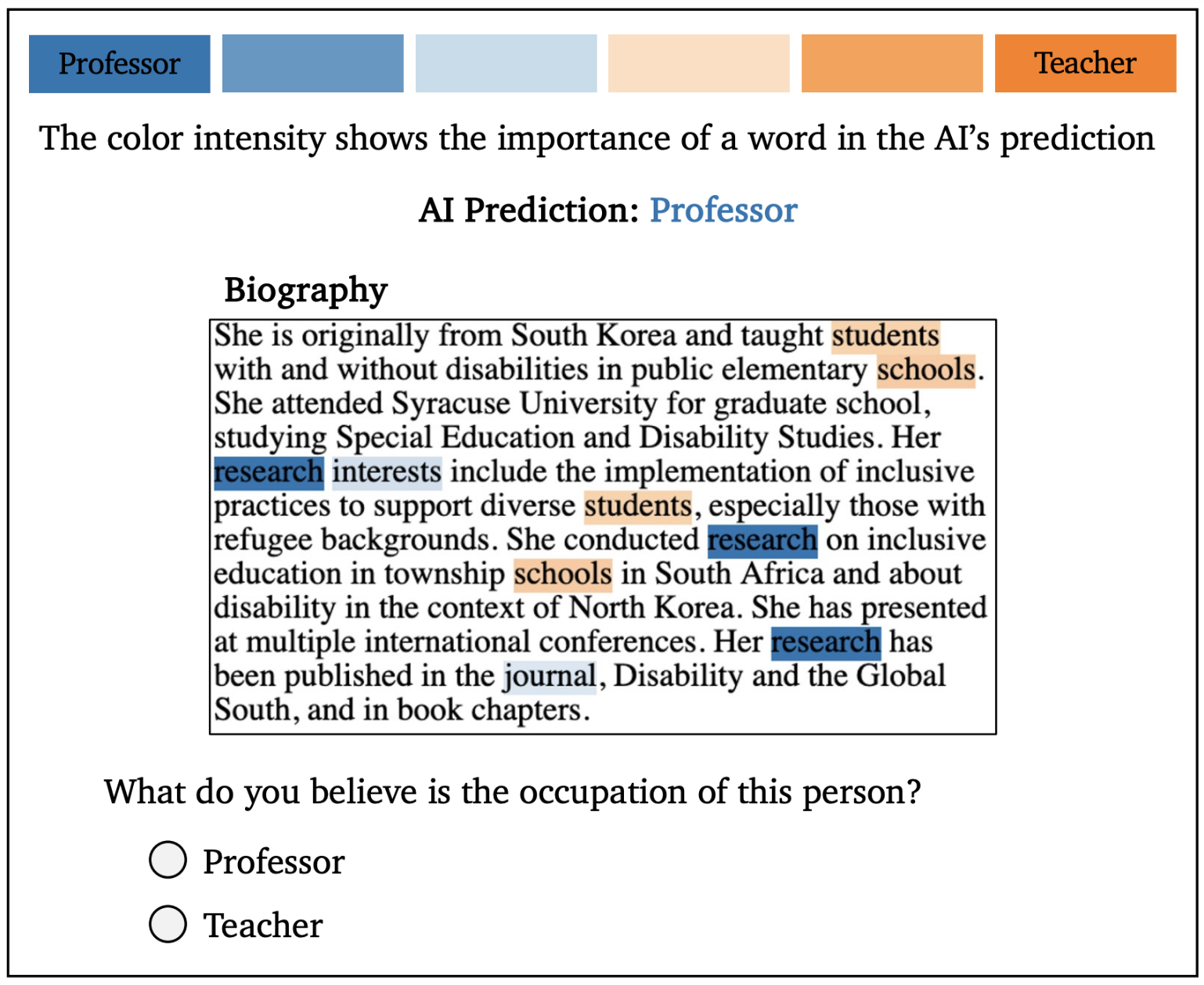}
        \caption{\ScenOne~Scenario 1: Job Prediction with feature attribution heatmap on biography text.}
    \end{subfigure}
    \hfill
    \begin{subfigure}[b]{0.32\linewidth}
        \centering
        \includegraphics[width=\linewidth]{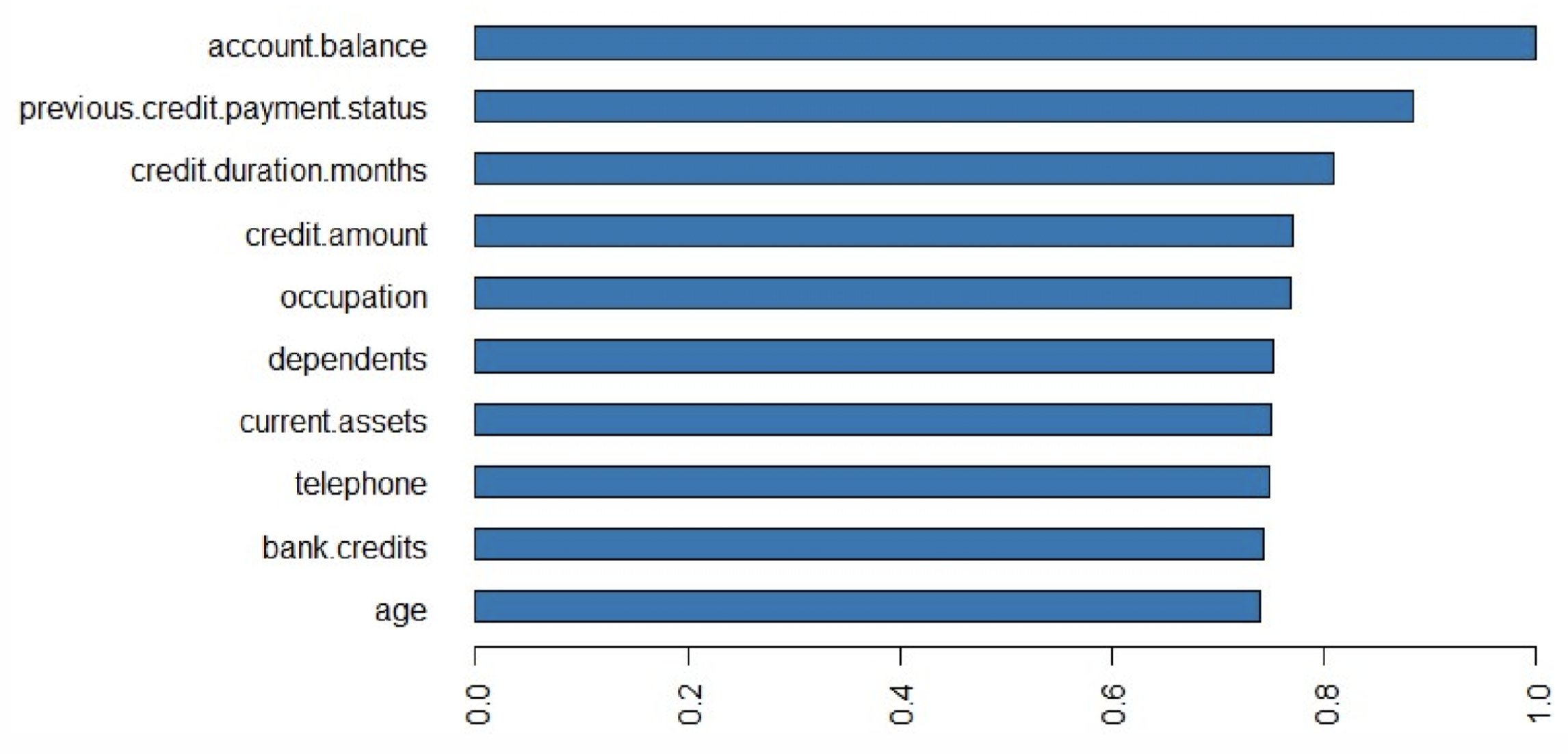}
        \caption{\ScenTwo~Scenario 2: Credit Risk with variable importance plot.}
    \end{subfigure}
    \hfill
    \begin{subfigure}[b]{0.32\linewidth}
        \centering
        \includegraphics[width=\linewidth]{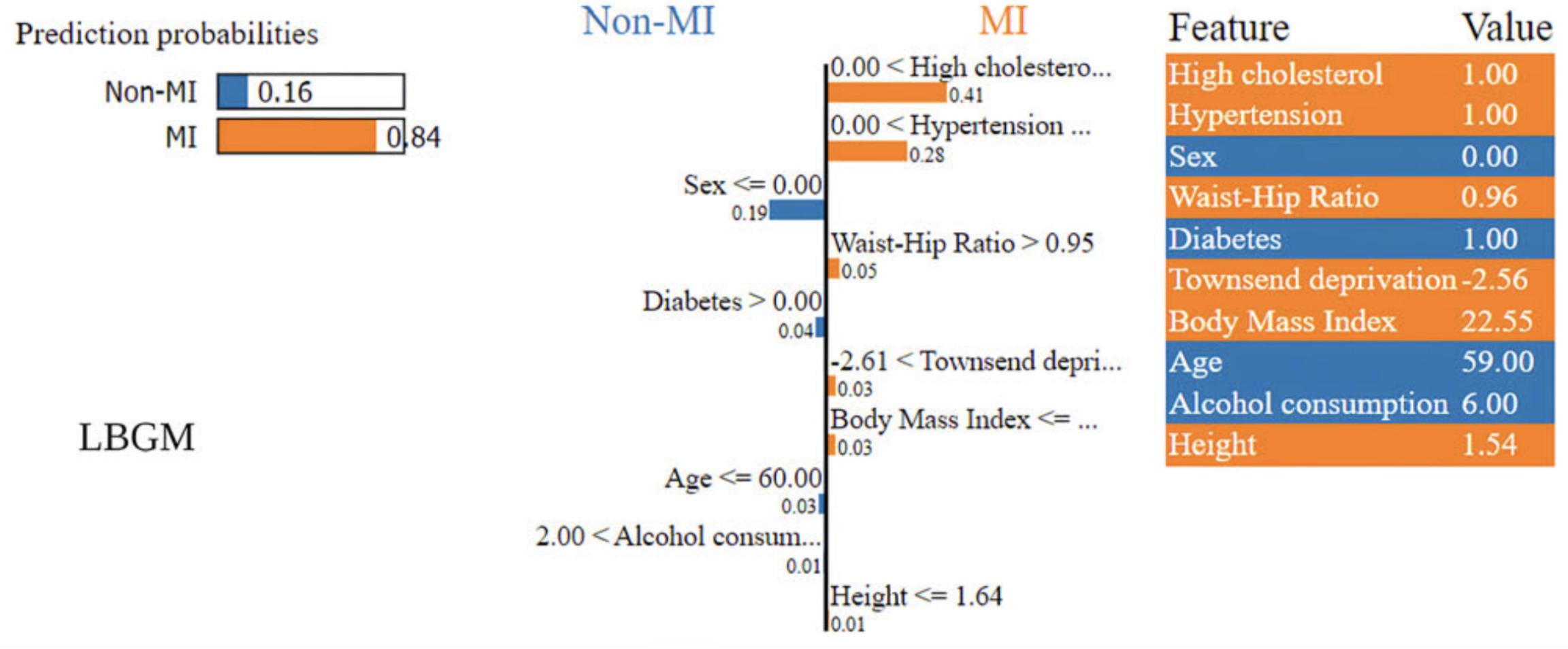}
        \caption{\ScenThree~Scenario 3: Heart Attack with \lime probability and feature explanations.}
    \end{subfigure}
    
    \vspace{0.5em}
    
    \begin{subfigure}[b]{0.32\linewidth}
        \centering
        \includegraphics[width=\linewidth]{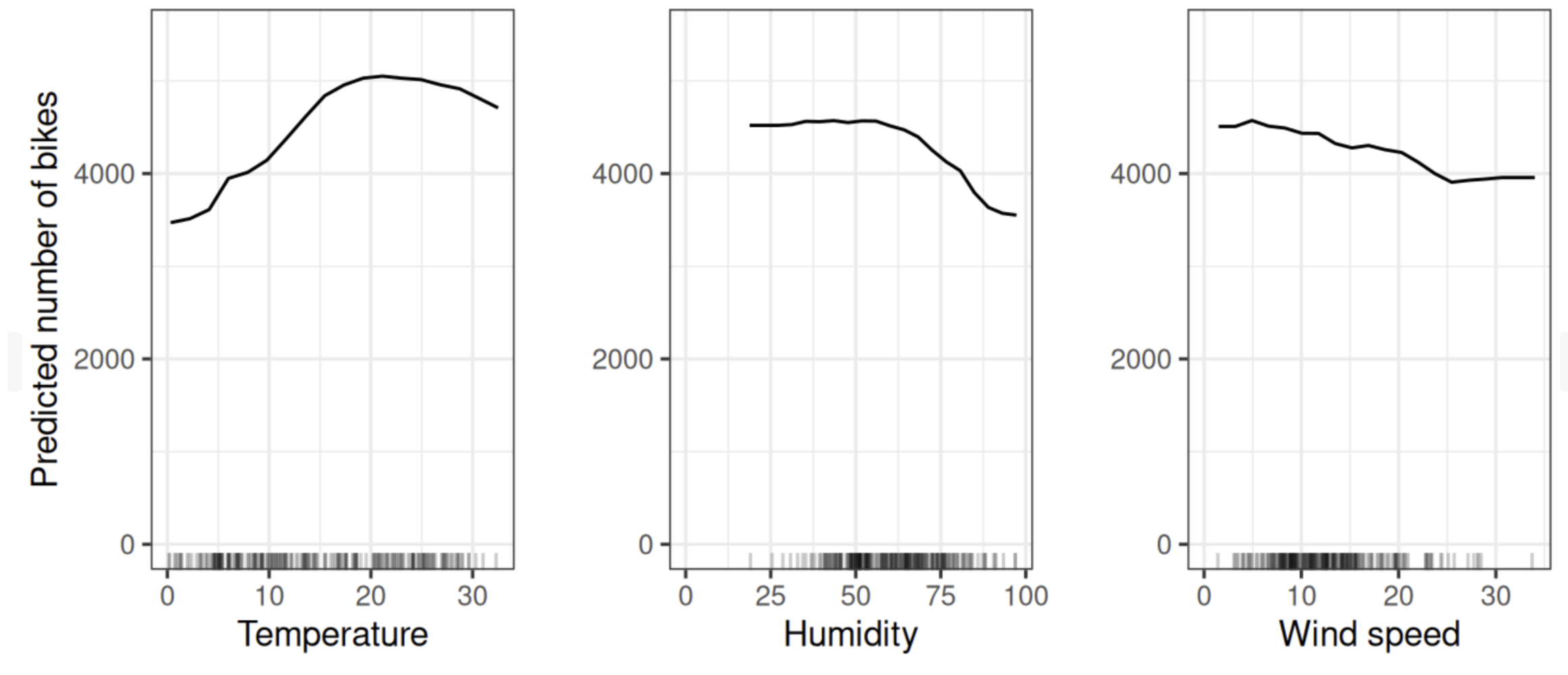}
        \caption{\ScenFour~Scenario 4: Bike Rental with \pdpsLong.}
    \end{subfigure}
    \hfill
    \begin{subfigure}[b]{0.32\linewidth}
        \centering
        \includegraphics[width=\linewidth]{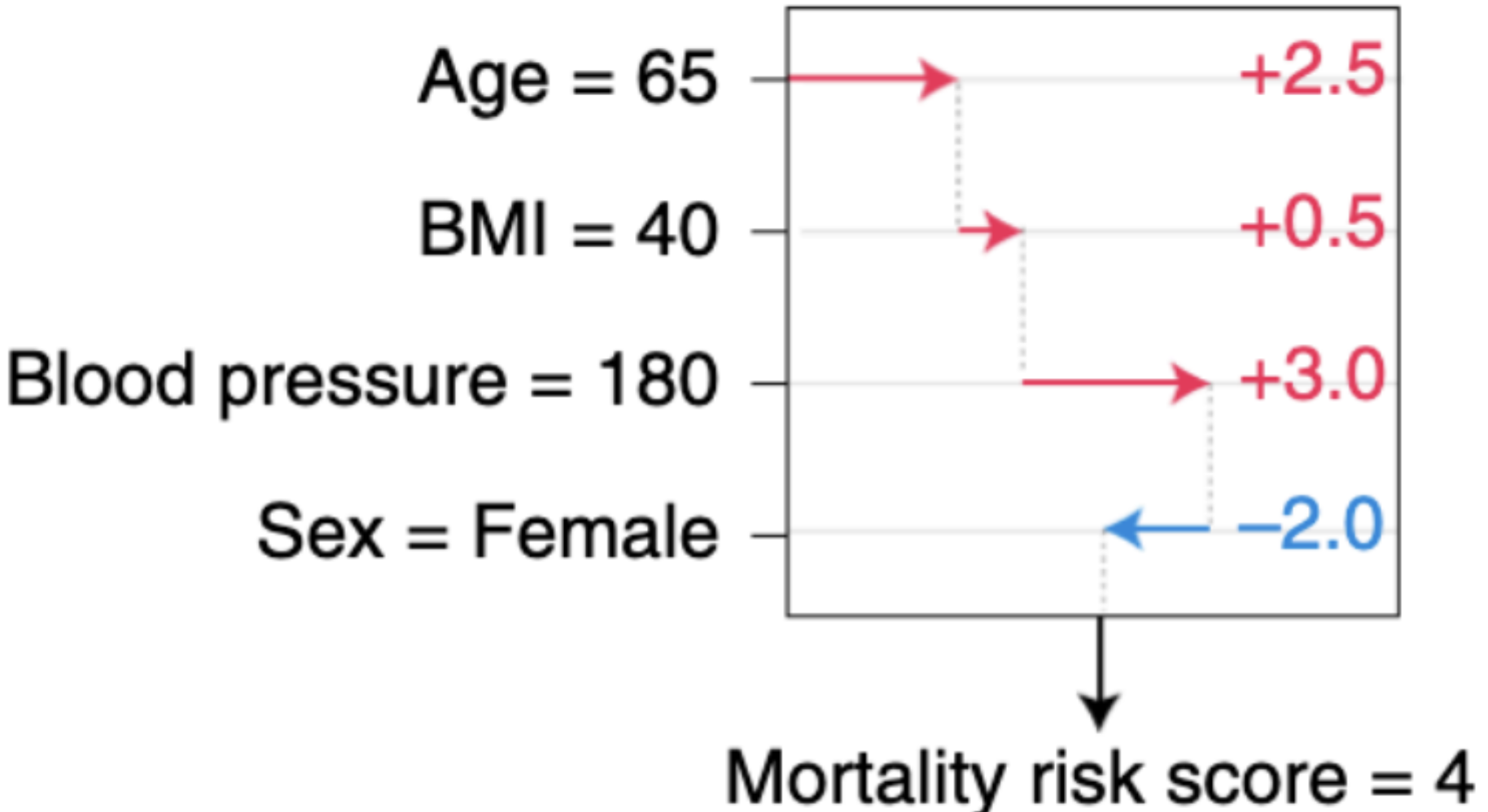}
        \caption{\ScenFive~Scenario 5: Mortality Risk with arrows showing how the attributes of a single patient shift the predicted risk score.}
    \end{subfigure}
    \hfill
    \begin{subfigure}[b]{0.32\linewidth}
        \centering
        \includegraphics[width=\linewidth]{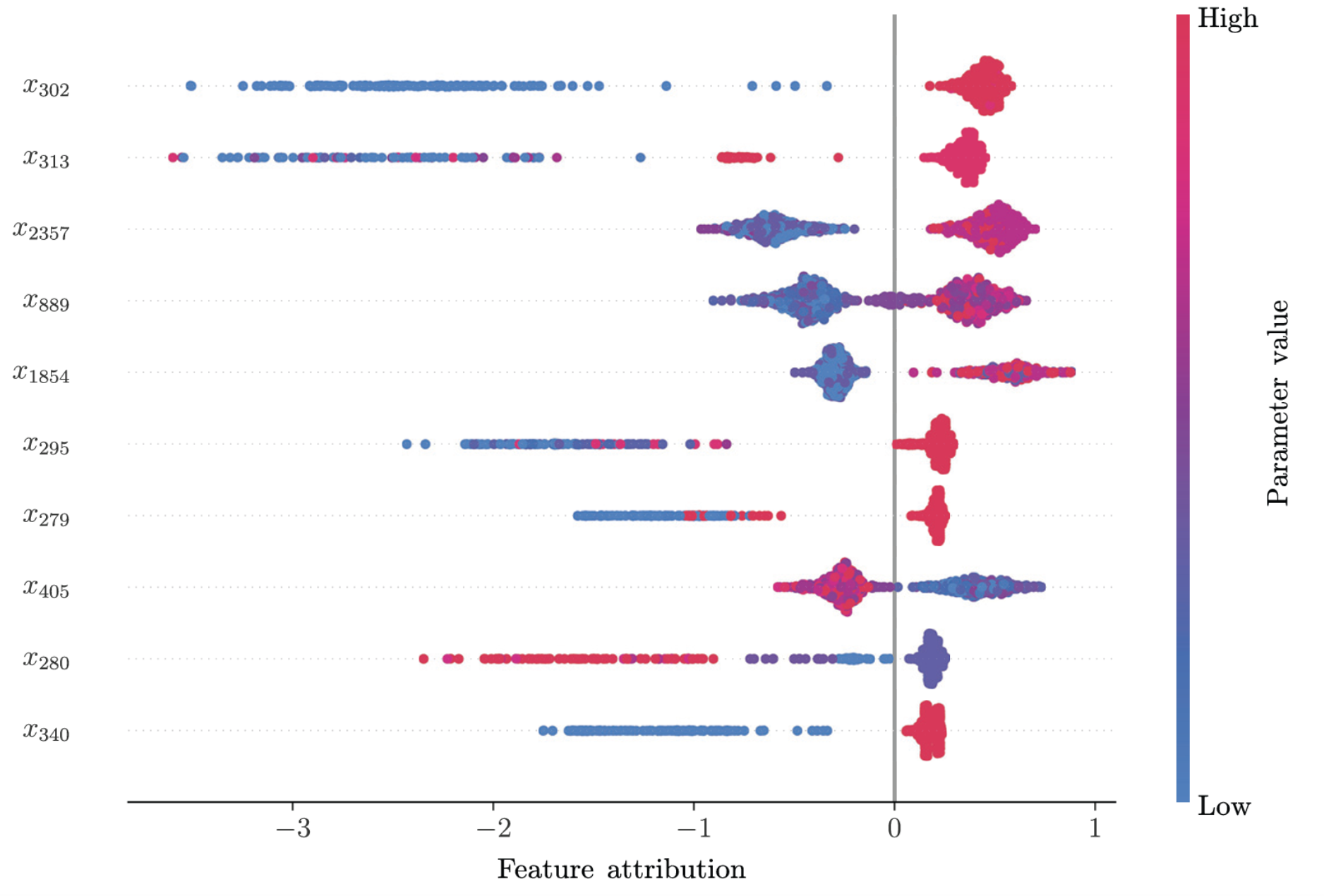}
        \caption{\ScenSix~Scenario 6: Quality Driver with \shap beeswarm plot.}
    \end{subfigure}
    
    \vspace{0.5em}
    
    \begin{subfigure}[b]{0.4\linewidth}
        \centering
        \includegraphics[width=\linewidth]{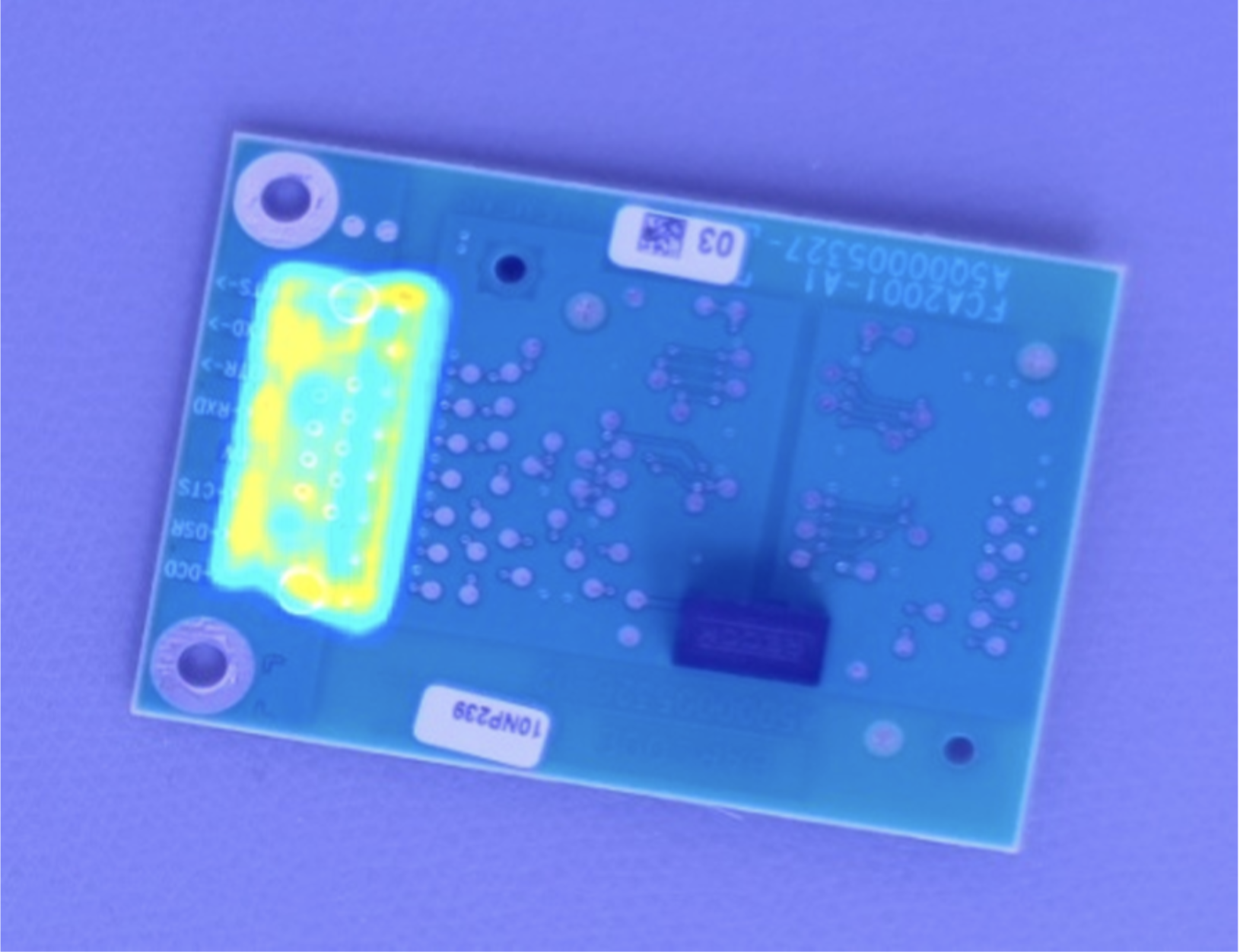}
        \caption{\ScenSeven~Scenario 7: Product Defect with saliency map heatmap on PCB image.}
    \end{subfigure}
    
    \caption{Seven \xai scenarios as presented to participants once with only the visualization (\NoExpl) and once with an \llm-generated textual explanation (\WithExpl). Each scenario used a different \xai visualization (feature attribution heatmap, variable importance plot, \lime, \pdpsLong, single-instance \shap attribution, \shap beeswarm plot, anomaly heatmap) applied across diverse domains (employment, finance, healthcare, transportation, manufacturing) and data modalities (text, tabular, image)..}
    \Description{Seven screenshots of the survey stimuli, one per scenario. In panel (a), Job Prediction, a biography paragraph has individual words shaded on a blue-to-orange scale, above it a colour legend running from Professor to Teacher and a one-line note that colour intensity shows a word's importance, together with the model's prediction and two answer options. Panel (b), Credit Risk, is a horizontal bar chart ranking ten applicant attributes by importance, with account balance highest, followed by previous credit payment status and credit duration in months. Panel (c), Heart Attack, is a \lime output combining predicted class probabilities for the two outcomes, a two-sided bar chart of features pushing towards or away from myocardial infarction, and a table of that patient's feature values. Panel (d), Bike Rental, shows three partial dependence line plots of the predicted number of bikes against temperature, humidity, and wind speed. Panel (e), Mortality Risk, is a single-instance attribution diagram in which four patient attributes shift a running risk score by labeled positive or negative increments to a final score. Panel (f), Quality Driver, is a \shap beeswarm plot of ten anonymized feature indices, each row a horizontal swarm of points colored from low to high parameter value and spanning negative to positive attribution. Panel (g), Product Defect, is a photograph of a printed circuit board with a blue-to-yellow anomaly heatmap overlay highlighting one rectangular region on the left of the board.}
    \label{fig:survey_screenshots}
\end{figure}

\begin{figure*}
    \centering
    \includegraphics[width=\linewidth]{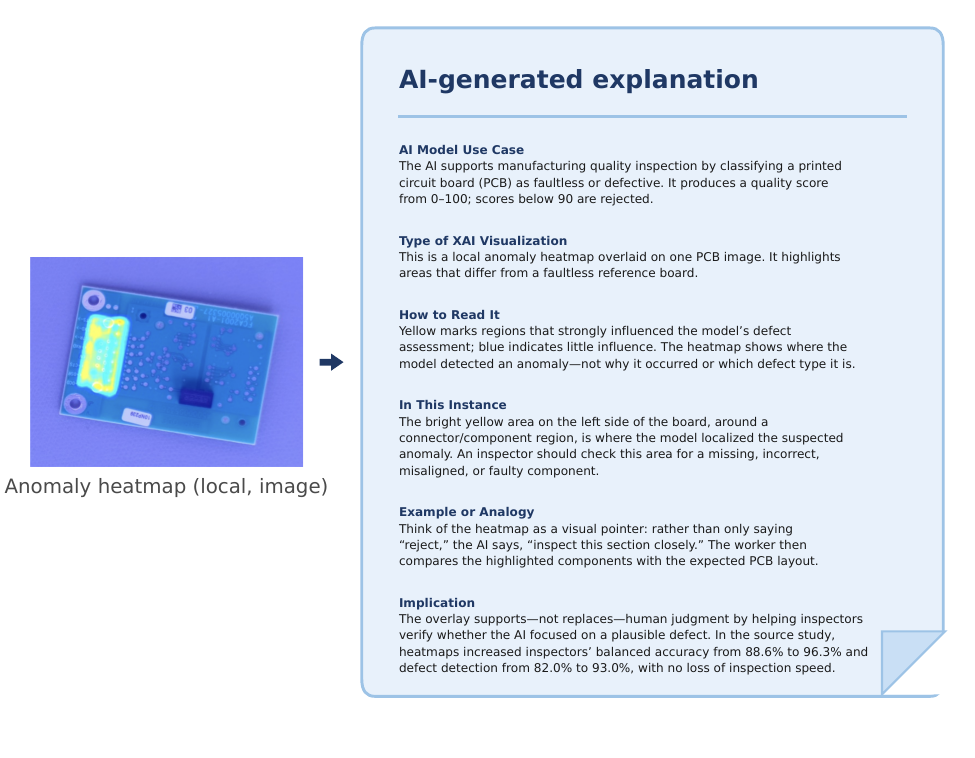}
    \caption{Example of \xai visualization with an \llm-generated explanation \WithExpl~condition for scenario  \ScenSeven~Product Defect detection with saliency maps over the printed circuit board (PCB) that participants saw, next to the LLM-generated explanation that accompanied it. The explanation follows the six-part-answer-structure prompting described in Figure~\ref{fig:prompt}. It states the model's use case, identifies the visualization type, states how to read the colored overlay, interprets the highlighted region for this instance, gives an analogy, and closes with the practical implication for a human inspector. This textual layer is the manipulation and the only difference between the two conditions.}
    \Description{A two-part figure. On the left is the stimulus image: a photograph of a green printed circuit board on a blue-tinted background, overlaid with an anomaly heatmap in which a bright yellow patch covers a rectangular connector region on the left third of the board while the rest of the board is blue. A label beneath it reads anomaly heatmap, local, image. An arrow points to the right, to a card headed AI-generated explanation, which reproduces the text participants read in six labeled sections. AI model use case: the AI supports manufacturing quality inspection by classifying a printed circuit board as faultless or defective and produces a quality score from 0 to 100, with scores below 90 rejected. Type of \xai visualization: a local anomaly heatmap overlaid on one board image, highlighting areas that differ from a faultless reference board. How to read it: yellow marks regions that strongly influenced the model's defect assessment and blue indicates little influence, and the heatmap shows where the model detected an anomaly, not why it occurred or which defect type it is. In this instance: the bright yellow area on the left of the board, around a connector or component region, is where the model localized the suspected anomaly, and an inspector should check that area for a missing, incorrect, misaligned or faulty component. Example or analogy: the heatmap is a visual pointer, so rather than only saying reject, the AI says inspect this section closely, and the worker then compares the highlighted components with the expected board layout. Implication: the overlay supports rather than replaces human judgment by helping inspectors verify whether the AI focused on a plausible defect, and it closes by citing the source study, stating that heatmaps increased inspectors' balanced accuracy from 88.6 percent to 96.3 percent and defect detection from 82.0 percent to 93.0 percent with no loss of inspection speed. The explanation therefore carries a performance claim in addition to its interpretation of the visualization.}
    \label{fig:explanationexample}
\end{figure*}

We used a counterbalanced within-subject experiment to mitigate order and carry over effect \cite{brooks2012counterbalancing}, which we configured in Qualtrics (\href{https://www.qualtrics.com/}{https://www.qualtrics.com/}). Hence, participant were exposed to two conditions in a randomized order (first all seven scenarios under \NoExpl, then \WithExpl or vice versa):

\begin{itemize}
    \item \textbf{Without-explanation condition (\NoExpl):} Participants viewed the description of the use case scenario, the origin of the underlying data, the name of the \xai visualization type (see figure ~\ref{fig:survey_screenshots}), and the raw \xai visualization output. The visualization was presented without any additional textual information, thus requiring participants to interpret the \xai output independently . This was a deliberate decision, because on of the key intentions of \xai to to be understandable, which our study challenges. 
    \item \textbf{With-explanation condition (\WithExpl):} Participants viewed the same \llm generated explanations, described in section \ref{sec:method_explanations})  (sse in example in figure \ref{fig:explanationexample})
\end{itemize}

\subsection{Statistical Analysis}
\label{sec:analysis}
The analysis will be performed against the results from the replication study with participant = $N = 128$. The results from the initial un-registered study will used section \ref{sec:validation} to compare difference against an older model and a seperate participant pool. We will calculate the following key metrics to validate H1, H2 and H3:

\textbf{Treatment effects.} To estimate the effect of the condition on perceived understanding (H1) and perceived trust (H2), we analyzed the primary outcomes $Y_{ij}$ (i.e., the variables \textit{Perceived Understanding} or \textit{Perceived Trust}) using linear mixed-effects models. For participant $i$ and rating $j$, we estimated the following model:
\begin{equation}
\label{eq:lmm}
Y_{ij} = \beta_0 + \beta_1 \, \textit{Explanation}_{ij}
+ \beta_2 \, \textit{Block}_{ij}
+ \beta_3 \, \textit{Sequence}_i
+ \sum_{k=1}^{6} \gamma_k \, \textit{Scenario}_{kj}
+ u_i + \varepsilon_{ij} ,
\end{equation}
where $\textit{Explanation}_{ij}$ indicates whether participants saw the visualization with (\WithExpl; =1) or without (\NoExpl; =0) an LLM-generated explanation; $\textit{Block}_{ij}$ indicates whether the rating was given during the participant's first or second pass through the seven scenarios; $\textit{Sequence}_i$ indicates which of the two counterbalanced presentation sequences (AB or BA) the participant was assigned to; $\textit{Scenario}_{kj}$ are six indicator variables for the seven scenarios, with \ScenFour~Bike Rental as the reference category; $u_i \sim \mathcal{N}(0, \sigma_u^2)$ is a random intercept for participants; and $\varepsilon_{ij} \sim \mathcal{N}(0, \sigma_\varepsilon^2)$ is the residual error term. Scenario therefore enters as a fixed effect (to capture between-scenario differences). The AB/BA design allows four observed \textit{Sequence} × \textit{Block} combinations, with treatment condition determined by \textit{Sequence} and \textit{Block}. Because a condition-specific carryover effect cannot be independently identified in this two-block AB/BA design without additional assumptions, we assess robustness using a first-block-only analysis (see Appendix~\ref{app:robustness}).

We fitted Eq.~\eqref{eq:lmm} using a restricted maximum likelihood method in Python 3.13 via the \texttt{statsmodels} package (v0.14.5, \texttt{MixedLM}) \cite{seaboldStatsmodelsEconometricStatistical2010} and report the estimated contrast $\beta_1$ for the explanation manipulation. For ease of reporting, we denote the estimated contrast by $\beta$ in the results, which is expressed in points of the five-point response scale, together with the paired within-subject effect size $d_z$ \cite{cohenStatisticalPowerAnalysis1988, lakensCalculatingReportingEffect2013}. Every likelihood-ratio test reported in this paper refits both the reduced and the full model by full maximum likelihood, because restricted maximum likelihood is not comparable across models that differ in their fixed-effect structure.

Paired $t$-tests and Wilcoxon signed-rank tests on within-subject aggregated scores are reported as a robustness check in Appendix~\ref{app:paired_tests}; the estimation details are given in Appendix~\ref{app:model_spec_detail}. Mixed-effects models are reported as Wald $z$-tests on the fixed-effect coefficients. Paired within-subject effect sizes are reported as Cohen's $d_z$ \cite{cohenStatisticalPowerAnalysis1988}, i.e, the mean of the participant-level change scores divided by their standard deviation. Internal consistency of ratings across the seven scenarios is assessed with Cronbach's $\alpha$ \cite{cronbachCoefficientAlphaInternal1951}). Two exploratory checks additionally use a Fisher exact test \cite{fisherInterpretation2Contingency1922, mehtaALGORITHM643FEXACT1986}) and a Mann--Whitney $U$-test \cite{mannTestWhetherOne1947, conroyWhatHypothesesNonparametric2012}). Following our preregistration, we use $\alpha =0.05$.

The scenario-specific and participant-level moderator analyses involve multiple hypothesis tests. Even when all corresponding null hypotheses are true, testing many effects increases the chance that some yield small $p$-values purely through sampling variation. We therefore apply the Benjamini-Hochberg procedure, which adjusts the results across each family of tests to control the expected proportion of the false discovery rate among the effects declared statistically significant \cite{benjaminiControllingFalseDiscovery1995}. We report the resulting FDR-adjusted $q$-values, with adjustment performed separately within each outcome.

\vspace{10pt}
\textbf{Mediation.} To examine H3, we tested whether \textit{Perceived Understanding} statistically mediates the relationship between LLM-generated explanations and \textit{Perceived Trust}. As both experimental conditions were observed for every participant, we followed the two-condition within-participant mediation framework of Montoya and Hayes \cite{montoyaTwoConditionWithinParticipant2017}. For each participant, we averaged \textit{Perceived Understanding} and \textit{Perceived Trust} across the seven scenarios separately within the \WithExpl{} and \NoExpl{} conditions and computed the corresponding within-participant condition differences. Thus, Eq.~\eqref{eq:lmm} provides the primary scenario-level analyses for H1 and H2, whereas H3 is evaluated at the participant level.

We use the conventional mediation notation $a$, $b$, $c$, and $c'$ \cite{baronModeratorMediatorVariable1986}, where $a$ denotes the effect of the condition on \textit{Perceived Understanding}, and $c$ the total effect of the condition on \textit{Perceived Trust}. The coefficient $b$ denotes the association between \textit{Perceived Understanding} and \textit{Perceived Trust} after accounting for the condition, while $c'$ denotes the remaining direct effect of the condition on \textit{Perceived Trust}. The indirect effect is quantified as $a \cdot b$. Following Montoya and Hayes \cite{montoyaTwoConditionWithinParticipant2017}, the model additionally includes each participant's average \textit{Perceived Understanding} across the two conditions, grand-mean centered across participants.

Inference for H3 focuses on the indirect effect $a \cdot b$. Because the sampling distribution of this product can be asymmetric, we estimated the corresponding uncertainty using a participant-level bootstrap rather than assuming a normal sampling distribution. We resampled participants with replacement 10{,}000 times, recomputed $a \cdot b$ in each resample, and report the resulting percentile-bootstrap 95\% confidence interval, following the within-participant mediation framework of Montoya and Hayes \cite{montoyaTwoConditionWithinParticipant2017}.
H3 is treated as supported when this interval excludes zero in the positive direction. 


\subsection{Robustness Check}
We assessed the stability of our results in two ways. First, as a robustness check, we re-estimated the registered models on the subset of the main analysis sample that passed all three attention and the honesty checks ($N = 75$ instead of 128; Appendix~\ref{app:robustness}), to rule out that inattentive or dishonest responding drove the effect differences. Second, as a validation, we compared the main analysis against an second study with explanations generated by GPT-4o ($N = 63$) that ran on the same instrument and the same seven scenarios, with the same attention-check restriction applied to that sample as well ($N = 41$)

\subsection{Ethics and Institutional Review}
\label{sec:ethics}

This study was conducted in accordance with ethical research standards and reviewed by the company's internal research review board to ensure compliance with scientific and ethical benchmarks. All participants provided informed consent before beginning the study and were informed of their right to withdraw at any time without penalty. Compensation exceeded minimum wage standards. No deception was used, and all data were collected and stored in compliance with GDPR regulations.

\section{Results}
\subsection{Descriptive Statistics}

Table~\ref{tab:descriptive_stats} presents descriptive statistics for \textit{Perceived Understanding} and \textit{Perceived Trust} ratings across experimental conditions and scenarios. The analysis includes 1{,}782 complete scenario-condition assessments from 128 participants; 10 of the 1{,}792 possible assessments contained a missing outcome response and were excluded. \textit{Perceived Understanding} was higher with explanations (\WithExpl; $M = 3.96$, $SD = 0.97$) than without explanations (\NoExpl; $M = 3.61$, $SD = 1.13$), a difference of $+0.35$ points on the five-point response scale. \textit{Perceived Trust} similarly increased from $3.76$ ($SD = 1.01$) without explanations to $4.01$ ($SD = 0.85$) with explanations, which corresponds to a difference of $+0.26$ points.

To summarize descriptive differences across scenarios, Table~\ref{tab:descriptive_stats} also reports scenario-level means pooled across both experimental conditions. here, the mean \textit{Perceived Understanding} ranged from $3.32$ ($SD = 1.27$) for \ScenSix~Quality Driver to $4.26$ ($SD = 0.66$) for \ScenOne~Job Prediction. The corresponding per-scenario distributions are shown in Appendix~\ref{app:supp_figures} (Figure~\ref{fig:scenarios}).

\begin{table}[H]
\centering
\caption{Descriptive statistics for key outcome measures.}
\label{tab:descriptive_stats}
\small

\makebox[\linewidth][c]{%
\begin{tabular}{@{}p{2.8cm}p{0.6cm}p{0.8cm}p{0.8cm}p{0.8cm}p{0.8cm}@{}}
\toprule
\textbf{Condition} & \textbf{Obs.} & 
\multicolumn{2}{c}{\textbf{\textit{Perceived Understanding}}} & 
\multicolumn{2}{c}{\textbf{\textit{Perceived Trust}}} \\
\cmidrule(lr){3-4} \cmidrule(lr){5-6}
  &  & $M$ & $SD$ & $M$ & $SD$ \\
\midrule
\textbf{By Explanation} \\
\NoExpl Without explanation & 891 & 3.61 & 1.13 & 3.76 & 1.01 \\
\WithExpl With explanation & 891 & 3.96 & 0.97 & 4.01 & 0.85 \\
\midrule
\textbf{By Scenario} \\
\ScenOne Job Prediction & 254 & 4.26 & 0.66 & 4.18 & 0.74 \\
\ScenTwo Credit Risk & 254 & 3.88 & 0.92 & 3.93 & 0.91 \\
\ScenThree Heart Attack & 254 & 3.70 & 1.01 & 3.86 & 0.85 \\
\ScenFour Bike Rental & 254 & 3.81 & 1.03 & 3.90 & 0.95 \\
\ScenFive Mortality Risk & 256 & 3.90 & 1.01 & 3.90 & 0.98 \\
\ScenSix Quality Driver & 256 & 3.32 & 1.27 & 3.65 & 0.99 \\
\ScenSeven Product Defect & 254 & 3.63 & 1.20 & 3.78 & 1.06 \\
\bottomrule
\end{tabular}%
}

\vspace{2pt}
\begin{tablenotes}
\scriptsize
\item \textit{Note:} All measures reported on a 5-point Likert scale.
\textit{Obs.} is the number of scenario assessments contributing to each row;
the total is 1{,}782 assessments from 128 participants. Scenario rows pool both conditions.
\end{tablenotes}
\end{table}
\Description{A descriptive statistics table with two sections. Column headers: Condition, Observations, Perceived Understanding (M, SD), Perceived Trust (M, SD). First section, By Explanation, has two rows: (1) Without explanation — 891 observations, Understanding M=3.61 SD=1.13, Trust M=3.76 SD=1.01; (2) With explanation — 891 observations, Understanding M=3.96 SD=0.97, Trust M=4.01 SD=0.85. Second section, By Scenario, has seven rows (one per scenario) with 254–256 observations each: Job Prediction (Understanding M=4.26 SD=0.66, Trust M=4.18 SD=0.74); Credit Risk (M=3.88/0.92, M=3.93/0.91); Heart Attack (M=3.70/1.01, M=3.86/0.85); Bike Rental (M=3.81/1.03, M=3.90/0.95); Mortality Risk (M=3.90/1.01, M=3.90/0.98); Quality Driver (M=3.32/1.27, M=3.65/0.99); Product Defect (M=3.63/1.20, M=3.78/1.06). All measures on 5-point Likert scale.}

\subsection{Hypothesis Testing}

Table~\ref{tab:hypothesis_summary} summarizes the results for hypotheses H1--H3; details are discussed below.

\begin{table}
\centering
\caption{Summary of Hypothesis Tests. \supported~= Supported. Significance: ***$p$ < .001; **$p$ < .01; *$p$ < .05.}
\label{tab:hypothesis_summary}
\small

\begin{tabular}{@{}cllcp{4.5cm}@{}}
\toprule
\textbf{Test} &
\textbf{Comparison} &
\textbf{Test Statistic} &
\textbf{$\boldsymbol{p}$-Value} &
\textbf{Finding} \\
\midrule

\multicolumn{5}{l}{\textit{Main Hypotheses}} \\

H1 &
\textit{Perceived Understanding}: \WithExpl~> \NoExpl &
$z$ = 8.57 &
<.001*** &
\supported~$\beta$ = 0.352, $d_z$ = 0.469 \\

H2 &
\textit{Perceived Trust}: \WithExpl~> \NoExpl &
$z$ = 7.06 &
<.001*** &
\supported~$\beta$ = 0.256, $d_z$ = 0.387 \\

H3 &
\textit{Perceived Understanding} mediates \textit{Perceived Trust} &
Indirect effect$^a$ &
CI excl.\ 0 &
\supported~($a \times b$ = 0.24) \\

\bottomrule
\end{tabular}

\vspace{0.3em}

\begin{flushleft}
\scriptsize
\textit{Note:}
For H1 and H2, $\beta$ denotes the estimated \textit{Explanation}
contrast $\beta_1$ from Eq.~\eqref{eq:lmm}, expressed in points of the
five-point response scale. Mixed-effects models with participant random
intercepts are reported as Wald $z$-tests; $d_z$ is the paired
within-subject standardized effect size.\\[0.5em]

$^a$Mediation was estimated at the participant level following the
two-condition within-participant framework of Montoya and Hayes
\cite{montoyaTwoConditionWithinParticipant2017}, using the conventional
$a$, $b$, $c$, and $c'$ path notation
\cite{baronModeratorMediatorVariable1986}. Inference is based on a
percentile bootstrap over 10{,}000 participant resamples. The indirect
effect is $a \times b$ = +0.239, 95\% CI [0.115, 0.380], against a total
effect of $c$ = +0.255; the remaining direct effect is not statistically
distinguishable from zero ($c'$ = +0.016, $p$ = .708). Because perceived
Perceived Understanding was measured rather than independently manipulated, the
indirect effect does not by itself establish a causal mechanism.
\end{flushleft}

\end{table}
\Description{A three-row hypothesis summary table with five columns: Test, Comparison, Test Statistic, p-Value, and Finding. Row 1 (H1): tests whether Perceived Understanding with explanation is greater than without explanation, test statistic z=8.57, p<.001, finding is Supported with beta=0.352 and d_z=0.469. Row 2 (H2): tests whether Perceived Trust with explanation is greater than without, test statistic z=7.06, p<.001, finding is Supported with beta=0.256 and d_z=0.387. Row 3 (H3): tests whether Perceived Understanding mediates the relationship between explanation and Perceived Trust, test statistic is indirect effect, p-value shows confidence interval excludes zero, finding is Supported with indirect effect a times b equals 0.24. All three main hypotheses are supported. Effect sizes and significance thresholds are provided in the note below the table.}

\subsubsection{User Perceived Understanding (H1): Significantly Higher with Explanations}
\label{sec:results_h1}

Hypothesis H1 predicted a higher \textit{Perceived Understanding} when participants viewed \xai visualizations with LLM-generated textual explanations compared to \xai visualizations alone. The distribution of \emph{Perceived Understanding} across conditions is shown in Figure~\ref{fig:distributions}. Participants viewing visualizations with explanations (\WithExpl) showed a significantly higher \textit{Perceived Understanding} ($\beta$ = 0.352, $SE$ = 0.041, $z$ = 8.57, $p$ < .001, 95\% CI [0.27, 0.43], Cohen's $d_z$ = 0.469) compared to those viewing visualizations alone (\NoExpl). The mean \textit{Perceived Understanding} increased from 3.61 ($SD = 1.13$) to 3.96 ($SD = 0.97$), a gain of $+0.35$ scale points. \emph{Therefore, Hypothesis H1 was supported: LLM-generated explanations significantly improved \textit{Perceived Understanding} of AI decision-making among end-users.}

\subsubsection{User Perceived Trust (H2): Significantly Higher with Explanations}

\begin{figure}
    \centering
    \includegraphics[width=\linewidth]{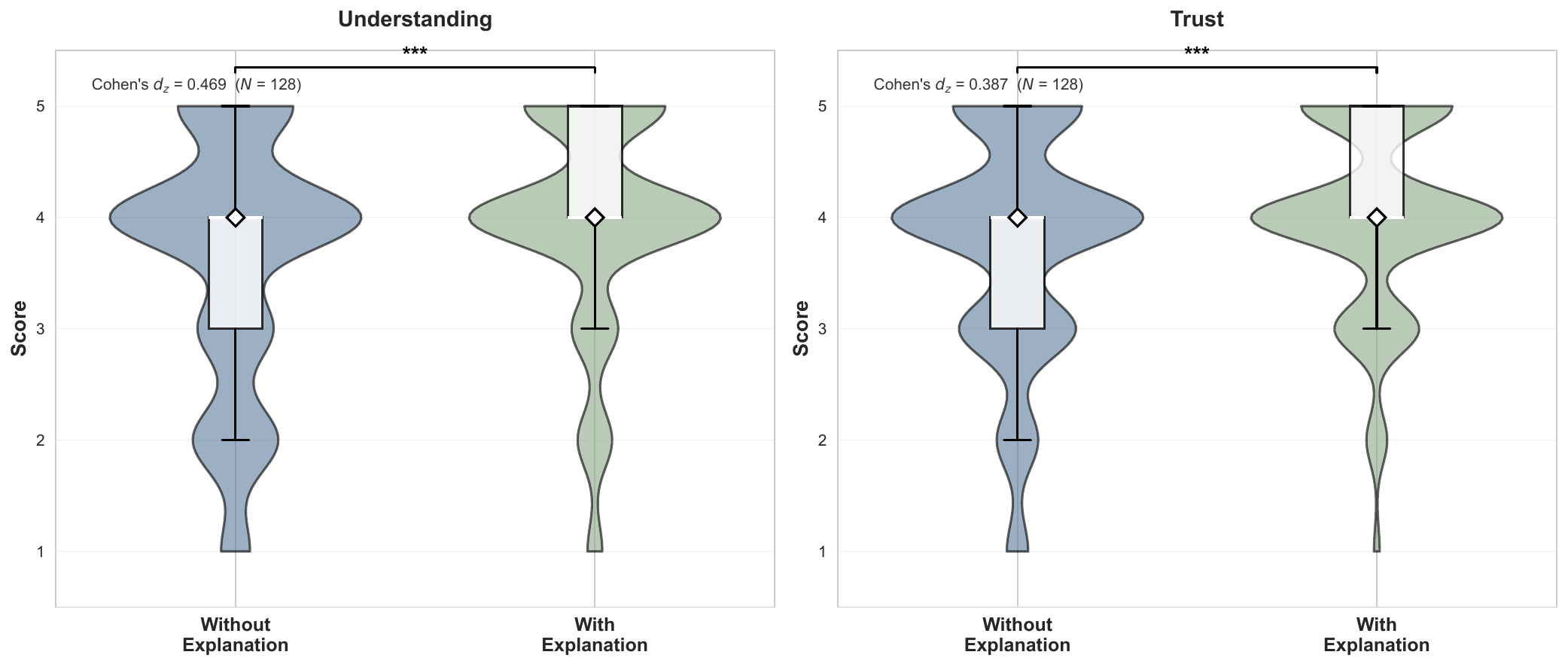}
    \caption{Violin plots showing the distribution of \textit{Perceived Understanding} and \textit{Perceived Trust} ratings with and without explanations. Both distributions shift upward with explanations (*** $p$ < .001). The annotated effect size is the paired within-participant $d_z$, that is the mean of the participant-level change scores divided by their standard deviation ($d_z$ = 0.469 for \textit{Perceived Understanding} and $d_z$ = 0.387 for \textit{Perceived Trust}; $N = 128$), which is the effect size the design supports and the one reported in Table~\ref{tab:hypothesis_summary}. A pooled-$SD$ Cohen's $d$ is not reported here because it divides the mean change by the between-participant spread, which is not the denominator of the paired test. Boxes span the interquartile range, whiskers extend to 1.5$\times$IQR, and the diamond marks the median. Because the response scale is discrete, the median is 4 in all four conditions, so the shift shows up in the interquartile range rather than in the median: for both outcomes the box moves from [3, 4] without explanations to [4, 5] with them.}
    \Description{Two side-by-side violin plots with box plots overlaid, one for Perceived Understanding and one for Perceived Trust, each comparing the without-explanation and the with-explanation condition on a score axis running from one to five. In both panels, the with-explanation distribution sits higher, with its mass concentrated at the top of the scale and a thinner lower tail, and each panel carries a significance bracket marked three asterisks for p below .001 together with its effect size, a paired Cohen's d subscript z of 0.469 for Perceived Understanding and 0.387 for Perceived Trust, both for 128 participants. Perceived Understanding rises from a mean of 3.61 with a standard deviation of 1.13 without explanations to a mean of 3.96 with a standard deviation of 0.97 with them, and Perceived Trust rises from 3.76 with a standard deviation of 1.01 to 4.01 with a standard deviation of 0.85. Because the response scale is discrete, the median is 4 in all four distributions, so the shift is visible in the boxes rather than the medians: for both outcomes the interquartile range moves from 3 to 4 without explanations up to 4 to 5 with them.}
    \label{fig:distributions}
\end{figure}

Hypothesis H2 predicted higher user \textit{Perceived Trust} when participants viewed \xai visualizations with LLM-generated textual explanations compared to visualizations alone (see Figure~\ref{fig:distributions}). Participants viewing visualizations with explanations (\WithExpl) showed significantly higher \textit{Perceived Trust} ($\beta$ = 0.256, $SE$ = 0.036, $z$ = 7.06, $p$ < .001, 95\% CI [0.18, 0.33], Cohen's $d_z$ = 0.387) compared to those viewing visualizations alone (\NoExpl). The mean \textit{Perceived Trust} increased from 3.76 ($SD = 1.01$) to 4.01 ($SD = 0.85$), which corresponds to a 6.8\% improvement. The main effects for both outcomes are also shown as a forest plot in Appendix~\ref{app:supp_figures} (Figure~\ref{fig:forest_plot}). \emph{Therefore, Hypothesis H2 was supported: LLM-generated explanations significantly improved the \textit{Perceived Trust} in \ai among end-users.}

\begin{figure}
    \centering
    \includegraphics[width=0.72\linewidth]{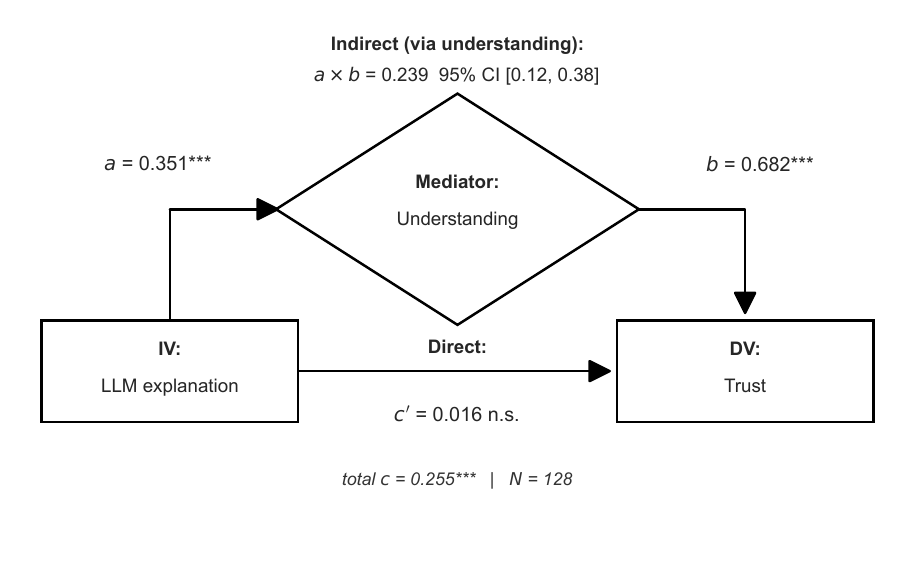}
    \caption{Decomposition of the mediation analysis for the explanation $\rightarrow$ \textit{Perceived Understanding} $\rightarrow$ \textit{Perceived Trust} pathway in the primary sample ($N = 128$), following the two-condition within-participant specification of Montoya and Hayes. Path $a$ is the effect of explanations on \textit{Perceived Understanding}, path $b$ the association between \textit{Perceived Understanding} and \textit{Perceived Trust} after accounting for condition and for the participant's grand-mean-centered average \textit{Perceived Understanding}, $c$ the total effect of explanations on \textit{Perceived Trust}, and $c'$ the remaining direct effect. The inferential result is the indirect effect $a \times b$ with its 95\% percentile bootstrap confidence interval, which excludes zero in the positive direction.}
    \Description{Mediation path diagram for the primary sample ($N = 128$). A box labeled LLM explanation represents the independent variable, a diamond labeled Perceived Understanding represents the mediator, and a box labeled Perceived Trust represents the dependent variable. The path from LLM explanation to Perceived Understanding is labeled $a = 0.351$ with $p < .001$, and the path from Perceived Understanding to Perceived Trust is labeled $b = 0.682$ with $p < .001$. A separate direct path from LLM explanation to Perceived Trust is labeled $c' = 0.016$ and is not statistically significant. The total effect of explanations on Perceived Trust is $c = 0.255$ with $p < .001$. Above the mediator, the indirect effect through Perceived Understanding is reported as $a \times b = 0.239$ with a 95 percent bootstrap confidence interval from 0.12 to 0.38.}
    \label{fig:mediation}
\end{figure}

\subsubsection{Mediation Analysis (H3): \textit{Perceived Understanding} Mediates Improvement in \textit{Perceived Trust}}
\label{sec:results_h3}

Hypothesis H3 predicted that a higher \textit{Perceived Understanding} is associated with improved \textit{Perceived Trust}, meaning that \textit{Perceived Understanding} mediates the relationship between explanations and \textit{Perceived Trust}. To test this, we conducted a mediation analysis examining the indirect pathway from the condition for \emph{Explanations} (\WithExpl vs. \NoExpl) $\rightarrow$ \textit{Perceived Understanding} $\rightarrow$ \textit{Perceived Trust}. 

The results showed that \textit{Explanations} significantly increased \textit{Perceived Understanding} ($a = +0.351$, $t(127) = 5.31$, $p < .001$), and \textit{Explanations} also significantly increased \textit{Perceived Trust} ($c = +0.255$, $t(127) = 4.37$, $p < .001$). These are the H1 and H2 effects re-estimated on participant-level change scores, which is why they match $\beta$ = 0.352 and $\beta$ = 0.256 while their test statistics are different from above. In the regression of the \textit{Perceived Trust} change on the \textit{Perceived Understanding} change, the change in \textit{Perceived Understanding} remained a significant predictor of the change in \textit{Perceived Trust} ($b = +0.682$, $SE = 0.052$, $p < .001$), while the direct effect of explanations, the intercept of that regression, was no longer statistically distinguishable from zero ($c' = +0.016$, $SE = 0.042$, $p = .708$; bootstrap 95\% CI [$-0.064$, $+0.104$]). The indirect effect was $a \times b = +0.239$ with a percentile bootstrap 95\% CI of [$+0.115$, $+0.380$] over 10{,}000 participant resamples. Because that interval excludes zero in the positive direction, H3 is supported (see Figure~\ref{fig:mediation}).

\subsection{Exploratory Analyses}
Table~\ref{tab:exploratory_summary} summarizes the key exploratory analyses; the corresponding results are reported in detail in the following.

\begin{table}
\centering
\caption{Summary of Key Exploratory Analyses. Significance: ***$p$ < .001; **$p$ < .01; *$p$ < .05.}
\label{tab:exploratory_summary}
\small

\begin{tabular}{@{}cllcp{4.5cm}@{}}
\toprule
\textbf{Test} &
\textbf{Comparison} &
\textbf{Test Statistic} &
\textbf{$\boldsymbol{p}$-Value} &
\textbf{Finding} \\
\midrule

\multicolumn{5}{l}{\textit{Exploratory Analyses}} \\

&
Scenario moderates effects &
$\chi^2$(6) = 5.93$^b$ &
.431 &
No detectable variation \\

&
Local vs.\ global (no diff.) &
$\chi^2$(1) = 0.06 &
.800 &
No difference \\

&
Order moderates \textit{Perceived Understanding} &
$\chi^2$(1) = 6.28 &
.012* &
Significant moderation \\

&
\xai-tool use frequency moderates both$^e$ &
$\chi^2$(1) = 19.27 &
<.001*** &
Larger gains for less frequent users \\

&
Prior \ai usage moderates \textit{Perceived Trust}$^e$ &
$\chi^2$(1) = 9.89 &
.002** &
Larger gains for frequent users \\

&
Education moderates \textit{Perceived Trust}$^e$ &
$\chi^2$(1) = 6.00 &
.014* &
Larger \textit{Perceived Trust} gains at higher degrees \\

&
\WithExpl~increases page interaction time &
$z$ = 1.68 &
.092 &
No reliable difference \\

&
\xai method$^c$ &
--- &
--- &
Inseparable from scenario \\

&
Data modality$^d$ &
--- &
--- &
Unbalanced; descriptive only \\

\bottomrule
\end{tabular}

\vspace{0.3em}

\begin{flushleft}
\scriptsize
\textit{Note:}
Interaction effects were evaluated by comparing models with and without
the corresponding \textit{Explanation}~$\times$~moderator term.
Likelihood-ratio tests were fitted using maximum likelihood. The page-time
model is reported using a Wald $z$-test. Benjamini--Hochberg adjustment is
applied to the exploratory moderator tests within each outcome; adjusted
$q$-values are reported in the corresponding Results sections.\\[0.5em]

$^b$Likelihood-ratio test comparing models with and without
\textit{Explanation}~$\times$~\textit{Scenario} interaction terms.
Per-scenario effects range from $\Delta$ = +0.189
(\ScenOne~Job Prediction) to $\Delta$ = +0.528
(\ScenSeven~Product Defect) and are positive in all seven scenarios.
See Section~\ref{sec:results_scenario} for details.\\

$^c$Each scenario used exactly one \xai method; therefore, a separate
method-level effect cannot be distinguished from the corresponding
scenario effect. The local-versus-global contrast is reported separately
above.\\

$^d$Data modality is reported descriptively because the design is
unbalanced across modalities (text $n=1$, image $n=1$, tabular $n=5$
scenarios), with the text and image modalities each represented by a
single scenario.\\

$^e$Participant-level moderators were entered as
\textit{Explanation}~$\times$~moderator interactions alongside the terms
of Eq.~\eqref{eq:lmm}. Education, prior \ai usage, and \xai-tool use
frequency enter on their full ordinal scales. The $\chi^2$ shown is for
\textit{Perceived Understanding} for \xai-tool use frequency and for \textit{Perceived Trust}
for prior \ai usage and education. \xai-tool use frequency also moderates
\textit{Perceived Trust} ($\chi^2$(1) = 20.51, $p$ < .001), whereas neither prior
\ai usage ($q$ = .054) nor education ($p$ = .694) moderates
\textit{Perceived Understanding} after adjustment. Benjamini--Hochberg-adjusted
results are reported in Sections~\ref{sec:results_education}
and~\ref{sec:results_ai_experience}.
\end{flushleft}

\end{table}
\Description{A nine-row exploratory analyses table with five columns: Test, Comparison, Test Statistic, p-Value, Finding. Row 1: Scenario moderates effects, chi-squared(6)=5.93, p=.431, no detectable variation. Row 2: Local vs. global XAI scope, chi-squared(1)=0.06, p=.800, no difference. Row 3: Presentation order moderates Perceived Understanding, chi-squared(1)=6.28, p=.012, significant moderation. Row 4: XAI-tool use frequency moderates both outcomes, chi-squared(1)=19.27, p<.001, larger gains for less frequent users. Row 5: Prior AI usage moderates Perceived Trust, chi-squared(1)=9.89, p=.002, larger gains for frequent users. Row 6: Education level moderates Perceived Trust, chi-squared(1)=6.00, p=.014, larger gains at higher degrees. Row 7: Explanations increase page interaction time, z=1.68, p=.092, no reliable difference. Row 8: XAI method, test statistics dashes (inseparable from scenario). Row 9: Data modality, test statistics dashes (unbalanced, descriptive only). All exploratory tests use likelihood-ratio or Wald tests with Benjamini-Hochberg adjustment within each outcome family.}

\subsubsection{Sensitivity by Scenario}
\label{sec:results_scenario}

\begin{figure}
    \centering
    \includegraphics[width=\linewidth]{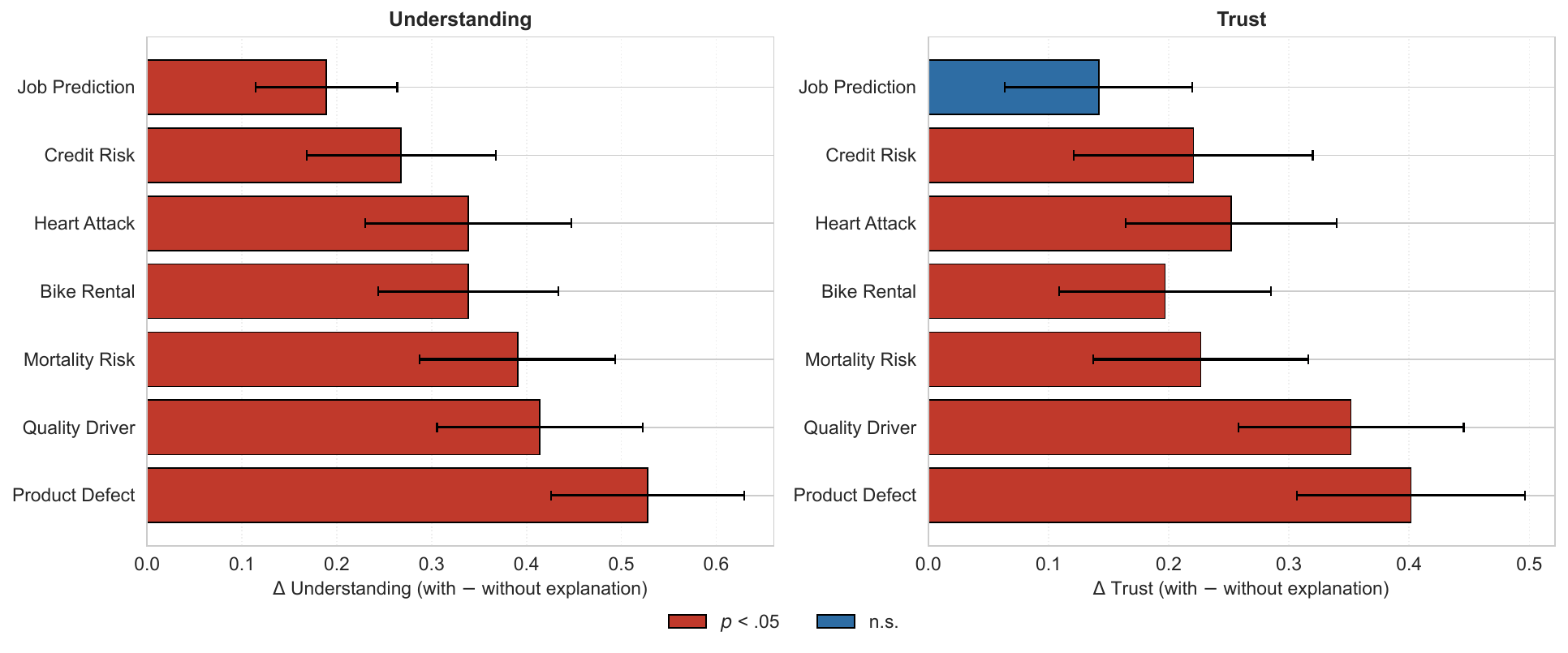}
    \caption{Sensitivity of the estimated effects by scenario for \textit{Perceived Understanding} (left) and \textit{Perceived Trust} (right). Bars show the participant-level mean change with error bars of $\pm 1$ standard error; coloring reflects paired $t$-tests on those change scores, red for statistically significant effects ($p$ < .05) and blue for non-significant ones. The coloring is unchanged by Benjamini--Hochberg adjustment, which is applied across the seven scenario tests within each outcome. A $\pm 1$ standard-error interval is narrower than a 95\% confidence interval, so it can exclude zero for an effect that is not significant (\ScenOne~Job Prediction for \textit{Perceived Trust}: $\Delta = +0.142$, $p$ = .072). All fourteen differences are positive, with \ScenSeven~Product Defect and \ScenSix~Quality Driver showing the strongest improvement on both outcomes.}
    \Description{Two side-by-side horizontal bar charts of the with-minus-without explanation difference, one for Perceived Understanding and one for Perceived Trust, with one bar per scenario in scenario order from top to bottom, error bars of plus or minus one standard error, and a color code distinguishing significant differences in red from non-significant ones in blue. All fourteen bars are positive. In the Perceived Understanding panel the differences are 0.19 for Job Prediction, 0.27 for Credit Risk, 0.34 for Heart Attack, 0.34 for Bike Rental, 0.39 for Mortality Risk, 0.41 for Quality Driver, and 0.53 for Product Defect, and every bar is marked significant. In the Perceived Trust panel they are 0.14 for Job Prediction, 0.22 for Credit Risk, 0.25 for Heart Attack, 0.20 for Bike Rental, 0.23 for Mortality Risk, 0.35 for Quality Driver, and 0.40 for Product Defect; here Job Prediction is the only bar marked non-significant, and it is also the smallest of the seven. Product Defect and Quality Driver are the two largest gains in both panels.}
    \label{fig:scenario_effects}
\end{figure}

The effectiveness of explanations showed only small variations across scenarios: the \textit{Explanation}~$\times$~\textit{Scenario} interaction was not significant for either outcome ($\chi^2$(6) = 5.93, $p$ = .431 for \textit{Perceived Understanding}; $\chi^2$(6) = 5.35, $p$ = .500 for \textit{Perceived Trust}). Marginal effects by scenario are shown in Figure~\ref{fig:scenario_effects} for \textit{Perceived Understanding} (left) and \textit{Perceived Trust} (right). Every one of the seven scenarios shows a positive effect on \textit{Perceived Understanding}, and all seven remain significant after Benjamini--Hochberg adjustment across the seven per-scenario tests for that outcome ($\Delta$ from $+0.189$ in \ScenOne~Job Prediction, $q$ = .013, to $+0.528$ in \ScenSeven~Product Defect, $q$ < .001). For \textit{Perceived Trust} the effect is likewise positive in all seven scenarios and is significant after adjustment in six; only \ScenOne~Job Prediction does not ($\Delta = +0.142$, $p$ = .072, $q$ = .072).

\subsubsection{Sensitivity by XAI Method}
By design, each scenario employed exactly one \xai method, so the effects of \xai method and scenario cannot be separated in this study. A formal interaction test for \xai method is therefore equivalent to the scenario-moderation test reported above. The explanation effect on \textit{Perceived Understanding} was positive across all seven \xai visualization types, ranging from $\Delta$ = 0.189 to $\Delta$ = 0.528 (Appendix~\ref{app:exploratory_detail}).

\begin{figure}[tbp]
    \centering
    \includegraphics[width=\linewidth]{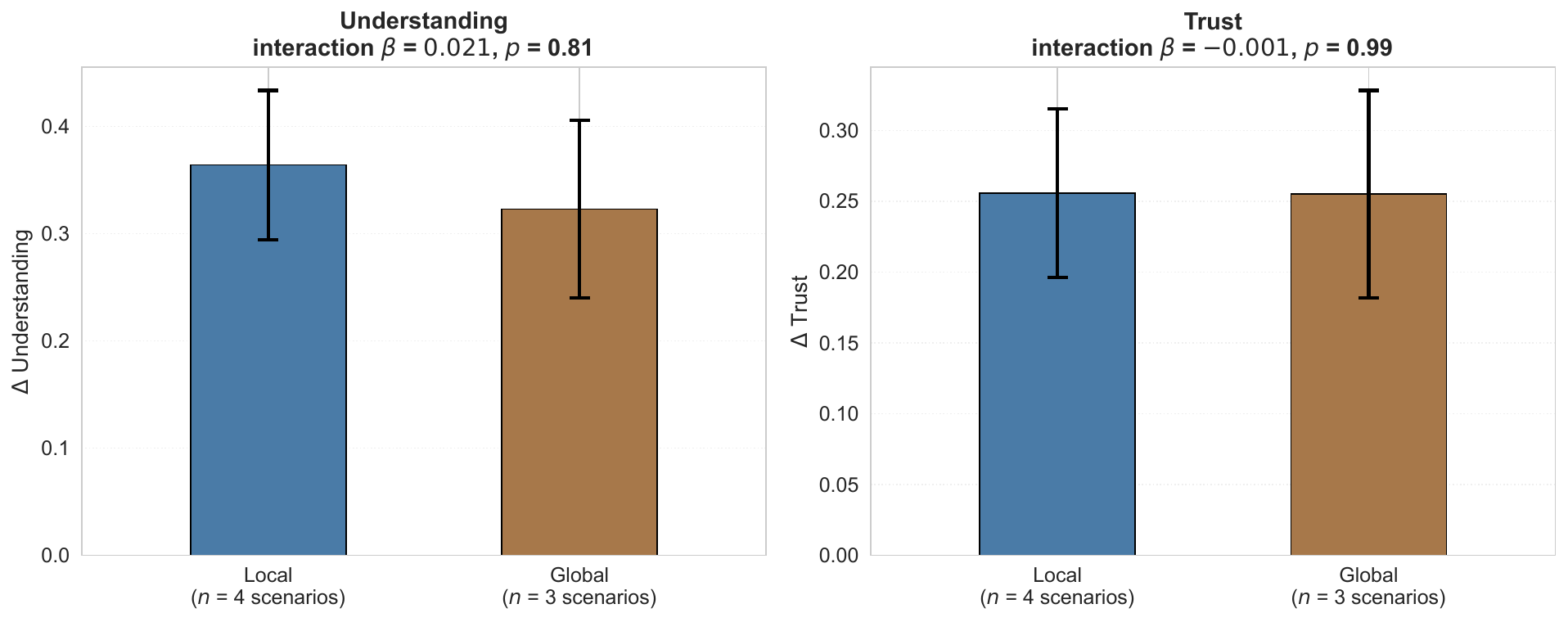}
    \caption{Mean difference ($\Delta$) in \textit{Perceived Understanding} and \textit{Perceived Trust} between the with- and without-explanation conditions for local \xai methods ($n=4$ scenarios) and global \xai methods ($n=3$ scenarios). The \textit{Explanation}~$\times$~\textit{Scope} interaction was not statistically significant for either outcome ($p$ = .80 and $p$ = .99, respectively), so no difference in effectiveness by \xai scope was detected. Since the interaction test is not an equivalence test, this does not establish that effectiveness is independent of scope.}
    \Description{Two side-by-side bar charts comparing the explanation benefit for local against global explanation scenarios, one panel for Perceived Understanding and one for Perceived Trust, each with error bars and the number of scenarios per group, four local and three global. In the Perceived Understanding panel the mean gain is 0.36 for local scenarios and 0.32 for global ones, a marginal difference relative to the error bars, and the panel title reports an interaction coefficient of 0.022 with p equal to .800. In the Perceived Trust panel the two bars are the same height at 0.26, and the panel title reports an interaction coefficient of minus 0.001 with p equal to .990. Neither interaction is significant.}
    \label{fig:local_global}
\end{figure}

Figure~\ref{fig:local_global} further compares local versus global explanations. The point estimate for \textit{Perceived Understanding} was slightly larger for local explanations (local: $\Delta$ = 0.364 vs. global: $\Delta$ = 0.323), and the two scopes were practically identical for \textit{Perceived Trust} (local: $\Delta$ = 0.256 vs. global: $\Delta$ = 0.255). Neither difference is statistically distinguishable: the \textit{Explanation}~$\times$~\textit{Scope} interaction was not significant for either outcome (\textit{Perceived Understanding}: $\beta$ = +0.022, $\chi^2$(1) = 0.06, $p$ = .80; \textit{Perceived Trust}: $\beta$ = $-$0.001, $\chi^2$(1) < 0.01, $p$ = .99).

\subsubsection{Sensitivity by Data Modality}
\label{sec:results_modality}

\begin{figure}[tbp]
    \centering
    \includegraphics[width=\linewidth]{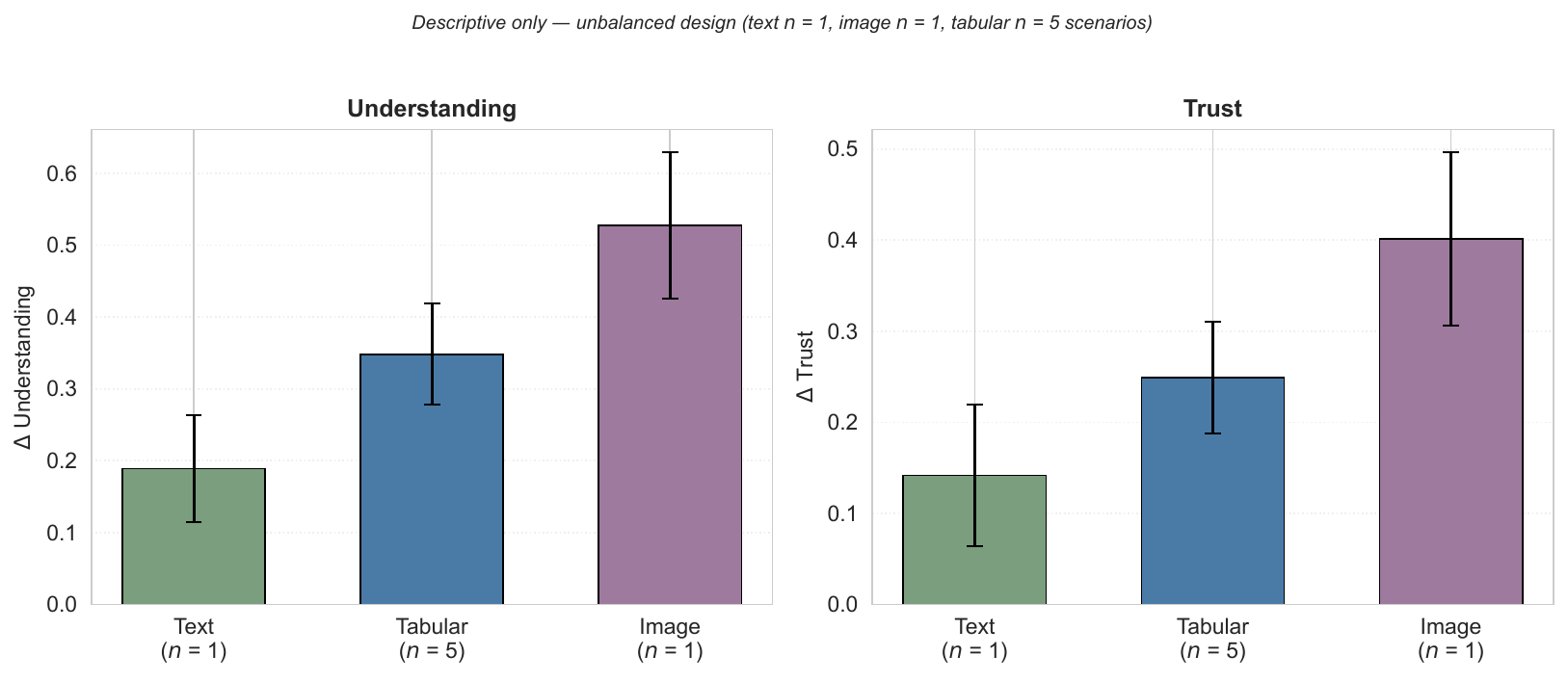}
    \caption{Mean difference ($\Delta$) in \textit{Perceived Understanding} and \textit{Perceived Trust} between the with- and without-explanation conditions across data modalities (text, tabular, image). The plotted mean differences are descriptive; the modality interaction is tested separately in Section~\ref{sec:results_modality}. The design is unbalanced (text $n=1$, image $n=1$, tabular $n=5$ scenarios), with the text and image modalities each represented by a single scenario, so these contrasts cannot be generalized to modality effects.}
    \Description{Two side-by-side bar charts comparing the explanation benefit across data modalities (text, tabular, and image), one panel for Perceived Understanding and one for Perceived Trust, with error bars and the number of scenarios per modality shown below each bar. In the Perceived Understanding panel, the mean gain is 0.19 for text, 0.35 for tabular, and 0.53 for image; in the Perceived Trust panel, it is 0.14 for text, 0.25 for tabular, and 0.40 for image. In both panels, the image modality has the largest observed mean difference, followed by tabular and text. The design includes one text scenario, five tabular scenarios, and one image scenario.}
    \label{fig:modality}
\end{figure}

Figure~\ref{fig:modality} examines data modality as a potential moderator. The \textit{Explanation}~$\times$~\textit{Modality} interaction is not significant for either outcome ($\chi^2$(2) = 4.61, $p$ = .100 for \textit{Perceived Understanding}; $\chi^2$(2) = 3.69, $p$ = .158 for \textit{Perceived Trust}). Because the design includes only one text scenario and one image scenario against five tabular ones, each modality estimate is inseparable from the particularities of an individual scenario, so we report modality differences descriptively (see also Appendix~\ref{app:exploratory_detail}).

\subsubsection{Page Interaction Time}
\label{sec:results_time}

Providing explanations did not significantly change how long participants spent on a scenario page. On log-transformed page interaction times, the explanation contrast was $\beta$ = 0.052 ($SE$ = 0.031, $z$ = 1.68, $p$ = .092, 95\% CI [$-0.009$, $+0.113$]), which corresponds to a 5.3\% longer page time in the with-explanation condition on the raw scale. This is a difference in the expected direction, but one that cannot be separated from zero.
A detailed breakdown across all scenarios is provided in Appendix~\ref{app:response_time}.

\subsubsection{Moderation by Education Level}
\label{sec:results_education}

Education level moderated the effect on \textit{Perceived Trust}, but not the effect on \textit{Perceived Understanding}. An ordinal moderator test confirms this pattern ($\chi^2$(1) = 0.16, $p$ = .694 for \textit{Perceived Understanding}; $\chi^2$(1) = 6.00, $p$ = .014, $q$ = .038 for \textit{Perceived Trust}). The positive interaction indicates that the explanation-related increase in \textit{Perceived Trust} was larger at higher education levels. The group means and the associations with baseline rating level are provided in Appendix~\ref{app:supp_figures} (Figure~\ref{fig:education}). As this analysis was exploratory and part of a larger set of moderator tests, the finding should be interpreted cautiously and requires further data collection and analysis.

\subsubsection{Moderation by AI Experience}
\label{sec:results_ai_experience}

Prior experience with \ai did moderate explanation effectiveness, and the two experience measures we collected pointed in opposite directions. After adding \textit{\ai usage frequency} as an ordinal predictor to the regression, the \textit{Explanation}~$\times$~\textit{\ai usage} interaction was positive for both outcomes and remained significant for \textit{Perceived Trust} after Benjamini--Hochberg correction across the exploratory moderator tests ($\beta = +0.189$, $\chi^2$(1) = 9.89, $p$ = .002, $q$ = .007). For \textit{Perceived Understanding}, the interaction was significant at the unadjusted level but not after Benjamini--Hochberg correction ($\beta = +0.159$, $\chi^2$(1) = 5.40, $p$ = .020, $q$ = .054). Participants who use \ai systems more frequently in everyday life therefore gained somewhat more \textit{Perceived Trust} from the explanations than infrequent users.

We further examined whether prior \xai tool use specifically influenced the effectiveness of explanations. Here, the \textit{Explanation}~$\times$~\textit{\xai tool use frequency} interaction, with use frequency entered as an ordinal predictor, was negative and significant for both outcomes ($\beta = -0.156$, $\chi^2$(1) = 19.27, $q$ < .001 for \textit{Perceived Understanding}; $\beta = -0.133$, $\chi^2$(1) = 20.51, $q$ < .001 for \textit{Perceived Trust}). The use frequency scale runs from \textit{never} to \textit{very frequently}. This indicates that participants with less prior exposure to interpretability methods showed larger explanation effects. Additionally, \xai tool use frequency was positively associated with the effect on \textit{Perceived Understanding} ($r = 0.38$, $p$ < .001) and \textit{Perceived Trust})($r = 0.27$, $p$ = .002) but negatively associated with the size of the \textit{Perceived Understanding} gain ($r = -0.23$, $p$ = .010). \emph{Overall, the strongest explanation effects were observed among participants reporting less frequent prior use of \xai tools.}

\subsubsection{Perceived Helpfuness}
\label{sec:qualitative}

Perceived helpfulness was positively associated with overall {Understanding} ($r = 0.32$, $p < .001$) and {Trust} ($r = 0.28$, $p = .002$), but not with the corresponding within-participant gains ($r = -0.01$, $p = .89$ and $r = 0.10$, $p = .29$, respectively).

\subsubsection{Analysis of Qualitative Feedback}

To characterize which concepts participants mentioned in their qualitative feedback,, we summarized the open-ended responses from all 128 participants using a descriptive Regex-based coding scheme with key words (see Appendix~\ref{app:keyword_dictionary}). Overall, 85 respondents (66.4\%) mentioned \textit{Understanding} or clarity, 55 (43.0\%) mentioned \textit{Trust} or confidence, and 7 (5.5\%) expressed skepticism or no change; 102 respondents (79.7\%) mentioned either \textit{Understanding} or \textit{Trust}.

\subsection{Robustness Check}

We performed several robustness checks, which led to conclusive findings. First, to rule out that inattentive responding biased our results, we re-estimated the main models on the attention-verified subset of the same sample, restricted to the $N = 75$ participants who passed all three attention checks. Results were very similar: \textit{Perceived Understanding} increased by $+0.378$ (subset) vs. $+0.352$ (analyzed), \textit{Perceived Trust} by $+0.278$ (subset) vs. $+0.256$ (analyzed), all $p < .001$. The mediation shows the same pattern: in the subset, we obtained $a = +0.364$, $b = +0.790$, $c = +0.269$, $c' = -0.019$, and an indirect effect $+0.287$ (95\% CI [$+0.114$, $+0.497$]), against $a = +0.351$, $b = +0.682$, $c = +0.255$, $c' = +0.016$, indirect effect $+0.239$ (95\% CI [$+0.115$, $+0.380$]) in the full sample. Ratings were also consistent across the seven scenarios (Cronbach's $\alpha = 0.806$ for \textit{Perceived Understanding}, $0.842$ for \textit{Perceived Trust}). 

Second, we performed a leave-one-scenario-out check by refitting the registered model seven times while dropping one scenario each time, which leaves the explanation contrast between $+0.323$ and $+0.379$ for \textit{Perceived Understanding} and between $+0.232$ and $+0.275$ for \textit{Perceived Trust} (all $p < .001$), so no single scenario drives the aggregate effect.

Third, we also refitted the registered model with a participant random slope on \textit{Explanation}, which leaves the estimates unchanged but widens their confidence intervals by about 60\% without changing the support for H1 or H2 (Appendix~\ref{app:random_slope}).

\subsection{Generalizability}
\label{sec:validation}

To assess whether the observed effects were specific to the \llm used in the main study, we compared the above results with an independent validation conducted in October 2025 with a separate sample of $N = 63$ participants from Prolific. The validation used the same survey instrument, seven \xai scenarios, experimental conditions, and attention checks, but the explanations were generated with GPT-4o (\texttt{chatgpt-4o-latest}) via the ChatGPT web interface. 

The validation showed the same overall pattern as the main study. First, \textit{Perceived Understanding} increased by $+0.433$ ($SE = 0.065$, $p < .001$, 95\% CI [$0.31$, $0.56$]), and \textit{Perceived Trust} by $+0.299$ ($SE = 0.057$, $p < .001$, 95\% CI [$0.19$, $0.41$]). Second, restricting the pilot to the $N = 41$ participants who passed all three attention checks left the pattern essentially unchanged (\textit{Perceived Understanding}: $+0.429$, $SE = 0.082$, $p < .001$, 95\% CI [$0.27$, $0.59$]; \textit{Perceived Trust}: $+0.265$, $SE = 0.070$, $p < .001$, 95\% CI [$0.13$, $0.40$]). Third, the mediation pathway is also robust: in the full pilot sample, $a = +0.433$, $b = +0.587$, $c = +0.299$, $c' = +0.045$ (n.s.), indirect effect $+0.254$; in subset passing attention/honesty check, $a = +0.429$, $b = +0.474$, $c = +0.265$, $c' = +0.062$ (n.s.), indirect effect $+0.203$. Overall, the direction of the treatment effect and the mediation structure (i.e., explanations raise \textit{Trust} predominantly by way of \textit{Understanding}) are consistent between the validation and the main analysis.

\section{Discussion}
\label{sec:discussion}

\subsection{Summary of Findings}

\xai visualizations aim to make \ai model behavior more understandable to humans, yet prior research shows that they can remain difficult for end-users to interpret \cite{Wangetal2022, martensBewareExplanationsAI2025}. This motivated our research on analyzing whether multimodal \llms can bridge this gap by providing an additional textual explanation layer that supports users in understanding \xai visualizations and affects their trust in the corresponding \ai model. 

\subsubsection{LLM Explanations Improve Perceived Understanding of XAI Visualizations}

RQ1 asked how \llm-generated textual explanations of \xai visualizations influence end-users' \textit{Perceived Understanding} of how an \ai system makes decisions. The increase of 0.352 points supports H1 and shows that adding a textual explanation can improve \textit{Perceived Understanding} relative to presenting the same \xai visualization alone.

This finding extends prior \xai research in several ways. Prior research has repeatedly shown that \xai visualizations can be difficult to interpret and unintuitive for users \cite{Wangetal2022}, especially for end-users without domain knowledge \cite{martensBewareExplanationsAI2025}. Prior work further suggests that textual explanations can help users understand \ai models \cite{miller2019explanation, liao2020questioning, kulesza2015principles, zhang2021relatable}. Our results add systematic empirical evidence that this also applies when a multimodal \llm interprets an \xai visualization and provides the resulting explanation as an additional textual layer.

Prior work on \llm-generated \xai explanations has mainly established the technical feasibility of our approach \cite{Mavrepisetal2024, nguyenLangXAI2024, Paraschouetal2025, zytekExplingoExplainingAI2024, Zyteketal2024, martensTellMeStory2025, kruschelChallengingPerformanceInterpretabilityTradeOff2025}, while the behavioral implications are unclear. To fill this gap, we examined whether an \llm-generated explanation placed beside a rendered \xai visualization can improve \textit{Perceived Understanding} relative to that same visualization alone. This is particularly relevant for end-users, such as managers, decision-makers, and domain professionals, whereas much existing \xai research focuses on engineers and technical users who design or validate models \cite{lipton2018mythos, doshivelez2017rigorous}. Explanations are increasingly applied in managerial and operational contexts to support decisions \cite{senonerUsingExplainableArtificial2022}, making the ability of non-technical users to interpret these outputs an important human-centered \xai problem.

\subsubsection{LLM Explanations Increase Perceived Trust of XAI Visualizations}

RQ2 asked how \llm-generated explanations of \xai visualizations affect \textit{Perceived Trust} in AI-assisted decisions among end-users. The estimated increase of 0.256 points supports H2 and shows that adding the textual explanation increased \textit{Perceived Trust} relative to the visualization alone.

This result adds to a literature in which the relationship between explanations and trust remains mixed. Prior work has found that explanations can increase or decrease trust depending on what they reveal about the model \cite{bansal2021does, buccinca2021trust, senonerExplainableAIImproves2024a}. Explanations can support trust when they help users assess whether predictions are based on meaningful model behavior \cite{lai2023towards}, but can also reduce trust when they expose limitations or questionable model reasoning \cite{bansal2021does, green2019principles}. So while our results shows that the specific \llm-generated explanations used in this study produced a positive trust effect across the seven tested \xai scenarios, other data suggests that there might be a more complex relationship between explanation design and \textit{Perceived Trust}.

\subsubsection{Perceived Understanding of XAI Visualizations Is Closely Linked to the Trust Benefit}

Building on H1 and H2, RQ3 asked to what extent \textit{Perceived Understanding} mediates the relationship between \llm-generated explanations and end-users' \textit{Perceived Trust}. The indirect effect of 0.24 (95\% bootstrap CI [0.12, 0.38]) supports H3 and provides evidence on the relationship between these two outcomes.

This finding adds to ongoing debates about whether explanations inherently increase user trust and confidence \cite{bansal2021does, buccinca2021trust, lai2023towards}. Explanations have, for example, been found not to improve \textit{Perceived Understanding} or \textit{Perceived Trust} calibration on unfamiliar tasks \cite{Wangetal2021, Wangetal2022, bobekUsercentricEvaluationExplainability2025}. In our setting, however, the positive trust effect was accompanied by a positive indirect effect through \textit{Perceived Understanding}. This supports the relevance of how users integrate explanations into their own mental models \cite{kulesza2015principles, liao2020questioning}, in addition to the visual or algorithmic qualities of interpretability tools \cite{lipton2018mythos}.

This finding has two explanations. The textual explanation could increase trust because (a)~an logical-sounding explanation could create the ''illusion'' of understanding \cite{chromikIllusionExplanatoryDepth2021} by making the \ai model decision appear more credible or because (b) it truly helps users form a better understanding of the information represented by the \xai visualization. Both interpretation can be supported by our findings that changes in \textit{Perceived Trust} were closely associated with changes in \textit{Perceived Understanding}. 

\subsubsection{Users With Less XAI Experience Benefit More}

The explanation effect differed by users' prior experience with \xai tools. Participants with less prior exposure to \xai tools showed larger gains in both \textit{Perceived Understanding} and \textit{Perceived Trust} ($p < .001$ for both outcomes), while the overall positive effect was not limited to inexperienced users. This aligns with prior work showing that users' background and expertise influence how \xai and \ai explanations are perceived \cite{ehsan2021who, Wangetal2021}.

One explanation is that users with less \xai experience have more room to benefit because they find the original visualization harder to interpret. A different explanation could be that experienced users may already know how to fully understand \xai visualizations, leaving less additional information for the textual layer to provide. A third potential explanation could be overconfidence of participants in their understanding. A similar descriptive pattern appears across scenarios: the two scenarios with the lowest baseline \textit{Perceived Understanding}, \ScenSix~Quality Driver (beeswarm plot, $M = 3.11$ without explanation) and \ScenSeven~Product Defect (saliency map, $M = 3.37$), also showed the largest gains ($\Delta = +0.414$ and $\Delta = +0.528$). Despite these, we observed that the effect of explanation is true for all level of \xai experience. 

These findings suggest that the value of an additional explanation layer may be particularly high where the gap between the technical artifact and the reader is larger \cite{belloThreelevelFrameworkLLMenhanced2025}.This is also compatible with cognitive load theory in which additional information can support users in processing complex information \cite{sweller1988cognitive}. For human-centered \xai, the implication is that explanation effectiveness depends not only on the technical explanation method, but also on the user who needs to interpret it.

\subsubsection{LLMs Can Act as an Interpretive Layer Between XAI and Users}

Our findings also provide a broader perspective on \xai and multimodal \llm-generated explanations as socio-technical systems. In our setting, the \llm does not explain the predictive model directly but interprets an \xai artifact that aims to represent the model's behavior and prediction. This introduces two distinct factor that influence the fidelity and effect of an explanation: (a) whether the \xai artifact faithfully represents the model, and (b) whether the \llm faithfully translates that artifact into natural language \cite{quExplainingUnintuitiveFeatures2025, zytekExplingoExplainingAI2024}.

When a multimodal \llm produces an explanation of an \xai artifact and humans subsequently interpret this explanation, automated explanation becomes a process involving the observed \ai model, the \xai visualization that aims to explain the \ai model, the multimodal \llm model that aim to interpret the \xai visualization and produces \xai explanation, and the user who perceive both the \xai visualization (including \ai model output) and \llm-generated \xai explanation. This shifts the perspective away from framing \xai as a static, one-directional artifact that aims is a direct our of the \ai model to the human \cite{green2019principles}. We therefore frame \llm-generated \xai explanation as boundary artifact \cite{starInstitutionalEcologyTranslations1989, leeBoundaryNegotiatingArtifacts2007, boundaryobjectCHI} between technical representations of model behavior and the users who need to interpret them. In this view, multimodal \llms can provide an additional interpretive layer without replacing the underlying \xai method.

\subsection{Practical Recommendations for Human-Centered XAI Design}
\label{sec:implications}

Our findings translate into four design principles for designing human-centered \xai systems.

\textbf{Principle 1: Pair technical \xai visualizations with concise natural-language interpretation.}
Augmenting \xai visualizations with textual explanations improved both \textit{Perceived Understanding} and \textit{Perceived Trust}. Our results suggest that dashboards, \ai interfaces, and decision support tools can benefit from using an additional textual explanation to support sense-making instead of relying on the visualization alone.

\textbf{Principle 2: Prioritize explanation support where users or visualizations require more interpretive effort.}
\llm-generated \xai explanation are not a one-size-fits-all solution to democratize \xai. In our study, participants with less prior exposure to \xai tools benefited more, and the two scenarios with the lowest baseline \textit{Perceived Understanding} also showed the largest gains. In HCI and \xai practice, one should therefore tailor the design and  consider the user's prior \xai experience and the accessibility of the underlying visualization when deciding where an additional explanation layer is most useful. 

\textbf{Principle 3: Provide task and model context when generating explanations.}
Our prompt design followed prior work on prompt and context engineering \cite{meiSurveyContextEngineering2025,huaContextEngineering202025,zhangAGENTICCONTEXTENGINEERING2026,girayPromptEngineeringChatGPT2023,linHowWriteEffective2024} and on LLM-based explanation generation \cite{Zyteketal2024}. We developed the prompt iteratively and manually compared alternative prompting methods against the expert-written reference explanations and selected the context-enriched configuration, which produced the closest outputs in these comparisons. We therefore suggest that practitioners implementing similar systems should therefore consider supplying as much context (e.g. information about the prediction task, model type, data domain, and \xai visualization) to the \llm to provide guidance for answering structures and to avoid potential hallucinations.

\textbf{Principle 4: Consider potential risks and tradeoffs.}
In our experiment, page interaction time was 5.3\% higher with explanations, but this difference was not statistically significant ($p = .092$). Prior research has further shown that explanations can have negative effects, including loss of trust, over-reliance, and an illusion of understanding \cite{chromikIllusionExplanatoryDepth2021, kimFosteringAppropriateReliance2025, schemmerAppropriateRelianceAI2023, poursabzi2021manipulating}. Our mediation result further shows that changes in \textit{Perceived Trust} are closely linked to \textit{Perceived Understanding}. Especially in high-stakes settings, explanation layers should therefore be evaluated not only through perceived outcomes but also through measures such as comprehension checks, error-detection tasks, and behavioral reliance.

\subsection{Limitations and Future Work}
\label{sec:limitations}
As with any empirical study, our findings should be considered in light of several limitations.

\emph{Sample and population.}
Our findings depend on the participant sample. Our sample covered a broad age range (19--72 years), included a mixed gender composition, and varied substantially in prior \xai experience. The validation sample ($N = 63$) showed similarly broad demographic coverage, with participants ranging from 19 to 74 years of age and a somewhat lower level of prior \xai-tool use. Both samples include highly educated and frequent users of \ai, so future studies should examine whether the effects extend to less educated populations. Nevertheless, the main effects remained consistent across robustness and validation checks. Together, these results provide evidence for the robustness and generalizability of the findings across samples, participant characteristics, analytical specifications, scenarios, and \llms.

\emph{Outcome measures and explanation format.}
Our primary outcomes capture \textit{Perceived Understanding} and \textit{Perceived Trust}; future work can complement these measures with objective comprehension, error-detection, and behavioral reliance measures. Participants also evaluated short, statically presented textual explanations, whereas interactive or conversational explanations may produce different user responses \cite{bertrandSelectiveMutableDialogic2023}. The present study therefore establishes the effects of a concise textual explanation layer on users' perceived outcomes, while future work may evaluate alternative interaction formats and outcome measures.

\emph{Model and prompt.}
Our results depend on the prompt configuration and \llms used to generate the explanations. Nevertheless, our studies used two different frontier multimodal \llms (i.e., GPT-4o and GPT-5.6-Sol) and showed robust overall pattern of positive effects of \llm-generated \xai explanations, which thus provides evidence for the generalizability across LLMs. Future work can examine additional models and prompt configurations, including adaptive and context-aware explanations tailored to different users and decision settings (see \cite{nimmoUserCharacteristicsExplainable2024} for inspiration).

\section{Conclusion}

\xai visualizations are increasingly used to communicate \ai model behavior and predictions, yet they often remain difficult for end-users to interpret. In response, our study demonstrates that augmenting \xai visualizations with \llm-generated textual explanations can improve both \textit{Perceived Understanding} and \textit{Perceived Trust}. Across a pre-registered within-subject experiment ($N = 128$) and a independent validation ($N = 63$), we observed the same overall positive pattern across seven scenarios spanning diverse applied decision contexts, including employment, credit, healthcare, forecasting, and quality management, using explanations generated by two different multimodal \llms. Our findings provide evidence for researchers and system designers seeking to make established \xai methods more accessible without replacing the underlying explanation technique. As \ai systems become increasingly embedded in everyday and professional decision-making, providing users with accessible ways to interpret existing \xai outputs will be important for supporting meaningful human oversight and interaction with \ai.

\bibliographystyle{ACM-Reference-Format}
\bibliography{bib}

\begin{appendices}

\section{Additional Analysis and Visualizations}
\label{app:additional_analysis}

\subsection{Research Design Overview}

\begin{figure}[H]
    \centering
    \includegraphics[width=0.78\linewidth]{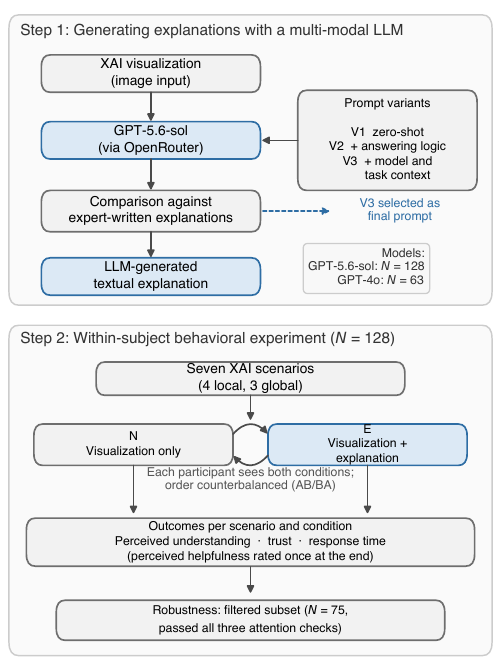}
    \caption{Research design. Our study consists of two steps: (1)~we generate explanations using context-engineered prompting with a multimodal \llm to create textual descriptions of \xai visualizations (GPT-5.6-sol, as shown; the validation stimuli were generated with GPT-4o); and (2)~we perform a within-subject behavioral experiment ($N = 128$) where participants evaluate seven \xai scenarios in two conditions (i.e., with vs. without LLM-generated explanations) to measure effects on Perceived Understanding and Perceived Trust. Each participant sees both conditions, with the presentation order counterbalanced (AB/BA). The three prompt variants of step~(1) are described in Section~\ref{sec:method_explanations}, and the robustness subset of step~(2) in Appendix~\ref{app:robustness}.}
    \Description{A two-step flow diagram. Step one, generating explanations with a multimodal \llm, runs from an \xai visualization supplied as image input, through GPT-5.6-sol accessed via OpenRouter, to a comparison against expert-written explanations, and then to the LLM-generated textual explanation; a box to the side lists the three prompt variants, zero-shot, plus answering logic, and plus model and task context, and notes that the third was selected as the final prompt. Step two, headed as a within-subject behavioral experiment with 128 participants, runs from the seven \xai scenarios, four local and three global, through the two conditions, visualization only and visualization plus explanation, which every participant saw in counterbalanced AB or BA order shown as a loop between the two condition boxes, to the outcomes recorded per scenario and condition, Perceived Understanding, Perceived Trust and response time, with perceived helpfulness rated once at the end, and finally to a robustness subset of the 75 participants who passed all three attention checks.}
    \label{fig:researchdesign}
\end{figure}

\subsection{Supplementary Result Figures}
\label{app:supp_figures}

The figures below supplement results already reported numerically in the main text (Tables~\ref{tab:descriptive_stats} and~\ref{tab:hypothesis_summary}).

\begin{figure}[H]
    \centering
    \includegraphics[width=\linewidth]{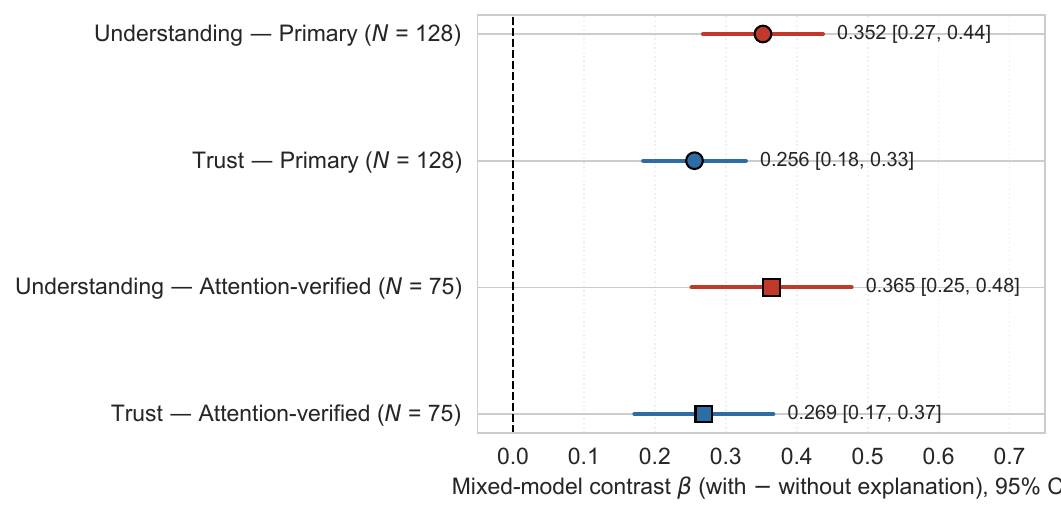}
    \caption{Forest plot of the mixed-model contrast (with $-$ without explanation) with 95\% confidence intervals, for both samples: in the primary sample ($N = 128$), \textit{Perceived Understanding} $\beta$ = 0.352 [0.27, 0.43] and \textit{Perceived Trust} $\beta$ = 0.256 [0.18, 0.33]; in the attention-verified subset ($N = 75$), \textit{Perceived Understanding} $\beta$ = 0.378 [0.27, 0.48] and \textit{Perceived Trust} $\beta$ = 0.278 [0.18, 0.37]. All four effects are statistically significant ($p$ < .001) and no interval crosses zero. All four estimates come from the pre-registered specification of Eq.~\eqref{eq:lmm}.}
    \Description{A forest plot with four rows showing the mixed-model contrast between the with-explanation and without-explanation conditions with 95 percent confidence intervals, each printed beside its row. For the primary sample of 128 participants, Perceived Understanding is 0.352 with an interval from 0.27 to 0.43 and Perceived Trust is 0.256 with an interval from 0.18 to 0.33. For the attention-verified sample of 75 participants, Perceived Understanding is 0.378 with an interval from 0.27 to 0.48 and Perceived Trust is 0.278 with an interval from 0.18 to 0.37. All four point estimates are positive and all four intervals lie clearly to the right of the zero reference line. The Perceived Understanding contrasts are larger than the Perceived Trust contrasts, and within each outcome the primary and attention-verified estimates are close to one another.}
    \label{fig:forest_plot}
\end{figure}

\begin{figure}[H]
    \centering
    \includegraphics[width=\linewidth]{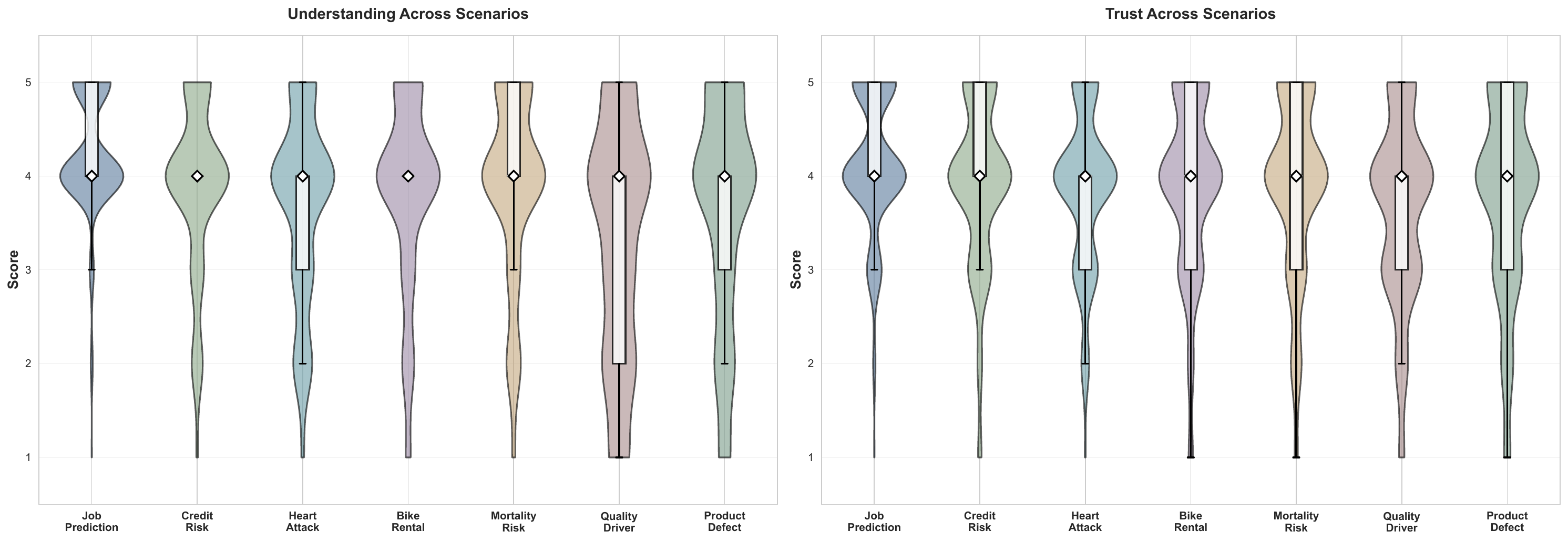}
    \caption{Distribution of \textit{Perceived Understanding} and \textit{Perceived Trust} assessments across all seven scenarios, pooled across both conditions for the primary sample ($N = 128$). Boxes span the interquartile range, whiskers extend to 1.5$\times$IQR, and the diamond marks the median. \ScenSix~Quality Driver shows notably lower \textit{Perceived Understanding} ratings ($M = 3.32$), while \ScenOne~Job Prediction shows the highest levels for both outcomes ($M = 4.26$ and $M = 4.18$). Because the response scale is discrete, all seven scenarios share a median of 4 on both outcomes, so the differences between them appear in the spread rather than in the median.}
    \Description{Two panels of violin plots with box plots overlaid, Perceived Understanding across scenarios and Perceived Trust across scenarios, each showing the distribution of ratings on a one-to-five scale for all seven scenarios, pooling both conditions. All fourteen distributions share a median of four, marked by a diamond, so the scenarios differ in spread rather than in central tendency. Quality Driver stands out by the width of its lower tail in the Perceived Understanding panel, with much more mass in the lower half of the scale than any other scenario, and its interquartile range reaches down to 2; Job Prediction is the most tightly concentrated at the top, with an interquartile range from 4 to 5. The corresponding scenario means, reported in the table of descriptive statistics, run from 3.32 for Quality Driver up to 4.26 for Job Prediction on Perceived Understanding, and from 3.65 for Quality Driver up to 4.18 for Job Prediction on Perceived Trust.}
    \label{fig:scenarios}
\end{figure}

\begin{figure}[H]
    \centering
    \includegraphics[width=\linewidth]{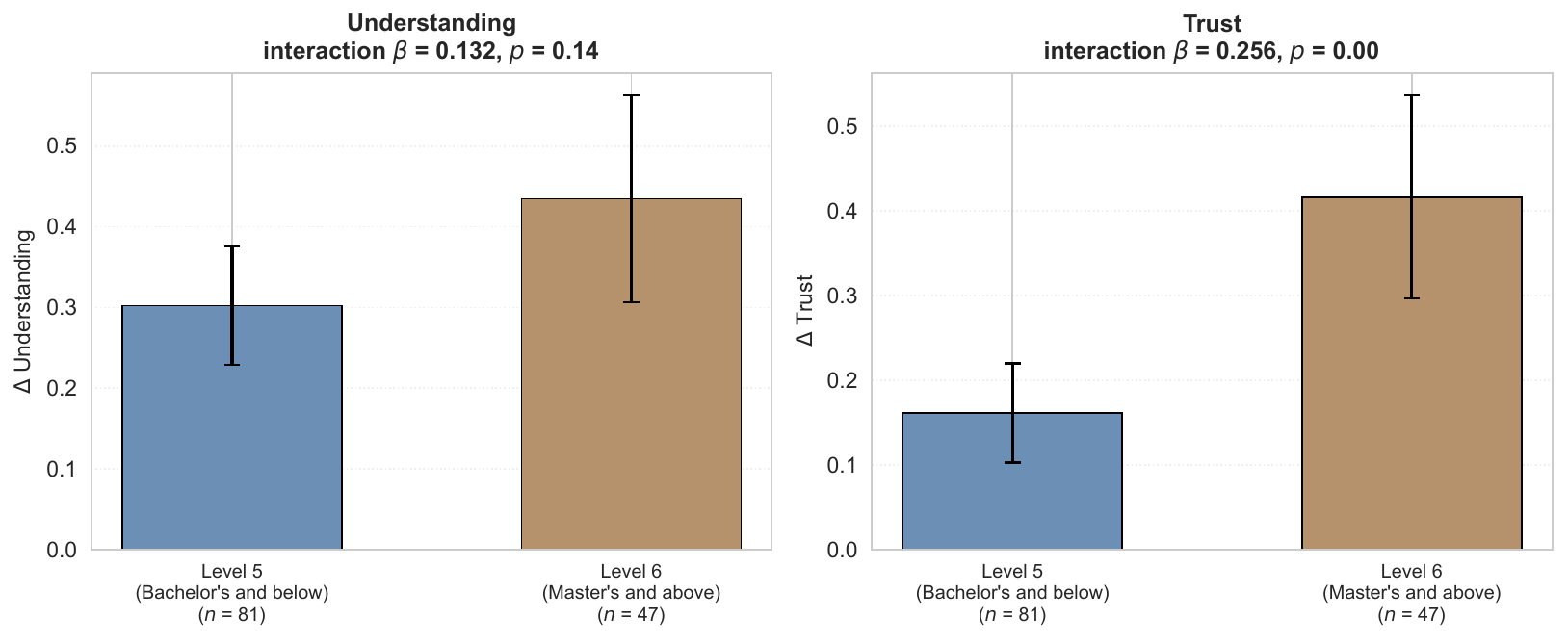}
    \caption{Change in \textit{Perceived Understanding} and \textit{Perceived Trust} from adding an explanation, by education level (Bachelor's degree and below, $n = 81$; Master's degree or higher, $n = 47$, of whom five reported a doctoral and two a professional degree such as a JD or MD). Bars show the participant-level mean change with error bars of $\pm 1$ standard error, and the panel titles report the \textit{Explanation}~$\times$~\textit{Education} interaction from the pre-registered specification of Eq.~\eqref{eq:lmm}. Both groups gained comparably in \textit{Perceived Understanding} ($\Delta = +0.436$ for Master's or higher vs.\ $+0.304$ for Bachelor's and below), whereas the gain in \textit{Perceived Trust} was more than twice as large in the more educated group ($\Delta = +0.418$ vs.\ $+0.162$); because education was not part of the planned design, we treat this as exploratory.}
    \Description{Two side-by-side bar charts comparing the explanation benefit for two education groups, bachelor's degree and below against master's degree and above, one panel for Perceived Understanding and one for Perceived Trust, each with error bars and the group sizes, 81 participants in the lower-education group and 47 in the higher one. In the Perceived Understanding panel the mean gain is 0.30 for bachelor's degree and below against 0.44 for master's degree and above, with an interaction coefficient of 0.128 and p equal to .132 reported in the panel title. In the Perceived Trust panel it is 0.16 against 0.42, with an interaction coefficient of 0.255 and p below .001. In both panels the higher-education group shows the larger mean gain, and the gap is much wider for Perceived Trust than for Perceived Understanding; the Perceived Understanding interaction is not significant while the Perceived Trust interaction is.}
    \label{fig:education}
\end{figure}

\subsubsection{Exploratory Trajectory and Response Visualizations}

\begin{figure}[H]
    \centering
    \includegraphics[width=\linewidth]{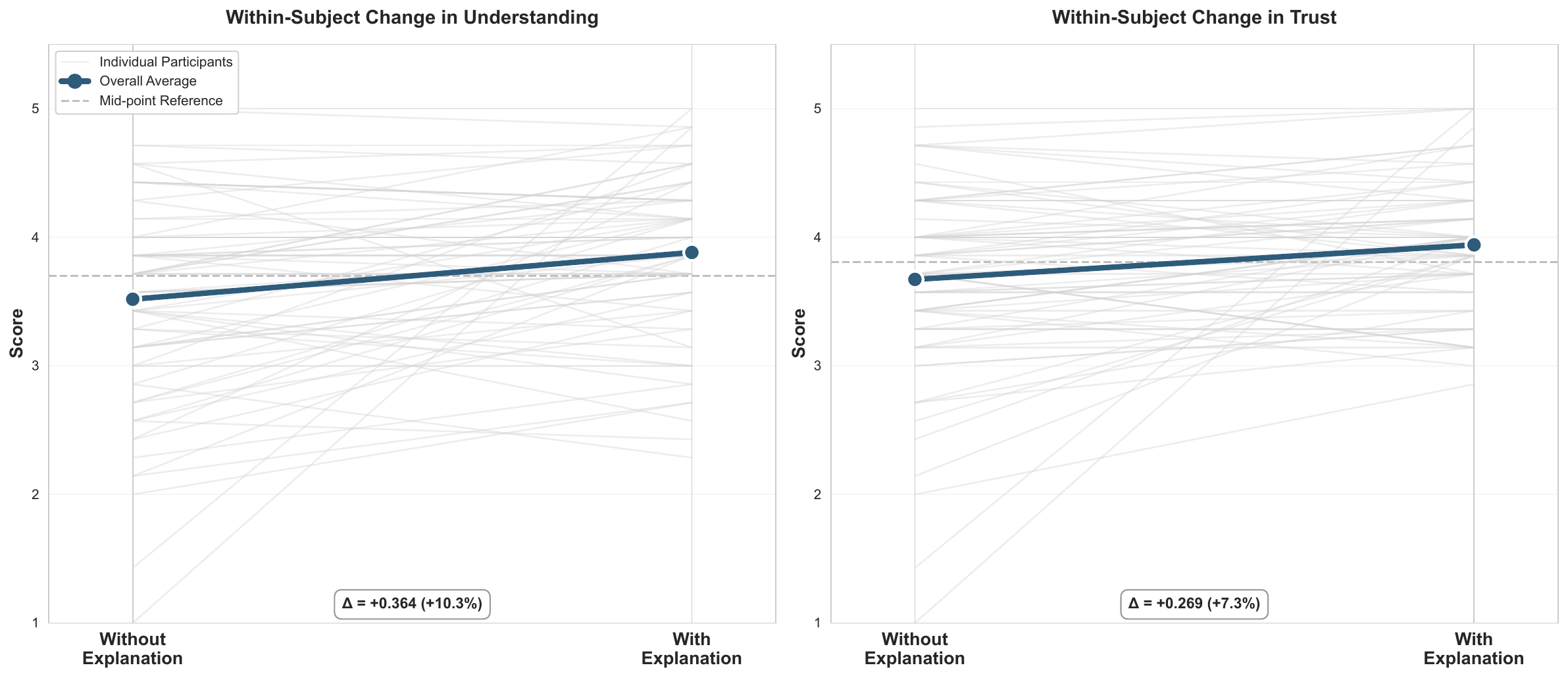}
    \caption{Individual trajectories of \textit{Perceived Understanding} and \textit{Perceived Trust} responses across all participants ($N = 128$), showing the effect of adding explanations. Gray lines represent individual participant trajectories; the thick blue line shows the overall average change. Most participants show upward trajectories (improvement with explanations), though the distribution is heterogeneous, consistent with the random-slope variance reported in Appendix~\ref{app:random_slope}.}
    \Description{Two panels of individual within-subject trajectory plots, one for Perceived Understanding and one for Perceived Trust, each plotting participant scores on a one-to-five axis moving from the without-explanation condition to the with-explanation condition. Thin gray lines show each of the 128 participants' individual change; a dashed horizontal mid-point reference line marks the without-explanation group average, and a thick dark blue line connects the overall average score in each condition. In the Understanding panel the average rises from about 3.53 to 3.89, an increase of 0.364 points (about +10.3\%); in the Trust panel it rises from about 3.70 to 3.94, an increase of 0.269 points (about +7.3\%). Most gray lines slope upward from left to right, but the slopes vary considerably, with some participants flat or declining, illustrating the participant-to-participant heterogeneity in the size of the explanation effect.}
    \label{fig:trajectories_within}
\end{figure}

\begin{figure}[H]
    \centering
    \includegraphics[width=\linewidth]{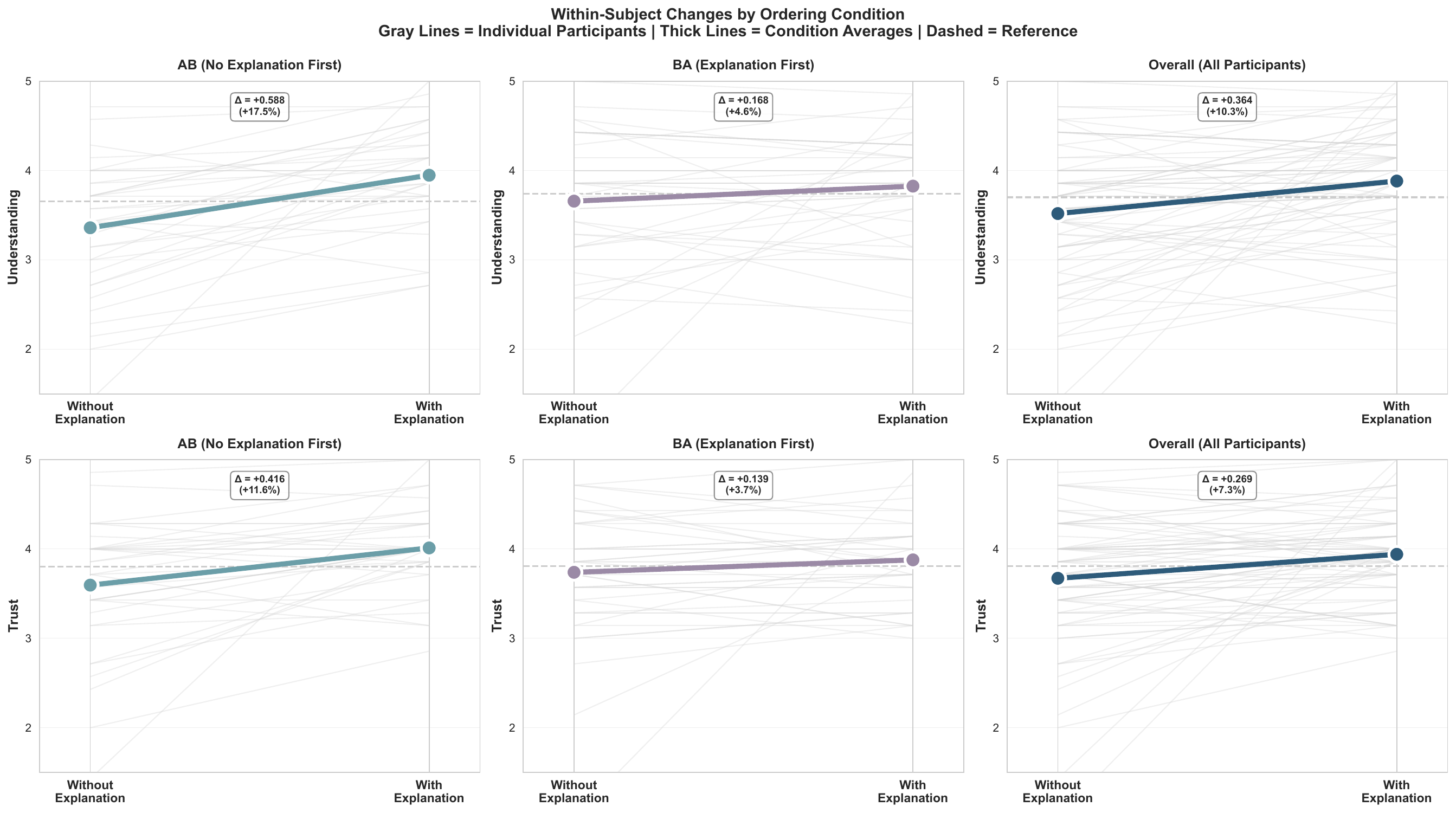}
    \caption{Within-subject trajectories split by presentation order (AB sequence: no-explanation first, $n = 64$; BA sequence: explanation first, $n = 64$). Gray lines show individual trajectories; colored thick lines show condition averages. The AB sequence (upper row) shows larger improvements than the BA sequence (middle row), consistent with a learning effect: participants who first interpreted visualizations unaided gained more from subsequently seeing explanations. Right column shows the overall average trajectory across all participants.}
    \Description{A grid of six within-subject trajectory plots, arranged in two rows (Understanding on top, Trust on bottom) and three columns (the AB sequence, where the no-explanation condition came first; the BA sequence, where the explanation came first; and an Overall panel pooling all participants), each plotting scores on a one-to-five axis from without-explanation to with-explanation. Thin gray lines show individual participant trajectories and a thick colored line shows the condition average, teal for AB, purple for BA, dark blue for Overall, with a dashed reference line at the without-explanation average. For Understanding, the average increases by 0.588 points (+17.5\%) in the AB sequence, by only 0.168 points (+4.6\%) in the BA sequence, and by 0.364 points (+10.3\%) overall; the AB panel's average line is visibly steeper than the BA panel's, which is nearly flat. For Trust, the corresponding increases are 0.416 points (+11.6\%) for AB, 0.139 points (+3.7\%) for BA, and 0.269 points (+7.3\%) overall, showing the same pattern of a much steeper rise when participants saw the unaided visualization first.}
    \label{fig:trajectories_by_order}
\end{figure}

\begin{figure}[H]
    \centering
    \includegraphics[width=\linewidth]{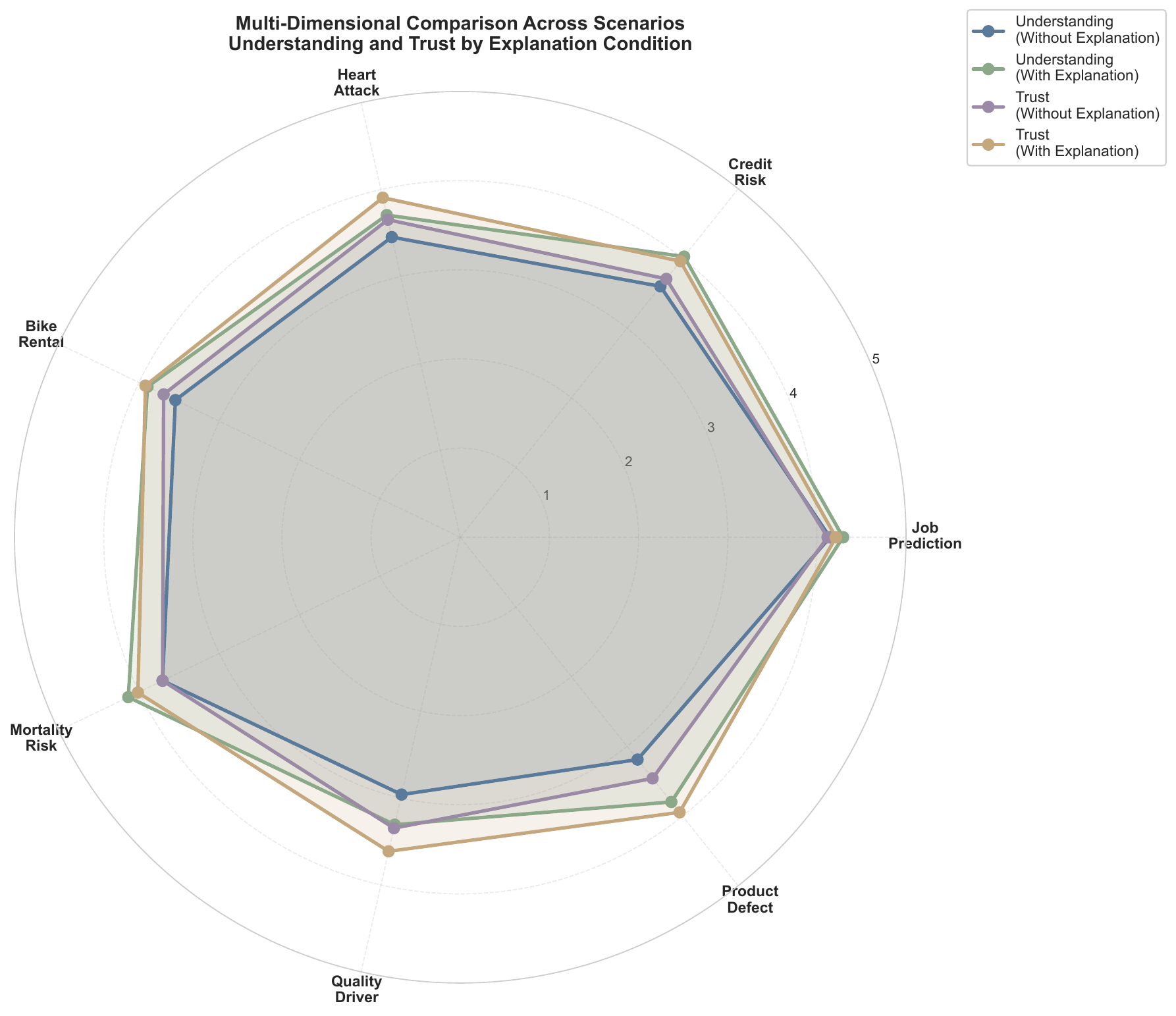}
    \caption{Radar (spider) plot comparing mean \textit{Perceived Understanding} and \textit{Perceived Trust} across all seven scenarios. Green/teal areas represent the with-explanation condition; blue/purple areas represent the without-explanation condition. Larger areas indicate higher average scores. \ScenSix~Quality Driver and \ScenSeven~Product Defect show the most dramatic improvements with explanations on both outcomes.}
    \Description{A seven-axis radar (spider) chart with one axis per scenario, arranged as a heptagon: Job Prediction, Credit Risk, Heart Attack, Bike Rental, Mortality Risk, Quality Driver, and Product Defect, each on a radial scale from one to five. Four overlaid polygons trace mean scores per scenario: blue for Understanding without explanation, green for Understanding with explanation, purple for Trust without explanation, and tan for Trust with explanation. At most vertices the green and tan (with-explanation) lines sit slightly outside the blue and purple (without-explanation) lines, indicating higher average scores with explanations. The gap between the with- and without-explanation lines is narrow at Job Prediction, where all four lines nearly coincide near the outer edge of the chart, reflecting an already-high baseline. The gap is widest at Quality Driver and Product Defect, where the with-explanation lines extend noticeably further out than the without-explanation lines for both Understanding and Trust, consistent with those two scenarios showing the largest gains.}
    \label{fig:radar_scenarios}
\end{figure}

\begin{figure}[H]
    \centering
    \includegraphics[width=\linewidth]{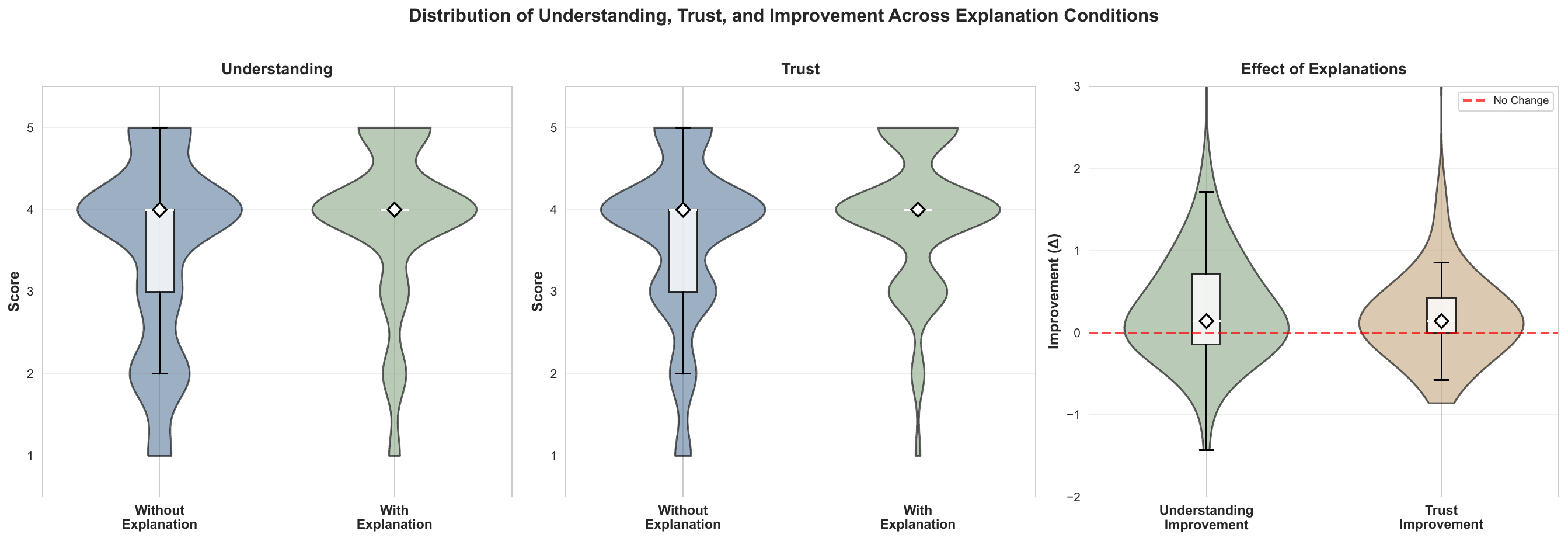}
    \caption{Combined view of the \textit{Perceived Understanding} and \textit{Perceived Trust} score distributions from Figure~\ref{fig:distributions}, alongside the distribution of participant-level improvement scores ($\Delta$ = with $-$ without explanation) for both outcomes (primary sample, $N = 128$). The improvement panel visualizes, at the participant level, the same contrasts reported in Table~\ref{tab:hypothesis_summary} (\textit{Perceived Understanding} $\beta$ = 0.352; \textit{Perceived Trust} $\beta$ = 0.256); most of each improvement distribution lies above the zero (no-change) reference line, consistent with those significant positive mean differences.}
    \Description{A three-panel figure titled Distribution of Understanding, Trust, and Improvement Across Explanation Conditions. The left panel is a violin plot with a box plot overlaid for Perceived Understanding, comparing the without-explanation and with-explanation conditions on a one-to-five score axis; the median in both conditions is four, marked by a diamond, and the interquartile range moves from three to four without explanation up to four to five with explanation. The middle panel shows the same comparison for Perceived Trust, with the same median-of-four pattern and the same upward shift in the interquartile range. The right panel, titled Effect of Explanations, shows two violin-and-box distributions of the per-participant improvement score on an axis running from about minus two to plus three: one for Understanding Improvement and one for Trust Improvement, with a dashed red horizontal line at zero labeled No Change. Both improvement distributions are shifted above the zero line, with their median diamonds sitting slightly above zero and most of each interquartile box also above zero, though each distribution has a long lower tail reaching into negative values for participants whose ratings did not improve or slightly declined; the Understanding improvement distribution appears somewhat more spread out than the Trust improvement distribution.}
    \label{fig:combined_distributions}
\end{figure}

\subsection{Page Interaction Time Analysis}
\label{app:response_time}

Table~\ref{tab:response_time_detailed} presents page interaction time (the duration on each scenario page before submitting ratings) by scenario for the main sample; Table~\ref{tab:response_time_validation} reports the same breakdown for the October 2025 validation sample (Section~\ref{sec:validation}, $N = 63$).

\begin{table}[H]
\centering
\caption{Page Interaction Time by Scenario, Main Sample ($N = 128$)}
\label{tab:response_time_detailed}
\small
\begin{tabular}{@{}lcccc@{}}
\toprule
\textbf{Scenario} & \multicolumn{2}{c}{\textbf{Median Time (s)}} & \textbf{Diff.} & \textbf{\% Chg.} \\
\cmidrule(lr){2-3}
 & \NoExpl & \WithExpl & (s) &  \\
\midrule
\ScenOne~Job Prediction & 60.8 & 43.0 & $-17.8$ & $-29.3\%$ \\
\ScenTwo~Credit Risk & 28.4 & 29.0 & $+0.6$ & $+2.3\%$ \\
\ScenThree~Heart Attack & 23.0 & 26.6 & $+3.6$ & $+15.7\%$ \\
\ScenFour~Bike Rental & 17.0 & 19.4 & $+2.5$ & $+14.5\%$ \\
\ScenFive~Mortality Risk & 18.0 & 19.3 & $+1.3$ & $+7.4\%$ \\
\ScenSix~Quality Driver & 15.6 & 17.9 & $+2.3$ & $+14.8\%$ \\
\ScenSeven~Product Defect & 16.0 & 14.8 & $-1.2$ & $-7.3\%$ \\
\midrule
\textbf{Overall} & 21.4 & 22.7 & $+1.4$ & $+6.4\%$ \\
\bottomrule
\end{tabular}
\end{table}
\Description{A seven-scenario response time table for the main sample (N=128) with columns: Scenario, Median Time Without explanation, Median Time With explanation, Difference in seconds, Percent change. Job Prediction: 60.8s without vs 43.0s with, diff -17.8s, -29.3\%. Credit Risk: 28.4s vs 29.0s, +0.6s, +2.3\%. Heart Attack: 23.0s vs 26.6s, +3.6s, +15.7\%. Bike Rental: 17.0s vs 19.4s, +2.5s, +14.5\%. Mortality Risk: 18.0s vs 19.3s, +1.3s, +7.4\%. Quality Driver: 15.6s vs 17.9s, +2.3s, +14.8\%. Product Defect: 16.0s vs 14.8s, -1.2s, -7.3\%. Overall: 21.4s vs 22.7s, +1.4s, +6.4\%. Five of seven scenarios show longer times with explanations, two show shorter times. The overall increase was not statistically significant.}

\noindent\textit{Note:} Median times reduce the influence of extreme outliers and include rating entry, not reading time alone. All 129 respondents with at least one rating recorded complete timers for all seven scenarios (1{,}806 page views); one of these 129 is excluded from the rating models for lacking a complete outcome pair. Page times did \emph{not} rise with explanations in every scenario (five of seven longer, two shorter); the largest relative increases were \ScenThree~Heart Attack (+15.7\%) and \ScenSix~Quality Driver (+14.8\%), and the one substantial decrease was \ScenOne~Job Prediction ($-29.3\%$), which also had by far the longest baseline (60.8~s) as the only text-modality scenario. Overall median time rose from 21.4~s to 22.7~s (+6.4\%); the log-scale model estimate is not distinguishable from zero ($\beta = 0.052$, $SE = 0.031$, $z = 1.68$, $p$ = .092, main text). As pre-registered, we do not treat page time as an index of cognitive load, since the timer also runs during rating entry and the page was not tab-locked.

\begin{table}[H]
\centering
\caption{Page Interaction Time by Scenario, October 2025 Validation Sample ($N = 63$)}
\label{tab:response_time_validation}
\small
\begin{tabular}{@{}lcccc@{}}
\toprule
\textbf{Scenario} & \multicolumn{2}{c}{\textbf{Median Time (s)}} & \textbf{Diff.} & \textbf{\% Chg.} \\
\cmidrule(lr){2-3}
 & \NoExpl & \WithExpl & (s) &  \\
\midrule
\ScenOne~Job Prediction & 51.3 & 92.5 & $+41.2$ & $+80.2\%$ \\
\ScenTwo~Credit Risk & 27.0 & 50.7 & $+23.7$ & $+88.1\%$ \\
\ScenThree~Heart Attack & 18.8 & 48.5 & $+29.7$ & $+158.2\%$ \\
\ScenFour~Bike Rental & 17.1 & 37.9 & $+20.8$ & $+121.0\%$ \\
\ScenFive~Mortality Risk & 14.7 & 31.5 & $+16.8$ & $+114.2\%$ \\
\ScenSix~Quality Driver & 16.1 & 30.7 & $+14.6$ & $+90.1\%$ \\
\ScenSeven~Product Defect & 14.7 & 31.6 & $+16.9$ & $+115.3\%$ \\
\midrule
\textbf{Overall} & 21.4 & 41.1 & $+19.7$ & $+92.2\%$ \\
\bottomrule
\end{tabular}
\end{table}
\Description{A seven-scenario response time table for the validation sample (N=63, collected October 2025) with columns: Scenario, Median Time Without explanation, Median Time With explanation, Difference in seconds, Percent change. Job Prediction: 51.3s without vs 92.5s with, diff +41.2s, +80.2\%. Credit Risk: 27.0s vs 50.7s, +23.7s, +88.1\%. Heart Attack: 18.8s vs 48.5s, +29.7s, +158.2\%. Bike Rental: 17.1s vs 37.9s, +20.8s, +121.0\%. Mortality Risk: 14.7s vs 31.5s, +16.8s, +114.2\%. Quality Driver: 16.1s vs 30.7s, +14.6s, +90.1\%. Product Defect: 14.7s vs 31.6s, +16.9s, +115.3\%. Overall: 21.4s vs 41.1s, +19.7s, +92.2\%. All seven scenarios in the validation sample show substantially longer times with explanations, ranging from +80\% to +158\%, in sharp contrast to the main sample which ranged from -29\% to +16\%.}

\noindent\textit{Note:} Unlike the main sample, every scenario in the validation sample shows a substantially \emph{longer} median time with explanations ($+80\%$ to $+158\%$ per scenario, vs.\ $-29\%$ to $+16\%$ in the main sample). We report this divergence descriptively rather than explain it away: the validation set used the same instrument and platform, so the difference in magnitude and direction is a genuine discrepancy between sample and not a data artifact, and it should temper any claim that the null time effect in the main sample generalizes.

\subsection{Robustness Checks}
\label{app:robustness}

\subsubsection{Sample Comparison}

Table~\ref{tab:robustness_samples} compares the pre-registered primary sample ($N = 128$) against its attention-verified subset ($N = 75$), and adds the corresponding comparison for the October 2025 validation sample ($N = 63$; attention-verified $N = 41$; \ref{sec:validation}). As pre-registered, neither primary analysis excludes on attention-check failure; both subsets are sensitivity analyses only.

\begin{table}[H]
\centering
\caption{Robustness Check: Full vs.\ Attention-Verified Subsets, Main and Validation Samples}
\label{tab:robustness_samples}
\small
\begin{tabular}{@{}llcccc@{}}
\toprule
\textbf{Outcome} & \textbf{Sample} & \textbf{Effect} & $\boldsymbol{\beta}$ & $\boldsymbol{z}$ & $\boldsymbol{p}$ \\
\midrule
\multirow{4}{*}{Perceived Understanding} & Main ($N=128$) & Explanation & 0.352 & 8.57 & <.001 \\
 & Main, attn.-verified ($N=75$) & Explanation & 0.378 & 7.04 & <.001 \\
 & Validation ($N=63$) & Explanation & 0.433 & 6.69 & <.001 \\
 & Validation, attn.-verified ($N=41$) & Explanation & 0.429 & 5.24 & <.001 \\
\midrule
\multirow{4}{*}{Perceived Trust} & Main ($N=128$) & Explanation & 0.256 & 7.06 & <.001 \\
 & Main, attn.-verified ($N=75$) & Explanation & 0.278 & 5.73 & <.001 \\
 & Validation ($N=63$) & Explanation & 0.299 & 5.29 & <.001 \\
 & Validation, attn.-verified ($N=41$) & Explanation & 0.265 & 3.77 & <.001 \\
\bottomrule
\end{tabular}
\end{table}
\Description{A robustness check table comparing effect estimates across sample and attention-check subsets, with columns: Outcome, Sample, Effect type, beta coefficient, z-statistic, p-value. Four rows for Perceived Understanding: main full sample N=128 yields beta=0.352, z=8.57, p<.001; main attention-verified N=75 yields beta=0.378, z=7.04, p<.001; validation full N=63 yields beta=0.433, z=6.69, p<.001; validation attention-verified N=41 yields beta=0.429, z=5.24, p<.001. Four rows for Perceived Trust: main N=128 yields beta=0.256, z=7.06, p<.001; main (filtered) N=75 yields beta=0.278, z=5.73, p<.001; validation N=63 yields beta=0.299, z=5.29, p<.001; validation (filtered) N=41 yields beta=0.265, z=3.77, p<.001. All eight effect estimates are positive and statistically significant, indicating stable results across samples and attention-check inclusion/exclusion.}

\noindent\textit{Note:} Main-sample rows use the pre-registered specification of Eq.~\eqref{eq:lmm}; validation-sample rows use the corresponding participant-level model on the October 2025 data. Within each sample, the attention-verified estimate is close to full-sample estimate; in no case does restricting to attentive participants change the sign or significance of the effect. This stability indicates the results are not driven by inattentive responding, and that retaining all analyzable participants, as pre-registered, is the more conservative choice.

\subsubsection{Internal Consistency of Measurements}

Cronbach's alpha \cite{cronbachCoefficientAlphaInternal1951} was computed separately for \textit{Perceived Understanding} and \textit{Perceived Trust}, treating the seven scenarios as items ($n = 128$ participants with a complete set of seven scenario scores): $\alpha = 0.806$ for \textit{Perceived Understanding}, $0.842$ for \textit{Perceived Trust} (attention-verified subset: $0.799$, $0.798$, $n = 74$). We report these because several analyses aggregate ratings across scenarios (the participant-level mediation, Section~\ref{sec:results_h3}; the associations with self-report covariates, Section~\ref{sec:results_ai_experience}), which is only meaningful if a participant's ratings hold together across scenarios. Each construct was measured with a single item per scenario and condition, so these coefficients index cross-scenario consistency rather than multi-item scale reliability, and moderate rather than high values are expected given that the scenarios deliberately span different contexts, \xai methods, and modalities. Values above $0.80$ nonetheless indicate a fairly stable person-level component, consistent with the explanation benefit not differing detectably between scenarios (Figure~\ref{fig:scenario_effects}).

\subsubsection{Order and Carryover Effects}

Because \textit{Explanation} is the exclusive-or of \textit{Sequence} and \textit{Block} in the AB/BA crossover, \textit{Block} is dropped from the interaction model to avoid saturation. Sequence moderated \textit{Perceived Understanding} ($\beta = -0.206$, $SE = 0.082$, $\chi^2$(1) = 6.28, $p$ = .012, $q$ = .049) but not \textit{Perceived Trust} ($\beta = -0.062$, $\chi^2$(1) = 0.74, $p$ = .389); the coefficient is negative because \textit{Sequence} is coded with BA (explanation first) as the contrast level. In the AB sequence (no explanation first, $n = 64$), \textit{Perceived Understanding} was $+0.455$ points higher with explanations ($SE = 0.086$); in the BA sequence (explanation first, $n = 64$), the same contrast was $+0.247$ points ($SE = 0.099$). This asymmetry is consistent with a learning effect: participants who first interpreted visualizations unaided gained more from subsequently seeing explanations. The contrast remains positive in both sequences, and the overall main effect remains significant when \textit{Block} and \textit{Sequence} are controlled as in Eq.~\eqref{eq:lmm}.

\subsubsection{First-Block-Only Analysis (Carryover-Free)}
\label{app:first_block}

Because \textit{Explanation}, \textit{Block}, and \textit{Sequence} saturate the four sequence~$\times$~period cells, differential carryover cannot be separated from the treatment effect in the full data. As pre-registered, we therefore also estimated the effect on each participant's \emph{first} block only, a between-participant comparison free of carryover and within-person contrast effects (\textit{Explanation} + \textit{Scenario} with participant random intercepts; seven observations per participant, so these are by construction our least precise estimates): \textit{Perceived Understanding} $\beta = +0.349$, $SE = 0.121$, $z = 2.88$, $p$ = .004, 95\% CI [$+0.112$, $+0.586$] ($M = 3.56$ vs.\ $3.91$); \textit{Perceived Trust} $\beta = +0.095$, $SE = 0.112$, $z = 0.84$, $p$ = .400, 95\% CI [$-0.126$, $+0.315$] ($M = 3.83$ vs.\ $3.92$).

\textit{Perceived Understanding} reproduces at almost its full-sample magnitude under this strictest control for carryover; \textit{Perceived Trust} is not detectable in a purely between-participant comparison. We do not read this as evidence against H2: the analysis discards half the data and all within-person contrasts, so its standard error is roughly three times that of the primary model, and the main estimate ($\beta = 0.256$) lies comfortably inside this interval. The lower precision, not a changed point estimate, accounts for the loss of significance. What the analysis does establish is that our evidence for H2 rests on the within-participant contrast and is not independently confirmed between participants (Section~\ref{sec:limitations}). On the attention-verified subset: $\beta = +0.464$, $p$ = .004 (\textit{Perceived Understanding}); $\beta = +0.281$, $p$ = .055 (\textit{Perceived Trust}).

\subsubsection{Random-Slope Sensitivity}
\label{app:random_slope}

The registered model constrains the explanation effect to be identical across participants, which can understate the treatment contrast's standard error if the effect varies. We refitted \texttt{Explanation + Block + Sequence + Scenario + (1 + Explanation | Participant)} on the primary sample ($N = 128$, 1{,}782 assessments); both models converged. The treatment estimate is essentially unmoved while its $SE$ grows by about 60\%: \textit{Perceived Understanding} $+0.352$ ($SE = 0.041 \rightarrow 0.066$, 95\% CI [$+0.222$, $+0.481$]); \textit{Perceived Trust} $+0.255$ ($SE = 0.036 \rightarrow 0.059$, 95\% CI [$+0.140$, $+0.370$]). Both intervals still exclude zero ($p < .001$ for each), so H1 and H2 hold under either random-effect structure; the wider intervals coincide almost exactly with the participant-level paired-$t$ intervals reported below, which likewise impose no common slope.

The random slope itself is substantial ($SD = 0.490$ for \textit{Perceived Understanding}, $0.386$ for \textit{Perceived Trust}, of the same order as the mean effect), and correlates negatively with the random intercept ($r = -0.69$, $-0.76$): participants who rated low without an explanation gained the most. A likelihood-ratio test favors the random slope for both outcomes ($\chi^2$(2) = 101.5 and 117.0, both $p < .001$), read as a diagnostic rather than a model-selection rule given the boundary conservatism of variance tests. The registered model remains the primary analysis, but the between-participant variation is real and consistent with the moderation by prior \xai exposure (Section~\ref{sec:results_ai_experience}); a design powered to model that variation directly is left to future work.

\subsubsection{Aggregated Paired Tests}
\label{app:paired_tests}

The registered plan specified participant-aggregated paired $t$-tests and Wilcoxon signed-rank tests, imposing no distributional assumption and treating each participant as one unit ($N = 128$ pairs, one mean rating per condition per participant). \textit{Perceived Understanding}: mean change $+0.351$ ($SD = 0.748$, 95\% CI [$+0.220$, $+0.482$]), $t$(127) = 5.31, $p$ < .001, $d_z$ = 0.469; Wilcoxon $W$ = 1126.5, $p$ < .001. \textit{Perceived Trust}: mean change $+0.255$ ($SD = 0.660$, 95\% CI [$+0.140$, $+0.370$]), $t$(127) = 4.37, $p$ < .001, $d_z$ = 0.387; $W$ = 1194.0, $p$ < .001. Both tests agree with the registered mixed model ($+0.352$, $+0.256$).

The distribution is informative beyond its mean: the median change is $+0.143$ points on both outcomes, positive for 55.5\% of participants on \textit{Perceived Understanding} and 56.3\% on \textit{Perceived Trust}, exactly zero for 18.8\%/23.4\%, and negative for 25.8\%/20.3\%. The aggregate effect is therefore a distributional shift with a substantial minority showing no change or a decrease, consistent with the random-slope variance above.

\subsubsection{Ordinal Outcome Model (Not Pre-Registered)}
\label{app:ordinal}

As a robustness extension treating the five-point response only through category ordering, we fit a cumulative-logit (proportional-odds) model,
\begin{equation}
\mathrm{logit}\left(\Pr(Y_{ij} \le k)\right) = \tau_k - \left(\beta_1 \textit{Explanation}_{ij} + \beta_2 \textit{Block}_{ij} + \beta_3 \textit{Sequence}_i + \boldsymbol{\gamma}^\top \textit{Scenario}_{ij}\right),
\end{equation}
with $k = 1,\ldots,4$ cut points and \textit{Scenario} as six indicators against \ScenOne~Job Prediction, otherwise matching Eq.~\eqref{eq:lmm}. \texttt{statsmodels} provides no mixed-effects ordinal model, so no participant random intercept could be included; inference on $\beta_1$ instead uses a participant-level cluster bootstrap (1{,}000 resamples, fixed seed, all 14 rows per drawn participant retained together; all resamples converged for both outcomes).

\textit{Perceived Understanding}: $\beta_1 = 0.638$, odds ratio $1.89$, 95\% bootstrap CI [$1.52$, $2.38$] ([$0.420$, $0.866$] in log odds). \textit{Perceived Trust}: $\beta_1 = 0.474$, odds ratio $1.61$, 95\% CI [$1.31$, $1.94$] ([$0.272$, $0.663$] in log odds). Both intervals exclude one, so the main conclusion does not depend on treating the Likert response as interval-scaled. The proportional-odds assumption was accepted as an approximation and not formally tested; a single binary predictor gives little information for a parallel-slopes test, so this model is reported as a scale-assumption check rather than a competing primary specification.

\subsubsection{Further Pre-Registered Sensitivity Analyses}
\label{app:further_sensitivity}

Two further registered checks and two post hoc analyses (full detail in the analysis notebook):
\begin{itemize}
    \item \textbf{Leave-one-scenario-out} (pre-registered). Refitting Eq.~\eqref{eq:lmm} seven times, dropping one scenario each time, leaves the explanation contrast between $+0.323$ and $+0.379$ (\textit{Perceived Understanding}) and $+0.232$ and $+0.275$ (\textit{Perceived Trust}), all $p$ < .001; no single stimulus drives the aggregate effect.
    \item \textbf{Self-reported honesty} (pre-registered). Excluding nine participants who self-reported random responding or consulting external resources ($N = 119$) yields $\beta = +0.373$ (\textit{Perceived Understanding}) and $+0.256$ (\textit{Perceived Trust}), both $p$ < .001, i.e.\ unchanged or slightly larger.
    \item \textbf{Ordinal outcome model} (post hoc; full detail in Appendix~\ref{app:ordinal}). Odds ratios of $1.89$ [$1.52$, $2.38$] and $1.61$ [$1.31$, $1.94$] confirm the effect is not an artifact of treating the outcome as continuous \cite{liddellAnalyzingOrdinalData2018}.
    \item \textbf{Source-study performance cues} (post hoc stimulus-confound check). Three of seven explanation texts carried source-study performance cues (two positive, one negative; the \ScenSeven~Product Defect figures concern human inspection performance, not the model's own accuracy). Dropping the two positive-cue scenarios leaves $\beta = +0.305$/$+0.208$; dropping all three leaves $+0.334$/$+0.224$ (both $p$ < .001 throughout). The effect survives omitting explicit positive performance cues, though omitting scenarios also changes the stimulus set and every remaining explanation still shares the same structure, fluency, and framing.
\end{itemize}

\subsubsection{Completion Time}

Main sample ($N=128$): completion times ranged 4.9--35.8 minutes ($Mdn = 14.9$, $M = 15.4$, $SD = 5.3$; $Q_1 = 12.1$, $Q_3 = 18.2$), uncorrelated with overall \textit{Perceived Understanding} ($r = -0.08$, $p = .38$) or \textit{Perceived Trust} ($r = -0.09$, $p = .29$), or with the size of either explanation effect ($r = +0.13$, $p = .14$; $r = +0.07$, $p = .46$). The 18 participants (14.1\%) finishing in under 10 minutes showed comparable effect sizes to slower participants. In the October 2025 validation sample ($N=63$), completion times ranged 7.2--35.7 minutes ($Mdn = 15.1$, $M = 16.4$, $SD = 6.2$); 10 participants (15.9\%) finished in under 10 minutes, a comparable pattern to the main study.

\subsection{Qualitative Coding of Open-Ended Responses}
\label{app:keyword_dictionary}

The open-ended item read: \emph{``Based on the Explainations you saw for the XAI Visualization, did this change your Perceived Understanding of the AI model or your Perceived Trust in it? Please take at least 15 seconds to reflect on your answer and describe it here.''} All 128 analyzed participants responded (mean 133 characters, median 105).

The coding is a descriptive keyword count, not a thematic analysis: each response was lower-cased and tested for the substrings in Table~\ref{tab:keyword_dictionary} by plain case-insensitive substring matching, without stemming, lemmatization, or synonym expansion beyond the listed variants. Categories are not mutually exclusive or exhaustive, so percentages do not sum to 100\%, and matching is mechanical (counted regardless of negation), which is why a separate ``Skeptical / no change'' family captures explicit negations. The procedure is deterministic and exactly reproducible from the response texts, but carries the well-known validity limits of dictionary-based coding. A thematic analysis with independent coders is left to future work.

\begin{table}[H]
\centering
\caption{Keyword families for the descriptive coding of open-ended responses, with counts on the primary analyzed sample ($N = 128$). Matching is case-insensitive substring matching; categories overlap.}
\label{tab:keyword_dictionary}
\begin{tabular}{p{3.4cm}p{7.6cm}rr}
\toprule
\textbf{Category} & \textbf{Match strings} & \textbf{$n$} & \textbf{\%} \\
\midrule
Perceived Trust (general) & Perceived Trust, confidence, reliable, faith & 55 & 43.0 \\
Perceived Understanding & understand, understood, grasp, comprehend, knowledge & 81 & 63.3 \\
Clarity / clear & clear, clearer, clarity, clarif & 19 & 14.8 \\
Helpful & helpful, useful, beneficial, helped, assist & 21 & 16.4 \\
Confusion & confus, unclear, complicated, difficult & 7 & 5.5 \\
Visualization & visual, graph, chart, plot, diagram, colour, color & 39 & 30.5 \\
Explanation quality & explanation, explain, reason, justif & 66 & 51.6 \\
Decision making & decision, decide, predict, outcome & 31 & 24.2 \\
\midrule
\multicolumn{4}{l}{\textit{Derived combinations reported in the main text}} \\
Perceived Understanding or clarity & union of the two families above & 85 & 66.4 \\
Perceived Understanding or Perceived Trust & union of the three families above & 102 & 79.7 \\
Skeptical / no change & not Perceived Trust, less Perceived Trust, skeptic, sceptic, doubt, disPerceived Trust, no change, did not change, unchanged & 7 & 5.5 \\
\bottomrule
\end{tabular}
\end{table}
\Description{An eight-row keyword family coding dictionary for open-ended responses (N=128) with columns: Category, Match strings (substrings searched case-insensitively), count, and percentage of responses matching. Row 1 (Perceived Trust general): keywords trust, confidence, reliable, faith — 55 responses (43.0\%). Row 2 (Perceived Understanding): understand, understood, grasp, comprehend, knowledge — 81 (63.3\%). Row 3 (Clarity): clear, clearer, clarity, clarif — 19 (14.8\%). Row 4 (Helpful): helpful, useful, beneficial, helped, assist — 21 (16.4\%). Row 5 (Confusion): confus, unclear, complicated, difficult — 7 (5.5\%). Row 6 (Visualization): visual, graph, chart, plot, diagram, colour, color — 39 (30.5\%). Row 7 (Explanation quality): explanation, explain, reason, justif — 66 (51.6\%). Row 8 (Decision making): decision, decide, predict, outcome — 31 (24.2\%). Three derived rows show overlapping categories: Understanding or clarity union yielding 85 (66.4\%); Understanding or Trust union yielding 102 (79.7\%); Skeptical/no change yielding 7 (5.5\%). Matching is mechanical, case-insensitive substring-based, categories overlap, and percentages do not sum to 100\%.}

\begin{figure}[H]
    \centering
    \includegraphics[width=\linewidth]{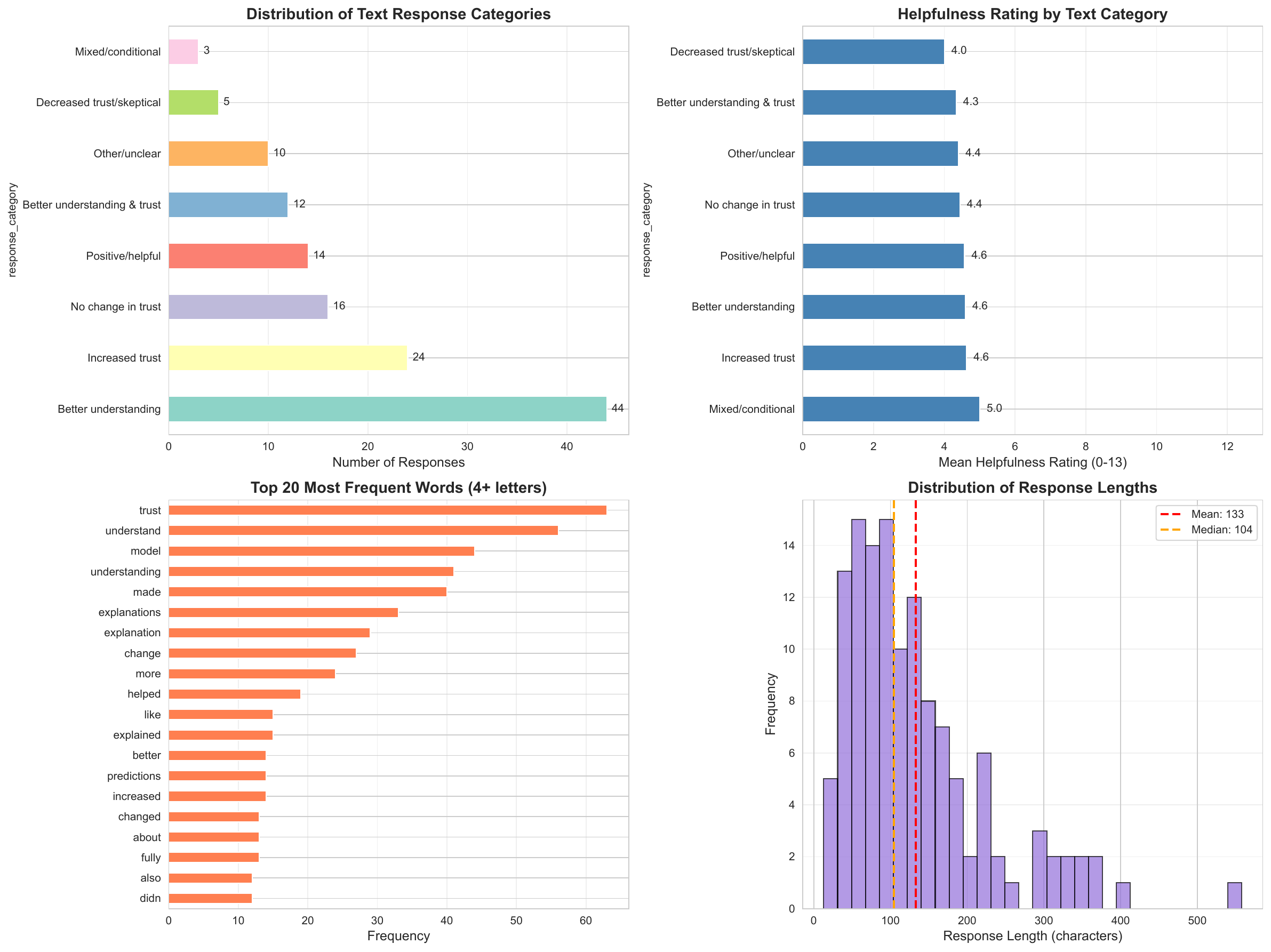}
    \caption{Qualitative coding of open-ended responses ($N = 128$ participants). Top-left panel shows the frequency of response categories; top-right panel shows the mean helpfulness rating (0--13 scale) by category; bottom-left panel shows the twenty most frequent words used; bottom-right panel shows the distribution of response lengths. Categories reflecting understanding and/or trust gains (better understanding, increased trust, or both) account for 80 responses (62.5\%); decreased-trust/skeptical and no-change-in-trust responses together account for 21 (16.4\%).}
    \Description{A four-panel figure summarizing qualitative coding of all 128 participants' open-ended text responses, each assigned to one mutually exclusive category. The top-left panel is a horizontal bar chart of response counts across eight categories: Better understanding (44), Increased trust (24), No change in trust (16), Positive/helpful (14), Better understanding and trust (12), Other/unclear (10), Decreased trust/skeptical (5), and Mixed/conditional (3). The top-right panel is a horizontal bar chart of the mean helpfulness rating (on a 0-to-13 scale) associated with responses in each category, ranging from 5.0 for Mixed/conditional down to 4.0 for Decreased trust/skeptical, with most other categories clustered between 4.3 and 4.6. The bottom-left panel is a horizontal bar chart of the twenty most frequent words of four or more letters used across responses, topped by trust, understand, model, and understanding, followed by words such as made, explanations, explanation, change, and helped. The bottom-right panel is a histogram of response length in characters, right-skewed with most responses between about 50 and 150 characters, a mean of 133 characters, a median of 104 characters, and a long tail of a few much longer responses extending out to roughly 550 characters.}
    \label{fig:text_responses}
\end{figure}

\section{Additional Study Details}
\label{app:additional_study_details}

\subsection{Participant Recruitment, Screening and Sample Composition}
\label{app:participants}

This section reports the screening qualifications, attention-check diagnostics, sample accounting, and demographic breakdown for the main sample; Section~\ref{sec:participants} gives the summary.

\subsubsection{Screening Qualifications}
\label{app:participants_screening}

Two qualifications apply. First, seven participants (5.5\% of the analyzed sample) self-reported a doctoral or professional degree despite the screener, so it did not fully exclude doctoral-level education. Second, prior \xai-tool use frequency was measured and spanned the full range, but a majority reported using explainable-AI tools weekly or more (Appendix~\ref{app:participants_demographics}). The sample therefore cannot be described as \xai-naive, and Section~\ref{sec:results_ai_experience} treats \xai-tool use frequency as a measured variable rather than something the recruitment filter controlled away.

\subsubsection{Attention Checks and Response Quality}
\label{app:participants_attention}

Data quality was addressed with three attention checks (one multiple-selection comprehension item; one logical item, whether 2010 came after 1990; one arithmetic item, whether $6 + 19 = 15$) and one honesty item asking whether participants had responded randomly or consulted external resources. As pre-registered, the primary analysis retains every participant with complete within-subject ratings regardless of attention-check outcome; the all-checks-passed subset and the honesty-based exclusion are reported separately as sensitivity analyses. Completion times and response patterns were additionally screened for inattentiveness; results were robust.

Excluding participants who failed any of the three checks removes 41.4\% of the sample ($N = 53$) and yields nearly identical coefficients (Appendix~\ref{app:robustness}). Failure concentrated in one item: 32.0\% ($n=41$) missed the 1990/2010 statement, vs.\ 14.1\% ($n=18$) on the comprehension item and 7.8\% ($n=10$) on the arithmetic item. Because a 32\% failure rate on a trivial item is more consistent with an instrument problem than with 41 inattentive participants, we examined it directly: failing it is associated with failing the other two (Fisher exact $p = .010$), but 27 of the 41 who failed it passed both others, their median completion time does not differ from those who passed (14.5 vs.\ 15.2 min, Mann--Whitney $p=.217$), and their explanation effect is no smaller (\textit{Perceived Understanding} $+0.416$ vs.\ $+0.320$, $p=.470$; \textit{Perceived Trust} $+0.246$ vs.\ $+0.259$, $p=.898$). The response options for this item cannot be reconstructed from the data export, so an answer-key or option-order error cannot be ruled out; we therefore treat the attention-verified subset as a sensitivity analysis, not the higher-quality sample, as pre-registered.

\subsubsection{Sample Accounting and Non-Response}
\label{app:participants_accounting}

Of 137 genuine submissions, eight consented but provided no scenario ratings (all left within 15--58 s of starting, before any stimulus was shown, and were never assigned a presentation sequence), leaving 129 respondents with at least one rating. A ten\textsuperscript{th} respondent skipped the \textit{Perceived Trust} item throughout the no-explanation block and contributes no complete outcome pair, leaving the analyzed $N = 128$. Ten further rows across the remaining sample have one item unanswered and are dropped by the complete-case rule, so the observation count is identical for both outcomes.

A formal comparison between the eight non-raters and the analyzed sample is not possible, since none of the eight reached the demographic block (age, gender, education, and \ai experience are missing for all). We therefore cannot empirically rule out non-response bias, though the attrition is small (8 of 137, 5.8\%) and occurred before any stimulus was shown, so it cannot reflect the explanation manipulation itself; because the design is within-subject, the estimated effects also do not depend on the two condition groups being demographically comparable.

\subsubsection{Demographic Breakdown and Participant-Level Associations}
\label{app:participants_demographics}

All participants reported their gender; none reported never having used \ai. \xai-tool use frequency skewed toward regular use (59.4\% weekly or more, 21.9\% occasionally, 18.8\% little or none), so the sample is experienced rather than mixed. \xai-tool use frequency was positively associated with the level of both outcomes ($r = 0.38$, $p<.001$ for \textit{Perceived Understanding}; $r = 0.27$, $p=.002$ for \textit{Perceived Trust}); prior \ai usage and age were not (all $p>.37$). Education was associated with \textit{Perceived Understanding} level ($r=0.25$, $p=.005$) but not \textit{Perceived Trust} ($p=.18$). Whether these variables also moderate the \emph{effect} of explanations is separate and reported in Sections~\ref{sec:results_education} and~\ref{sec:results_ai_experience}.

\subsection{Scenario Descriptions}
\label{app:scenario_detail}

\textbf{\ScenOne~(Job Prediction)} used a feature attribution heatmap for text classification, where word-level scores indicate which biography terms predict an occupation \cite{schoefferExplanationsFairnessAppropriate2024}. \textbf{\ScenTwo~(Credit Risk)} used a variable importance plot \cite{friedmanGreedyFunctionApproximation2001} ranking how strongly each attribute contributes to credit default predictions overall \cite{schmittAccuracyArtificialIntelligence2023}. \textbf{\ScenThree~(Heart Attack)} used \limeFull \cite{salihPerspectiveExplainableArtificial2025} bar charts of clinical factors (e.g., cholesterol, hypertension, BMI) contributing to predicted heart-attack risk. \textbf{\ScenFour~(Bike Rental)} used \pdpsFull \cite{kruschelChallengingPerformanceInterpretabilityTradeOff2025} showing predicted rental demand against temperature, humidity, and wind speed. \textbf{\ScenFive~(Mortality Risk)} used a single-instance \shap attribution diagram \cite{lundbergLocalExplanationsGlobal2020} in which one patient's attributes (age, BMI, blood pressure, sex) shift their mortality risk score. \textbf{\ScenSix~(Quality Driver)} used a \shap beeswarm plot \cite{senonerUsingExplainableArtificial2022} of feature effects across manufacturing quality predictions. \textbf{\ScenSeven~(Product Defect)} used a saliency-based anomaly heatmap \cite{senonerExplainableAIImproves2024a} highlighting image regions behind a detected defect.

We intentionally did not implement a full factorial design over use cases, \xai methods, and modalities: in practice certain methods are tied to specific model types and modalities (e.g., saliency maps for images, \shap/\lime for tabular data), so not all combinations would be realistic, and a full factorial would have substantially increased participant burden and fatigue.

\subsection{Model Specification and Estimation Detail}
\label{app:model_spec_detail}

Restricted maximum likelihood is used only to estimate single models; because likelihood-ratio tests compare models with different fixed-effect structures, for which restricted maximum likelihood is not comparable across models, every likelihood-ratio test in this paper refits both models by full maximum likelihood (\texttt{reml=False}). For the carryover analysis (Appendix~\ref{app:robustness}), the \textit{Explanation}~$\times$~\textit{Sequence} interaction is a linear combination of terms already in Eq.~\eqref{eq:lmm}, so \textit{Block} is omitted from that interaction model; all other contrasts come from Eq.~\eqref{eq:lmm} without an interaction term. Paired $t$-tests and Wilcoxon signed-rank tests used \texttt{scipy.stats} (SciPy v1.16.2) \cite{virtanenSciPyFundamentalAlgorithms2020} on within-subject aggregated scores (Appendix~\ref{app:paired_tests}).

For the mediation decomposition (Section~\ref{sec:results_h3}): the change in \textit{Perceived Trust} is regressed on the change in \textit{Perceived Understanding} together with the participant's mean-centered average \textit{Perceived Understanding} level, which guards $b$ against confounding by overall rating level; this level term is not distinguishable from zero ($d = 0.02$, 95\% CI [$-0.11$, $0.15$] primary; $d=-0.01$, 95\% CI [$-0.16$, $0.13$] attention-verified), so including it leaves the decomposition unchanged. The proportion of the total effect running through the indirect path follows from $c = c' + a\cdot b$ by construction and therefore describes the estimated decomposition rather than supplying independent evidence for it.

\subsection{Exploratory Moderator Detail}
\label{app:exploratory_detail}

The explanation benefit was positive across all \xai visualization types (feature attribution heatmaps, variable importance plots, \lime bar charts, \pdpsLong, single-instance \shap attributions, \shap beeswarm plots, saliency-based anomaly heatmaps), with per-method \textit{Perceived Understanding} gains from $\Delta = 0.189$ to $0.528$, suggesting the benefit is not confined to one \xai format (disentangling method from task would require applying multiple methods to the same scenario).

By data modality (Figure~\ref{fig:modality}, exploratory), the largest descriptive gain was for the single image scenario (\textit{Perceived Understanding} $\Delta=0.528$, \textit{Perceived Trust} $\Delta=0.402$), followed by tabular ($\Delta=0.348$, $0.249$) and the single text scenario ($\Delta=0.189$, $0.142$); the \textit{Explanation}~$\times$~\textit{Modality} interaction is not significant for either outcome ($\chi^2$(2) = 4.61, $p=.100$; $\chi^2$(2) = 3.69, $p=.158$). With only one text and one image scenario against five tabular ones, each modality estimate is inseparable from that scenario's idiosyncrasies, so we report these descriptively only.

On page interaction time (Section~\ref{sec:results_time}): per-scenario medians move in both directions ($-29.3\%$ in \ScenOne~Job Prediction to $+15.7\%$ in \ScenThree~Heart Attack), so added reading time is not a uniform cost. Participants gaining more \textit{Perceived Understanding} spent, if anything, slightly less additional time ($r=-0.18$, $p=.038$); the time change was unrelated to the \textit{Perceived Trust} change ($r=-0.03$, $p=.72$).

\section{GUIDE-LLM Reporting Checklist}
\label{app:guide_llm}

Tables~\ref{tab:guide_llm_core} and~\ref{tab:guide_llm_optional} report this study against the GUIDE-LLM checklist \cite{feuerriegelReportingChecklistLarge2026}. \llms were used in one bounded role: generating the textual explanations used as experimental stimuli. Scenario selection, survey construction, participant recruitment, and statistical analysis involved no \llm component.

\begin{table}
\centering
\footnotesize
\renewcommand{\arraystretch}{0.95}
\caption{GUIDE-LLM checklist, core items.}
\label{tab:guide_llm_core}
\begin{tabular}{p{0.24\linewidth}p{0.70\linewidth}}
\toprule
\textbf{Item} & \textbf{Response} \\
\midrule
A.1 Scope of \llm use & Stimulus generation: explanations of \xai visualizations (Section~\ref{sec:method_explanations}, Figure~\ref{fig:prompt}) using GPT-4o (validation, N=63) and GPT-5.6-sol (Main pre-registered, N=128). LLM judges graded candidate generations against ground-truth keys for model selection. Minor LLM assistance with language editing; all content authored and verified by authors. \\
A.2 Degree of automation & Human-in-the-loop. Generated explanations were read and compared against expert-written explanations by two of them before use as stimuli. Participants never interacted with the \llm. \\
B.1 Model name and version & Validation: GPT-4o (OpenAI), \texttt{chatgpt-4o-latest}, March 27, 2025 snapshot; accessed March--June 2025. Main: GPT-5.6-sol (OpenAI), \texttt{openai/gpt-5.6-sol}; accessed August 2026. Model size is not disclosed by the provider for either model. \\
B.2 Access and context mode & Validation: ChatGPT web interface, stateless, so no context carried over between explanations. Main: OpenRouter API, one request per generation, likewise stateless. In both studies each submission comprised the prompt text, the visualization image, and a text description of the documented model setting (\ml method, data, reported performance metrics) drawn from the source publication; no file was attached. \\
B.3 Configuration & Provider defaults; no sampling parameters modified. Validation: one generation per visualization. Main analysis: all stimuli regenerated under fixed configuration. Validation analysis (ChatGPT web): no exposed settings. Main generation (OpenRouter API): 8,000 token max only. \\
B.4 Customization & Base model. No fine-tuning, retrieval augmentation, automated prompt optimization, tool or function calling, web search, or agentic workflow. Model context was supplied as prompt text; no file attachment or retrieval system was involved. \\
B.5 Persistent memory & No. Memory was deactivated for all sessions. \\
C.1 Exact prompts & Yes. The stimulus-generation prompt is shown in Figure~\ref{fig:prompt} and described in Section~\ref{sec:method_explanations}. The prompt text was identical across all seven scenarios; scenario-specific information was supplied separately, not written into the prompt. The opening sentence is stated generically (``Given an \xai visualization''), which in the original sessions named a visualization type by way of example. \\
C.2 System-wide instructions & \emph{Validation:} the ChatGPT web interface, whose provider-default system prompt is not exposed to the user. \emph{Main explanations:} the OpenRouter API, to which our instructions were sent as a single user-role message with no system-role message of our own; any upstream provider- or route-side instruction is not exposed and cannot be ruled out. In both studies our own instructions were supplied as prompt content. \\
D.1 Personal or sensitive data & None processed by the \llm. All seven scenario visualizations are published figures from the studies in Table~\ref{tab:scenarios} with no identifiable records, prepared before participant recruitment. Participant data handling is described in Section~\ref{sec:ethics}. \\
E.1 Human validation & Yes. Three prompt configurations compared against expert explanations (accuracy, completeness, readability) by main author and scored by two LLM judges against round-truth explanation from experts; model with fewest errors selected. \\
E.2 Post-processing & None. Generated explanations were shown to participants verbatim, directly beneath the corresponding \xai visualization; no trimming, reformatting, or editing was applied. \\
F.1 Materials shared & Yes. Code and data are provided for review via the repository given in the Introduction's contribution footnote; the prompt is reproduced in Figure~\ref{fig:prompt}. \\
G.1 Competing interests & No funding or collaboration with LLM developers. One author employed by a technology company with AI interests; institutional review by internal board. Model access via standard commercial subscription, no in-kind support. \\
\bottomrule
\end{tabular}
\end{table}
\Description{A GUIDE-LLM core reporting checklist table with item codes and responses covering LLM governance and transparency. Column headers: Item, Response. Eight rows covering: A.1 Scope of LLM use (stimulus generation for explanations of XAI visualizations using GPT-4o for validation and GPT-5.6-sol for main study), A.2 Degree of automation (human-in-the-loop; explanations read and compared to expert references before use), B.1 Model name and version (GPT-4o March 2025, GPT-5.6-sol August 2026), B.2 Access and context mode (stateless, no persistent context, with model/task details supplied per submission), B.3 Configuration (provider defaults, no parameter modification except max token limit), B.4 Customization (base model only, no fine-tuning or retrieval augmentation), B.5 Persistent memory (none; deactivated), C.1 Exact prompts (yes, shown in Figure; identical across scenarios), C.2 System-wide instructions (ChatGPT defaults not exposed; OpenRouter with no custom system-role message), D.1 Personal or sensitive data (none processed by LLM), E.1 Human validation (yes; three prompts compared against expert explanations), E.2 Post-processing (none; explanations shown verbatim), F.1 Materials shared (yes; code, data, and prompt reproduced), G.1 Competing interests (no LLM developer funding or collaboration; one author employed at tech company with AI interests; institutional review; standard commercial access).}

\begin{table}[t]
\centering
\footnotesize
\caption{GUIDE-LLM checklist, optional items.}
\label{tab:guide_llm_optional}
\begin{tabular}{p{0.24\linewidth}p{0.70\linewidth}}
\toprule
\textbf{Item} & \textbf{Response} \\
\midrule
Justification for the \llm choice & Validation: GPT-4o for pilot and SOTA at the time. Main: GPT-5.6-sol selected based on rankig and from six candidates by lowest error rate on ground-truth keys. Both closed-weight; exact checkpoints not archivable (noted in limitations).\\
Rationale for the prompt design & Yes. The prompt uses a structured format (task description, required answer sections, output constraints) following a context-engineering logic (Section~\ref{sec:method_explanations}, Figure~\ref{fig:prompt}). \\
Comparison against other methods or \llms & Both. Three prompt configurations evaluated against expert explanations. Validation: Gemini tried informally; GPT-4o retained for accessibility. Main: six models compared on all seven visualizations, graded against ground-truth answers. \\
Training data leakage & Not directly applicable to the participant outcome evaluation, since outcomes are participant self-reports rather than benchmark data. Memorization of the publicly available source publications by the generating model cannot be ruled out and was not tested. \\
Risk of bias in \llm behavior & \xai visualizations can themselves be biased and that LLM-generated explanations may increase such bias, including through hallucination. No subgroup analysis of \llm behavior was conducted; participant-level moderation analyses are reported in the exploratory analyses. \\
Conversation transcripts & N/A. Participants did not interact with the \llm. A representative generated explanation, as shown to participants, is reproduced in Figure~\ref{fig:explanationexample}. \\
Ethical implications & Reported in Section~\ref{sec:ethics}: institutional review, informed consent, absence of deception, compensation, and GDPR-compliant data handling. \\
Computational resources & No dedicated infrastructure. Validation: ~20--30 ChatGPT web requests (7 scenarios × 3 prompts + informal comparisons). Main: 42 OpenRouter API generations (6 models × 7 visualizations) + judge passes; grading dominated costs. \\
\bottomrule
\end{tabular}
\end{table}
\Description{A GUIDE-LLM optional reporting checklist table providing extended transparency items. Column headers: Item, Response. Seven rows covering: Justification for LLM choice (GPT-4o for validation as SOTA at time, GPT-5.6-sol for main selected via lowest error rate on ground-truth keys from six candidates), Rationale for prompt design (structured format with task description, required sections, output constraints, following context-engineering logic), Comparison against other methods or LLMs (three prompt variants evaluated against expert references; Gemini tried informally; six models compared on all visualizations in main study), Training data leakage (not directly applicable to participant self-reports; memorization of published sources by LLM not tested), Risk of bias in LLM behavior (XAI visualizations themselves can be biased; no subgroup analysis of LLM behavior conducted), Conversation transcripts (not applicable; participants did not interact with LLM; representative example shown in Figure), Ethical implications (institutional review, informed consent, no deception, compensation, GDPR compliance), Computational resources (no dedicated infrastructure; validation ~20-30 ChatGPT requests; main 42 OpenRouter generations plus judge passes; grading dominated costs).}

\end{appendices}

\end{document}